\documentclass{nws}
\usepackage[hidelinks]{hyperref}
\usepackage{multirow}
\usepackage{multicol}
\usepackage{float}
\usepackage{graphicx}
\usepackage[framemethod=TikZ]{mdframed}
\usepackage{subcaption}
\usepackage{amssymb}
\makeatletter
\newenvironment{ProblemSpecBox}[1]{
  \protected@edef\@currentlabelname{#1}
  \protected@edef\@currentlabel{#1}
  \begin{mdframed}[
    innerlinewidth=0.5pt,
    innerleftmargin=10pt,
    innerrightmargin=10pt,
    innertopmargin = 10pt,
    innerbottommargin=10pt,
    skipabove=\dimexpr\topsep+\ht\strutbox\relax,
    frametitle={#1},
    frametitlerule=true,
    frametitlerulewidth=1pt]
}{
  \end{mdframed}
}
\makeatother

\title[Community structure evaluation]
{Community Structure: A Comparative Evaluation of Community Detection Methods}

 \author[V.L.~Dao, C.~Bothorel and P.~Lenca]
       {VINH LOC DAO, CECILE BOTHOREL and PHILIPPE LENCA \\
       IMT Atlantique, \\
        Lab-STICC CNRS UMR 6285\\
        F-29238, Brest, France\\
         \email{\{vinh.dao, cecile.bothorel, philippe.lenca\}@imt-atlantique.fr}}

\jdate{Nov 2019}
\pubyear{2019}
\pagerange{\pageref{firstpage}--\pageref{lastpage}}
\doi{}

\begin{document}

\label{firstpage}

\maketitle
%\maketitle[t] %%for Editorial
%\maketitle[T]  %% for {\sf EDITORIAL}
\begin{abstract}
	Discovering community structure in complex networks is a mature field since a tremendous number of community detection methods have been introduced in the literature. Nevertheless, it is still very challenging for practioners to determine which method would be suitable to get insights into the structural information of the networks they study. Many recent efforts have been devoted to investigating various quality scores of the community structure, but the problem of distinguishing between different types of communities is still open. In this paper, we propose a comparative, extensive and empirical study to investigate what types of communities many state-of-the-art and well-known community detection methods are producing. Specifically, we provide comprehensive analyses on computation time, community size distribution, a comparative evaluation of methods according to their optimisation schemes as well as a comparison of their partioning strategy through validation metrics. We process our analyses on a very large corpus of hundreds of networks from five different network categories and propose ways to classify community detection methods, helping a potential user to navigate the complex landscape of community detection. 
	
	\vspace{12pt}
	
\end{abstract}

\tableofcontents

\section{Introduction}

In network science, \textit{community detection} (sometimes called \textit{graph clustering})
%\footnote{The concept of graph clustering might refer to two different meanings existing in the literature. The first one implies a categorization of many graphs into different sets within which graphs share a common similar feature. The second one relates to the problem of partitioning nodes of a graph into densely connected groups. Here we means graph clustering in the latter case.} 
is one of the fundamental challenges to discovering the structure of networks on the mesoscopic level. However, it is an ill-defined problem. According to Arifin  \textit{et al.}  \cite{arifin:2017a}, \textit{``it does not have clear goals, solution paths or expected solution''}. There is no universal definition or closed form formula of what kind of objects one should be looking for~\cite{fortunato:2016a}, and consequently there is no golden standard to assess the quality of a community structure and the performance of a community detection algorithm. 

The most frequently found definition of community in the network science literature is derived from the mechanism of connection preference. It implies that \textit{a community is a group of nodes (a subgraph) in a graph where there must be more edges (denser) connecting them together than edges connecting the community with the rest of the graph}~\cite{radicchi:2004a,fortunato:2010a}. Newman defines a community as a \textit{``group of vertices with a higher-than-average density of edges connecting them"} \cite{newman:2006a}. Depending on the context, a community may be called a \textit{cluster}, a \textit{module}, a \textit{class} or a \textit{modular group}. This  is the most basic definition that sets the fundamental requirement for most of its derivative definitions. Many different variations of community could be found in~\cite{wasserman:1994a}, for instance, \textit{LS-set}, which is a set of nodes in a network such that each of its proper subsets has more ties to its complement within the set than outside; or \textit{k-core}, which is a subgraph in which each node is adjacent to at least a minimum number $k$ of the other nodes in the subgraph. However, in recent developments of community detection algorithms, there is no consensus of the quantity of edges in reality that could be considered as \textit{``many"}. Communities are just algorithmically defined, i.e. they are final products of the algorithm without any precise a priori definition~\cite{fortunato:2010a}. 

In practice, there are even more constraints, which are sometimes not explicitly expressed, than which appeared in the announcement of the problem. If one only looks for a partition of a graph that maximizes the number of internal edges and minimizes the number of external edges, then the graph itself can be considered as a big community with no external connection. Another solution is to leave the node having the smallest degree of a graph in one community, and all other nodes in another community. This solution could also maximize the ratio between external and internal edges. However, these monotonous solutions do not seem to seduce most (if not all) analysts considering using a community detection algorithm to detect communities. In fact, it is preferable to cluster a network into groups of relatively similar size
%\footnote{Community detection is identified in the research community as the search for \textit{natural groups} in networks without a given number of clusters. When the number and the size of clusters are specified, the problem is often referred as \textit{graph partitioning} or \textit{graph bisection} for a division into only two clusters.}
~\cite{newman:2010a}. This means that somehow, the relative size of communities is significant but this notion of size has not been explicitly announced. Besides, there are many other criteria that could be mentioned such as community complete mutuality, reachability, vertex degree distribution and the comparison of internal versus external cohesion~\cite{wasserman:1994a,fortunato:2010a}. 
There exists a subtle compromise between adding into a community new vertices---along with their edges--- and conserving the common property that defines the group. 
In fact, different community detection methods have different ways to define what defines the groups and how to consider those constraints or not. They produce different community structures. The following are the main reasons that could lead to these disagreements between detection methods:

\begin{itemize}
\item Different algorithms may implement different meanings of the notion of community.%; the discovered structure strongly depends on the assumptions made about expected community structure.
\item When two algorithms define the same concept of community, it may also mathematically and algorithmically be formalized in different ways (same objective but different objective functions) and hence leads us to different results.
\item Even when two algorithms have exactly the same objective function, the algorithmic mechanism they employ to find communities also decides what they are going to find, especially in heuristic searching approaches.
\item Initial configuration is also another important factor that affects the final result of an algorithm, many community detection methods are not deterministic.
\item Each method may include a consideration between obtaining an optimized solution and providing reasonable performances (in terms of calculation time, memory consumption, etc.). This trade-off may be considered differently across the methods.
\item Some algorithms are variable according to input data and will prove more or less efficient on some kinds of inputs than on others. 
\item Variations due to implementation factors could also impact the final result of an algorithm.
\item Finally, in some algorithms, there are tie-break situations where they have to choose randomly without any factor related to their final objectives. It may also significantly affect the result that one would get if the tie-break problems have been resolved in a different way. 
\end{itemize}

Due to the many reasons stated above, choosing the community detection method that corresponds well to a particular scenario or to an expectation of quality is not straightforward. This paper is not the first effort as a guide for choosing community detection methods. For the same purpose, readers can also refer to \cite{fortunato:2016a}, \cite{ghasemian:2018a}, \cite{jebabli2015overlapping}, and \cite{jebabli:2018a}. The difference with existing work is that we provide an exhaustive empirical evaluation on many state-of-the-art methods, using a large network dataset (more than 100 different real networks, of which only a few are synthetic) and using popular quality scoring functions of structural characteristics of communities. Hence, we are able to disclose a performance review of many different methods based solely on empirical experiments and measurements that help network practitioners to reveal the functionality of each method, and behaviours that they could exhibit in real world networks, independently of theoretical mechanisms. The objective of this work does not differ from existing work in the literature. It does not aim to bring readers an innovative factor in the community detection problem nor discuss every related challenging point of the community detection problem. However it provides an evaluation as well as an approach to examine community detection algorithms. This evaluation is very useful for network practitioners to obtain a quick notion of community structure quality before actually applying a method to their concrete dataset. These notions of quality are expected to guide users to appropriate choices of methods according to their quality criteria.

The paper is organized as follows: Section~\ref{sec:detectionmethods} introduces some popular and state-of-the-art community detection methods that will be analyzed in this paper as well as the benchmarking dataset employed in our experiments. Then, Section~\ref{sec:preliminaryanalayis} presents analyses of the most essential aspects of community detection performance including computation time, community size distribution as well as numbers of detected communities by each method. In Section~\ref{sec:goodnesssimilarity}, we address different structural quality aspects of community structures. This is followed by a comparative evaluation using many popular clustering validation metrics in Section~\ref{sec:validationproximity}, which are widely used in the context of community detection. The Sections~\ref{sec:preliminaryanalayis} to \ref{sec:validationproximity} are independent. They bring complementary material and criteria to describe the quality of community structures and compare the different methods from different perspectives. Each section is detailed, making this article a little long, but they end with a summary that allows readers  to get an overview and provide them with the necessary key lessons to be learned. Finally, we present some close results of related work that can be found in the literature in Section~\ref{sec:relatedwork} and conclude our study with some discussions and recommendations in Section~\ref{sec:conclusion}.

\section{Community detection methods and dataset}
\label{sec:detectionmethods}

\subsection{Community detection methods}
We present, in this section, some popular community detection methods that have been widely used and discussed in the literature. Note that in recent years, there are a large number of innovative methods which have been proposed to solve either generic or specific cases. However, an empirical and exhaustive analysis of all methods would be impractical, even unrealizable. To the best of our knowledge, we are trying to include the most important and representative methods among the community detection approaches.   

There are many possible theoretical taxonomies for community detection methods, depending on the final objective of each categorization. For instance, one could classify methods according to differences in searching mechanisms, objective functions, assumptions about the structure to be found, expected qualities, hypothesis models, or even the theoretical model employed, etc. Moreover, what makes the problem trickier is that many methods are not just some simple algorithms to resolve a specific problem, but instead are combinations of many different approaches in order to leverage as much algorithmic power as possible from each one. There is not a consensus on how different methods are similar and how they can be classified into different families whose functionality can be resumed in some simple words. Porter \textit{et al.}  use centrality based, local techniques, \textit{modularity} optimization,
%\footnote{First introduced by \cite{newman:2004a} to assess hierarchical clustering levels of a community detection algorithm, \textit{modularity} has became the most popular objective function in the context of community detection.}
 spectral clustering to describe communities in networks \cite{porter:2009a}. In \cite{fortunato:2010a,fortunato:2016a}, the authors group community detection methods into traditional data clustering methods, divisive algorithms, modularity-based methods, spectral algorithms, dynamic algorithms and methods based on statistical inference. Coscia \textit{et al.} classify community discovering into feature distance based, internal density, bridge detection, diffusion process, closeness based, structural pattern based, link clustering, meta clustering \cite{coscia:2011a}. In a context of Social Media, Papadopulos \textit{et al.} compare methods in substructure detection, vertex clustering, community quality optimization, divisive and model-based approaches \cite{papadopoulos:2011a}. Bohlin \textit{et al.} aggregate different approaches into three principle classes representing different network models: null models, block models and flow models \cite{bohlin:2014a}. Schaub \textit{et al.} classify methods into four perspectives: cut based, clustering internal density based, stochastic equivalent based and dynamic based showing four different facets of community structure \cite{schaub:2017a}. Finally, Ghasemian \textit{et al.} adopt an experimental classification \cite{ghasemian:2018a} as well as group community detection methods in distinct families based on their outputs on many real-world networks using a validation metric (a topic we will address in Section \ref{sec:similarityevaluation}).

In the following sections, we choose to classify community detection methods according to different theoretical approaches including edge removal, modularity optimization, spectral partitioning, dynamic process and statistical inference. Although every theoretical taxonomy may be questionable, this categorization is expected to support the empirical analyses in the next sessions to verify whether theoretical and conceptual closeness could engender quality closeness in practice.

\subsubsection{Edge removal based methods}

\textbf{Edge betweenness} (\textit{GN}) \cite{newman:2004a} detects communities by removing edges progressively according to their betweenness centrality scores. This method is based on the intuition that dense areas of a graph are loosely connected by a few edges that are located in the shortest paths between pairs of nodes. Removing these edges would reveal densely connected communities.\\
\textbf{Edge clustering coefficient} (\textit{RCCLP})  \cite{radicchi:2004a} suggests replacing the edge betweenness centrality of Girvan-Newman's method by an edge clustering coefficient, which requires less computation time and hence reduces the complexity. In this paper, we analyze two configurations of this method corresponding to triangular ($g=3$ denoted by \textit{RCCLP-3}) and quadrangular ($g=4$ denoted by \textit{RCCLP-4}) versions.   

\subsubsection{Modularity optimization methods}
\textbf{Greedy optimization} (\textit{CNM}) \cite{clauset:2004a} greedily maximizes the modularity function $Q$ by aggregating iteratively connected communities which induces a maximum increase or minimum decrease in modularity $\Delta Q$. \\
\textbf{Louvain method} \cite{blondel:2008a} adopts a two-step agglomerative process similar to the greedy optimization method. However, in each iteration of the first step, it allows nodes to move between communities until no additional gain in modularity can be obtained with these local switches. Then, a new graph whose vertices are the communities resulting from the first step is built and the process is repeated on the new graph to reduce computation time, leading to a hierarchical clustering. \\
\textbf{Spectral method} (\textit{SN})  \cite{newman:2006b} reformulates the idea of maximizing modularity as a spectral partitioning problem by constructing a \textit{modularity matrix} and using its leading eigenvector to spectrally partition networks into sub-networks. The community structure is identified using the eigenvectors of this matrix. Eigenvectors are used to project each node into low-dimensional node vectors, and the community structure is identified through clustering the node vectors (e.g. with the k-means  clustering method).\\
%identifies community structure by finding leading eigenvectors corresponding to largest eigenvalues of a modularity matrix. In this method,  the  problem  of  modularity  optimization  is \textit{translated} to  the  problem  of  vector partitioning of modularity matrix. \\

\subsubsection{Dynamic process based methods}
\textbf{Walktrap} \cite{pons:2005a} defines a pairwise \textit{dynamic distance} between nodes and then applies traditional hierarchical clustering to detect community structure. 
%The distance is formulated using the transition probability of a random walker based on the concept that nodes belonging to the same community tend to \textit{"see"} other nodes in the same way. 
The  distance  between  two  nodes  is  defined  in  terms  of a random walk process. The basic idea is that if two nodes are in the same community, they tend to \textit{"see"} other nodes in the same way, i.e. the probability of getting to a third node through a random  walk  should  not  be  very  different  for  these two nodes, with higher distance for nodes belonging to other communities.\\
\textbf{Infomod} \cite{rosvall:2007a} uses an information theoretic model where a \textit{signaler} tries to send the structure of a network over a limited capacity transmission channel to a \textit{receiver}. The network must be encoded in community structure in a way that minimizes the transferred information and the information loss. \\
\textbf{Infomap} \cite{rosvall:2009a} represents networks by a two-level structure description. Analogically, each node in a network is encrypted by a unique codeword composed of two parts: a prefix representing the community to which it belongs and a suffix representing the local code. Detecting community structure becomes equivalent to searching for the coding rule to minimize the average code length describing random walks on the network. 

\subsubsection{Statistical inference based methods}
\textbf{Stochastic Block Model} (\textit{SBM})  \cite{riolo:2017a}. The Stochastic Block Model was introduced by Holland \textit{et al.} \cite{holland1983stochastic}. Here we use the implementation by Riolo \textit{et al.} which uses a Monte Carlo sampling scheme to maximize a Bayesian posterior probability distribution over possible divisions of the network into communities. This probability implies an expected network model to be fitted from the observed network data. In this block model variant, the authors employ a new prior on the number of communities based on a queueing-type mechanism to calculate posterior probability. We analyze, in the following sections, both traditional $SBM$ and the \textit{degree-corrected} version \textit{DCSBM}, which had been proven to perform better in practice.\\ 
\textbf{Order statistics local optimization} (\textit{OSLOM}) \cite{lancichinetti:2011a} measures the statistical significance of a community by calculating the probability of finding a similar one in a null model. Following this concept, nodes are gradually aggregated into communities to find significant communities. Then nodes are considered to be swapped between communities in order to increase significance level. 

\subsubsection{Other methods}
\textbf{Spin glass model} (\textit{RB}) \cite{reichardt:2006a} is a method  relying  on  an  analogy  between  the  statistical  mechanics  of  complex  networks and physical spin glass models. It finds communities by fitting the ground state of a spin glass model. Instead of favoring only intra-community edges and penalizing inter-community edges like the traditional modularity measure, this model also favors inter-community non edges and penalizes intra-community non-edges. \\
\textbf{Label propagation} (\textit{LPA}) \cite{raghavan:2007a} exploits the topology of networks to infer community structure. It is closely related to the context of message passing paradigms or epidemic spreading. The principled idea of this method is based on the concept that nodes should belong to the community of most of their neighbors. Hence, they gradually update their memberships according to their incident nodes. \\
\textbf{Speaker-listener label propagation} (\textit{SLPA}) by Xie and Szymanski \cite{xie:2012a} modifies the propagation mechanism above by a new label update strategy. Also, instead of keeping only hard membership information, each node is equipped by a memory to contain the labels that it receives. Then, in the update phase, nodes transmit the membership to their neighbors according to the membership frequency in the memories. \\
\textbf{Mixing global and local information} (\textit{Conclude}) \cite{demeo:2014a} combines a dynamic distance with a modularity optimization process to identify community structure. Firstly, the authors define a new pairwise proximity function using random and non backtracking walks of finite length to determine distances between vertices. Then, the multi-level modularity optimization strategy of \textit{Louvain} method \cite{blondel:2008a} is combined with the defined distance to find community structure. \\

\begin{table}
%\centering
\caption{Community detection methods involved in the study.}
\label{tab:implementation}
    \begin{minipage}{\textwidth}
\begin{tabular}{lllll}
\hline\hline
Approach \hspace{20pt} & Publication \hspace{75pt} & Ref. label \hspace{10pt} & Time order \hspace{20pt} & Implementation\\
\hline\hline
\multirow{2}{*}{\shortstack{Edge \\ removal}} 
& \cite{girvan:2002a} & GN & $\mathcal{O}(nm^2)$ & igraph\footnote{Published at \url{http://igraph.org/} } \\ %\cline{2-4}                     
& \cite{radicchi:2004a} & RCCLP & $\mathcal{O}(m^4/n^2)$ &  Authors\footnote{Published at \url{http://homes.sice.indiana.edu/filiradi/resources.html}} \\ %\cline{1-4}
\noalign{\vspace {.5cm}}
\multirow{3}{*}{\shortstack{Modularity \\ optimization}}     
& \cite{clauset:2004a} & CNM & $\mathcal{O}(m\log^2(n))$ & igraph \\ %\cline{2-4}                     
& \cite{blondel:2008a} & Louvain & $\mathcal{O}(n\log (n))$ & Authors\footnote{Published at \url{https://sourceforge.net/projects/louvain/}} \\ %\cline{2-4}
& \cite{newman:2006b} & SN & $\mathcal{O}(nm\log (n))$ & igraph \\ %\cline{1-4}
\noalign{\vspace {.5cm}}
\multirow{3}{*}{\shortstack{Dynamic \\ process}} 
& \cite{pons:2005a} & Walktrap & $\mathcal{O}(n)$ & igraph\\ %\cline{2-4}   
& \cite{rosvall:2007a} & Infomod & NA & Authors\footnote{Published at \url{http://www.tp.umu.se/~rosvall/code.html}} \\ %\cline{2-4}   
& \cite{rosvall:2009a} & Infomap & $\mathcal{O}(m)$ & Authors\footnote{Published at \url{http://www.mapequation.org/}}/igraph \\ %\cline{1-4}  
\noalign{\vspace {.5cm}}
\multirow{2}{*}{\shortstack{Statistical \\ inference}} 
& \cite{lancichinetti:2011a} & Oslom & $\mathcal{O}(n^2)$ & Authors\footnote{Published at \url{http://www.oslom.org/}} \\ %\cline{2-4}  
& \cite{riolo:2017a} & (DC)SBM & Parametric & Authors\footnote{Published at \url{http://www-personal.umich.edu/~mejn/}} \\ %\cline{1-4}  
\noalign{\vspace {.5cm}}
\multirow{4}{*}{\shortstack{Other \\ methods}} 
& \cite{reichardt:2006a} & RB & $\mathcal{O}(n^2log (n))$ & igraph\\ %\cline{2-4} 
& \cite{raghavan:2007a} & LPA & $\mathcal{O}(m)$ & igraph\\ %\cline{2-4} 
& \cite{xie:2012a} & SLPA & $\mathcal{O}(m)$ & Authors\footnote{Published at \url{https://sites.google.com/site/communitydetectionslpa/}} \\ %\cline{2-4} 
& \cite{demeo:2014a} & Conclude & $\mathcal{O}(n+m)$ & Authors\footnote{Published at \url{http://www.emilio.ferrara.name/code/conclude/}} \\ %\cline{1-4} 
\hline\hline
\end{tabular}
\vspace{-2\baselineskip}
\end{minipage}
\end{table}

Table \ref{tab:implementation} summarizes the methods presented previously, grouped by different approaches. Since community detection is receiving more and more attention in the network science community, there is a huge volume of work that has been published in recent years to evaluate different methods including both theoretical and empirical approaches. However, there is no formal and quantitative definition of community that is explicitly implemented inside algorithms. Therefore it is challenging to distinguish the topological differences of community structures using different methods, even when the associated concepts are quite theoretically discernible. Additionally, it is still not clear whether proximity in the assumption of community concept will engender a structural similarity of communities that could be detected. Our comparative analysis in the next sections will try to address these questions in more detail. 

\subsection{Experimental dataset}
\label{sec:categorizeddataset}
In this section, we describe some statistical properties of networks that will be included in the following analysis. 
%It is expected that networks in each category are spread in a wide range of structural measures. However, a
Available biological networks that have been published and analyzed widely are relatively small in comparison to the other networks of the other families. Besides, due to the complexity of the analysis process, we limit the domains of interest to 5 categories which are commonly researched and where numerous networks are available. We introduce a 6th category with various types of networks, each of which is under-represented, such as ecological networks for example. In this study, we consider 108 different networks, which is relatively large in comparison to many other studies. Many notable related works where some of these networks are also employed could be mentioned for quick reference: Orman \textit{et al.} use 6 networks to evaluate the structure of communities discovered by several detection techniques~\cite{orman:2012a}; Lancichinetti \textit{et al.} use 15 networks to characterize structural communities~\cite{lancichinetti:2010a}; Hric \textit{et al.} use 16 networks to reveal differences between structural communities and ground truth \cite{hric:2014a}; Leskovec \textit{et al.} use over 100 networks to analyze network community profile \cite{leskovec:2008a} and 230 networks to evaluate the goodness of ground-truth communities in social networks. Within this number, 225 samples of the Ning online social networking platform networks\footnote{https://www.ning.com/} are aggregated \cite{yang:2013a}. Let us mention, finally, the related work by Ghasemian \textit{et al.} which has introduced the large CommunityFitNet corpus\footnote{https://github.com/AGhasemian/CommunityFitNet} containing 572 real-world  networks. Table~\ref{tab:dataset} summarizes the composition of networks that have been analyzed in this section. 

\begin{table}
\centering
\caption[A summary of the dataset used in our analysis.]{A summary of network dataset used in this analysis where ``Size'' is the number of networks analyzed in each category, ``Nodes'' and ``Edges'' indicate the average number of nodes and edges of networks in each category respectively. $^*$The last row shows the total number of networks in the whole dataset. This dataset is collected from several sources including: \url{http://networkrepository.com} \protect{\cite{rossi:2015a}}, \url{http://konect.uni-koblenz.de}~\protect{\cite{jerome:2013a}}, \url{http://snap.stanford.edu}~\protect{\cite{snapnets}}}.
\label{tab:dataset}
\begin{tabular}[1.0\textwidth]{lllll}
\hline\hline
Category \hspace{10pt} & Size \hspace{10pt} & Nodes \hspace{10pt} & Edges \hspace{15pt} & Notable networks \\
\hline\hline
Biological & 7 & 1860 & 10763 & Yeast, brain, protein-protein interactions\\
Communication \hspace{5pt} & 9 & 39595 & 195032 & Email, forums, message exchanges\\
Information & 25 & 38358 & 159812 & Amazon, DBLP, citation $\&$ education webs\\
Social & 37 & 6888 & 49666 & Facebook, Youtube, Google Plus networks\\
Technological & 19 & 18431 & 48494 & Internet, AS Caida, Gnutella P2P networks\\
Miscellaneous & 11 & 4298 & 49033 & Ecology, power-grid, synthetic networks\\
\hline
Total$^*$ & 108 & & & \\
\hline\hline
\end{tabular}
\end{table}

\begin{figure}
\centering
\includegraphics[width=1\linewidth]{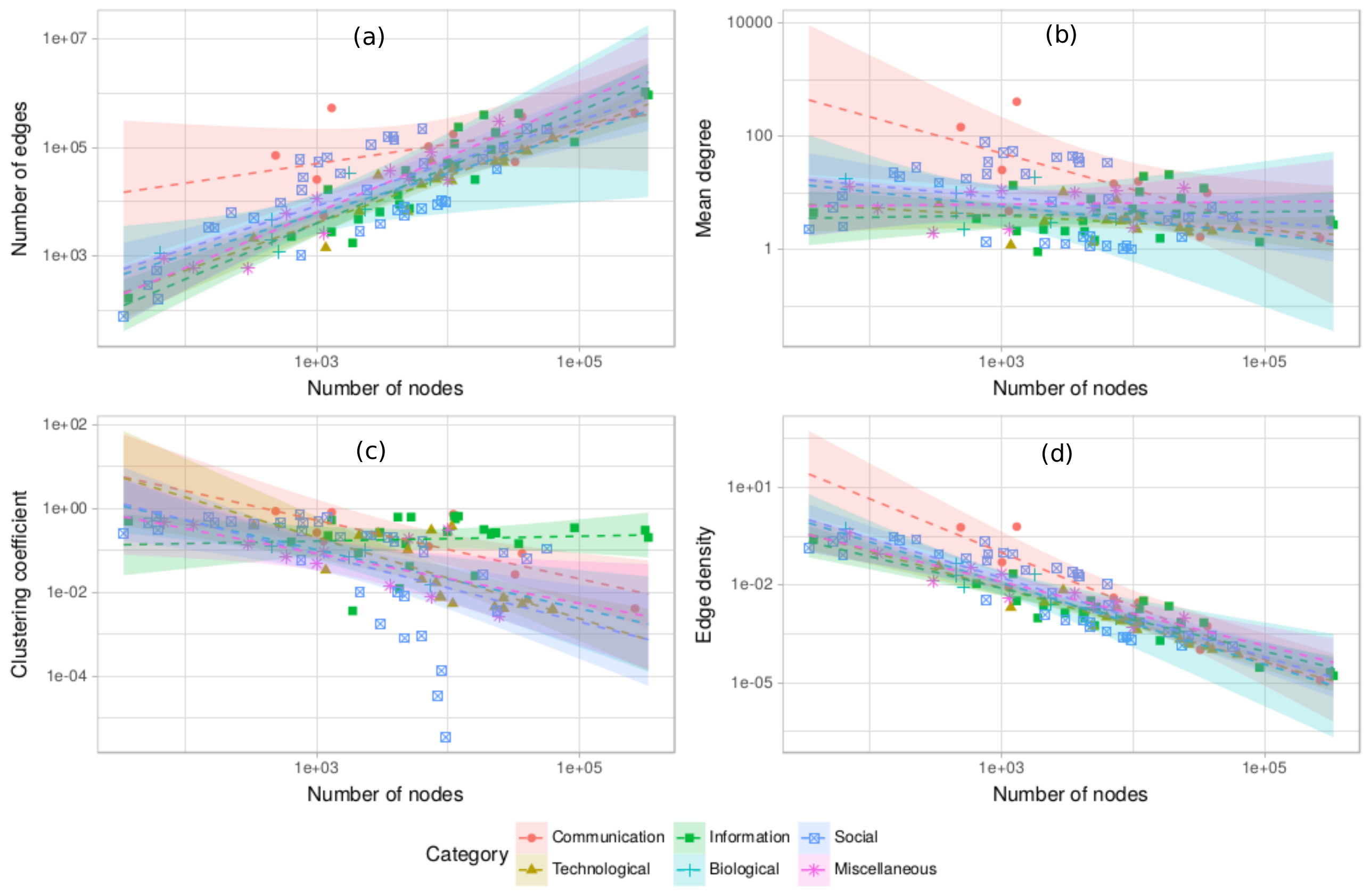}
\caption{From left to right, top to bottom, we illustrate structural characteristics of the 108 networks: (a) Number of edges as a function of the number of nodes, (b) Mean degree $\langle k \rangle$ as a function of the number of nodes, (c) Clustering coefficient according to the number of nodes, (d) Edge density as a function of the number of nodes. The colored backgrounds represent the $95\%$ confidence intervals of the relations estimated from the dataset using a linear regression model for the corresponding variables in each network category.}
\label{fig:network_dataset}
\end{figure}

Some notable structural characteristics of networks in the dataset are illustrated in Figure~\ref{fig:network_dataset}. It is noticeable that apart from biological networks which are relatively small, the other classes cover quite a wide range of numbers of nodes, edges, mean degree, clustering coefficient and edge density. Since real world networks are relatively sparse, the number of edges increases linearly according to the number of nodes and consequently, the edge densities decrease linearly by the number of nodes (since the number of possible connections increases quadratically by the number of nodes in a community). This sparsity property can easily be seen in Figure~\ref{fig:network_dataset}(a,d). Specifically, the number of edges increases linearly according to the number of nodes with equivalent rates among different network categories as can be deduced from the gradients of the linear estimates. From Figure~\ref{fig:network_dataset}(b), it can be seen that the average degree of the networks in the dataset varies principally between $1$ and $100$ edges per node except for 2 communication networks. Also, the majority of networks have an average degree from 10 to 20 connections. From a global point of view, networks in the dataset have a quite strong modular quality since most of them have relatively high clustering coefficients as shown in Figure~\ref{fig:network_dataset}(c). 

\section{Preliminary analysis of community detection methods}
\label{sec:preliminaryanalayis}
\subsection{Computation time performance}
\label{sec:timeanalysis}
Since computation time is a crucial factor to be considered in the selection of an algorithm, it is worth analyzing experimental performances to see how different community detection methods accomplish their task in real-world networks. We tested the official implementations of community detection methods introduced in Table \ref{tab:implementation} on the dataset collection presented in Table \ref{tab:dataset}. The implementations are provided officially either by their authors or by popular network analysis libraries which can be easily accessed by a large public.

We ran the implementations stated above to identify community structures on all the networks in the dataset. We measured the time needed for each implementation to compute each partition on each network. The default parameters configured by the implementations remained unchanged during the test. The calculations were executed on a server equipped by an Intel Xeon CPU E5-2650 with 32 cores of 2.60 GHz and a memory capacity of approximately 100 GBytes. However, due to the high complexity of some methods, only processes that finish within a practical time limit (less than 4 hours) are taken into account. However, for a reference purpose, we let some longer computations continue. For example \textit{Conclude} method took approximately 9 days to identify community structures on a network of 300 thousand vertices and 1 million edges; \textit{GN} method did not finish its calculation for networks of more than 4 thousand nodes and 40 thousand edges within 2 days. Consequently, the experiments that theoretically require too much time are neglected in the test. It is also worth noting that the calculations of communities on large-scale networks are also restrained by limited memory. Thus, calculations that should be finished within 4 hours but required too much memory cannot be shown here either. We repeat the calculations 5 times for each pair graph/method to reduce the fluctuation impact. Eliminating all the cases that do not satisfy our requirements, the final successful rate (number of partitions identified over the number of possible tests) ended at around $44.72\%$, mainly because of time/memory overflow.

\begin{figure}[ht]
 \centering
 \includegraphics[width=.9\linewidth]{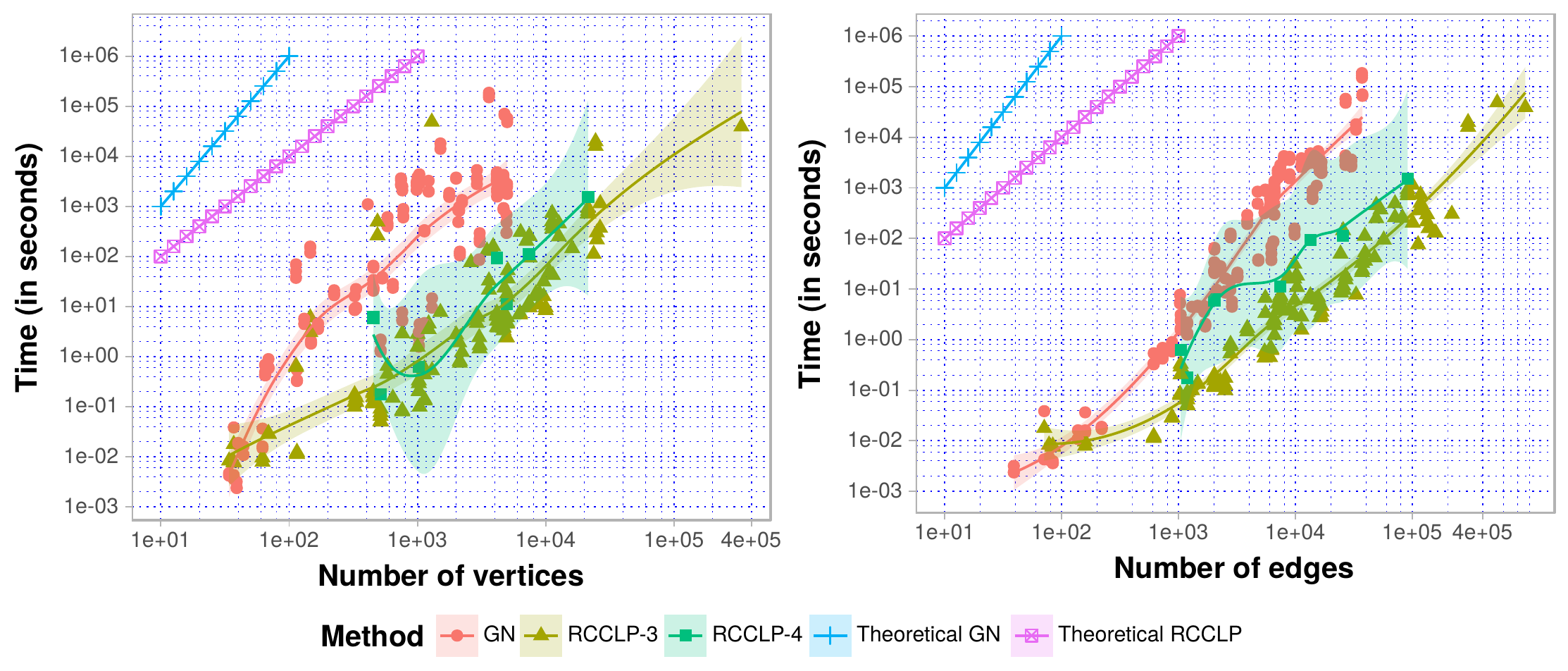}
  \caption{The execution time needed by GN, RCCLP-3 and RCCLP-4 methods to identify community structures on networks of the dataset.}
  \label{fig:time-gr1}
\end{figure}

In the following figures (from Figure \ref{fig:time-gr1} to Figure \ref{fig:time-all}) that illustrate the analyses of experimental time consumption, some conventions are commonly used. Points in the figures correspond to separated executions that have been measured. The solid lines with the same colors are estimated relations between computation time and network size (number of vertices and number of edges), using a local regression model \cite{cleveland:1979a}. The dark colored backgrounds around the regression curves represent $95\%$ confidence intervals of the model parameters. Besides, for reference purposes, we also show \textit{theoretical} execution time (i.e. in the worst case, in terms of number of calculations needed) for each algorithm. In our estimates, we plot this theoretical execution time by assigning $n = m$, as from the analysis of structural characteristics of the dataset shown above in Figure \ref{fig:network_dataset}(a), most networks are sparse, i.e. the number of edges ($m$) increases  linearly with the number of nodes ($n$). For simplicity of illustration, we grouped our results according to our classification (Table~\ref{tab:implementation}).

The first group of methods consists of centrality detection techniques to identify community structure. As we can see in Figure \ref{fig:time-gr1}, the \textit{GN} method cannot be accomplished in our test for networks of more than 4 thousand nodes or 30 thousand edges. The outcome is quite reasonable since the theoretical estimation for this method is $\mathcal{O}(nm^2)$, which grows quickly with the network size. It should be remembered that the primary goal of the \textit{RCCLP} method is to reduce the time complexity of the \textit{GN} method. We can easily observe that this objective is achieved since the \textit{RCCLP-3} reduces by an order of around $10^3$ times for graphs of 3 hundred nodes. \textit{RCCLP-3} can  function well with graphs up to millions of edges. However, when we used the same test with \textit{RCCLP-4}, the method rarely reached its terminus for large graphs as well as small graphs. As we can see in the figure, there are few dots at either side. The reason is that there are not many (they may even be absent) 4-step close paths on real world networks. As it is not very probable that such structures exist in small graphs, finding them in large graphs also requires a huge amount of time, \textit{RCCLP-4} shows a poor performance in our tests. Therefore, this configuration of the method is not recommended, as versions with $g > 4$ would logically perform poorly. It is also worth noticing that \textit{RCCLP-3} and \textit{RCCLP-4} are extremely memory consuming and are not suitable for limited resource devices. Finally, theoretical and practical time seem to find a consensus as the increments of time according to network size are quite consistent in the three cases.

\begin{figure}[ht]
 \centering
 \includegraphics[width=.9\linewidth]{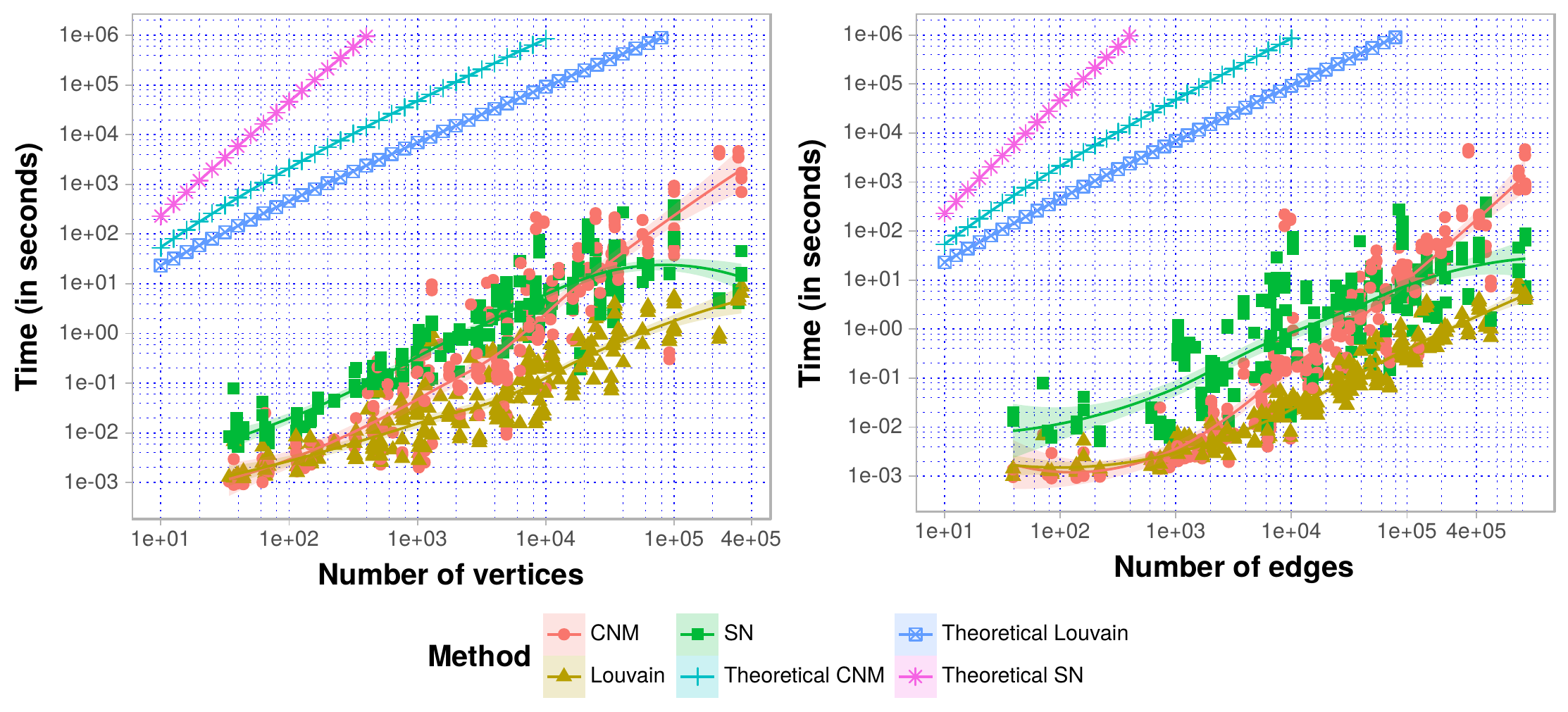}
  \caption{The execution time needed by CNM, Louvain and SN methods to identify community structures on networks of the dataset.}
  \label{fig:time-gr2}
\end{figure}

The next group includes methods using modularity optimization processes whose experimental measures are shown in Figure \ref{fig:time-gr2}. Practically, the three methods in this family require a reasonable time for calculating community structures. The longest experiment took less than 2 hours for a graph of 1 million edges. The \textit{Louvain} method is the fastest in this group; its computation increases approximately in linear time. It took only 9 seconds for the largest graph. Among the three methods, the optimization using the spectral approach is the most expensive. However, all three methods perform better than the methods in the edge removal group previously stated. The experimental results also verify theoretical estimates concerning the complexity of these methods.

\begin{figure}[ht]
 \centering
 \includegraphics[width=.9\linewidth]{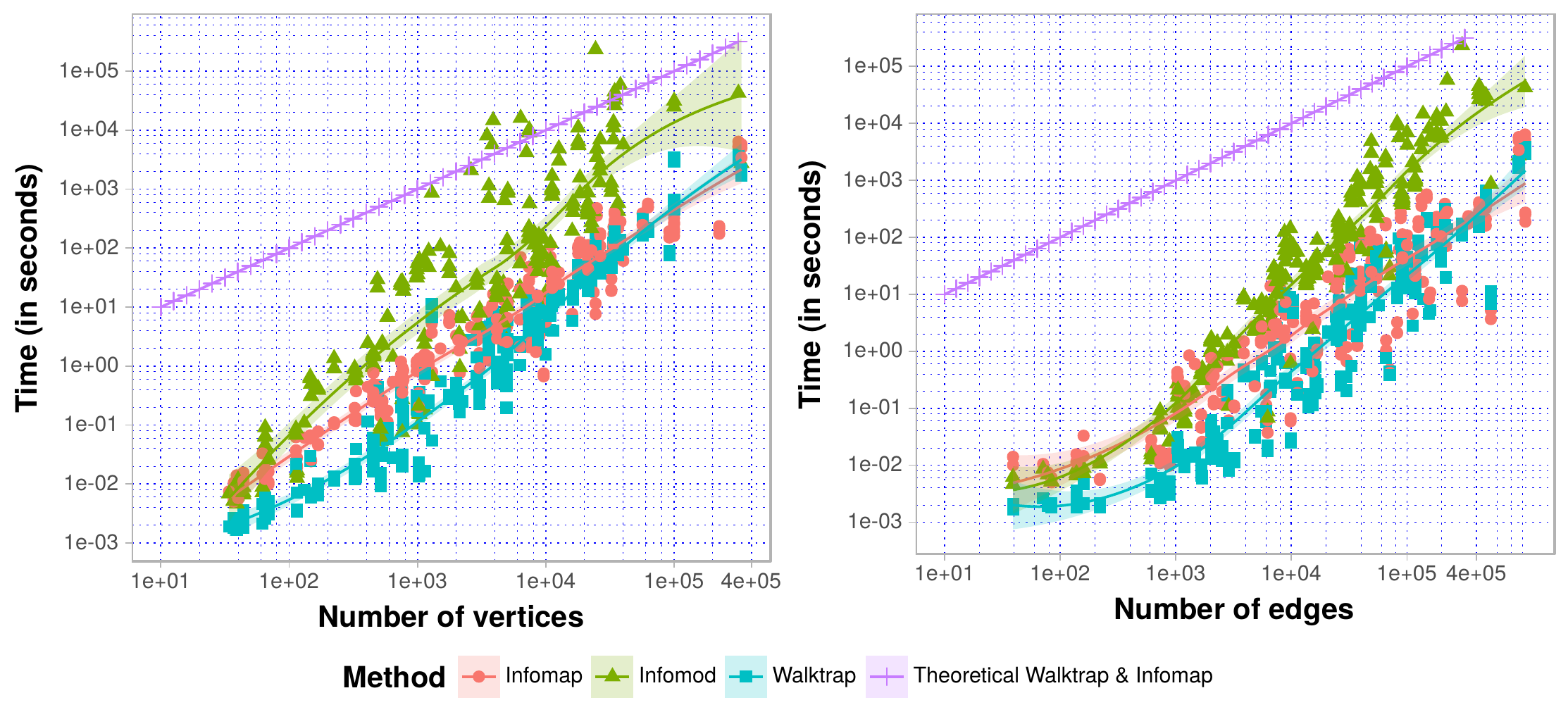}
  \caption{The execution time needed by Infomap, Infomod and Walktrap methods to identify community structures on networks of the dataset.}
  \label{fig:time-gr3}
\end{figure}

%Despite of some existing limitations such as the resolution limit \ the Newman-Girvan modularity

Similarly to the two previous groups, the computation time needed by methods in the dynamic process group is illustrated in Figure \ref{fig:time-gr3}. In terms of time consumption, this group performs better with respect to the first group, but generally worse than the modularity optimization group (except for the \textit{Walktrap} method for small and average size graphs). Among these three methods, \textit{Infomod} generally requires more calculation time than the others. At the same time, \textit{Walktrap} and \textit{Infomap} work asymptotically equally well with a slightly better rendition for \textit{Walktrap} in small and average size graphs.

\begin{figure}[ht]
 \centering
 \includegraphics[width=.9\linewidth]{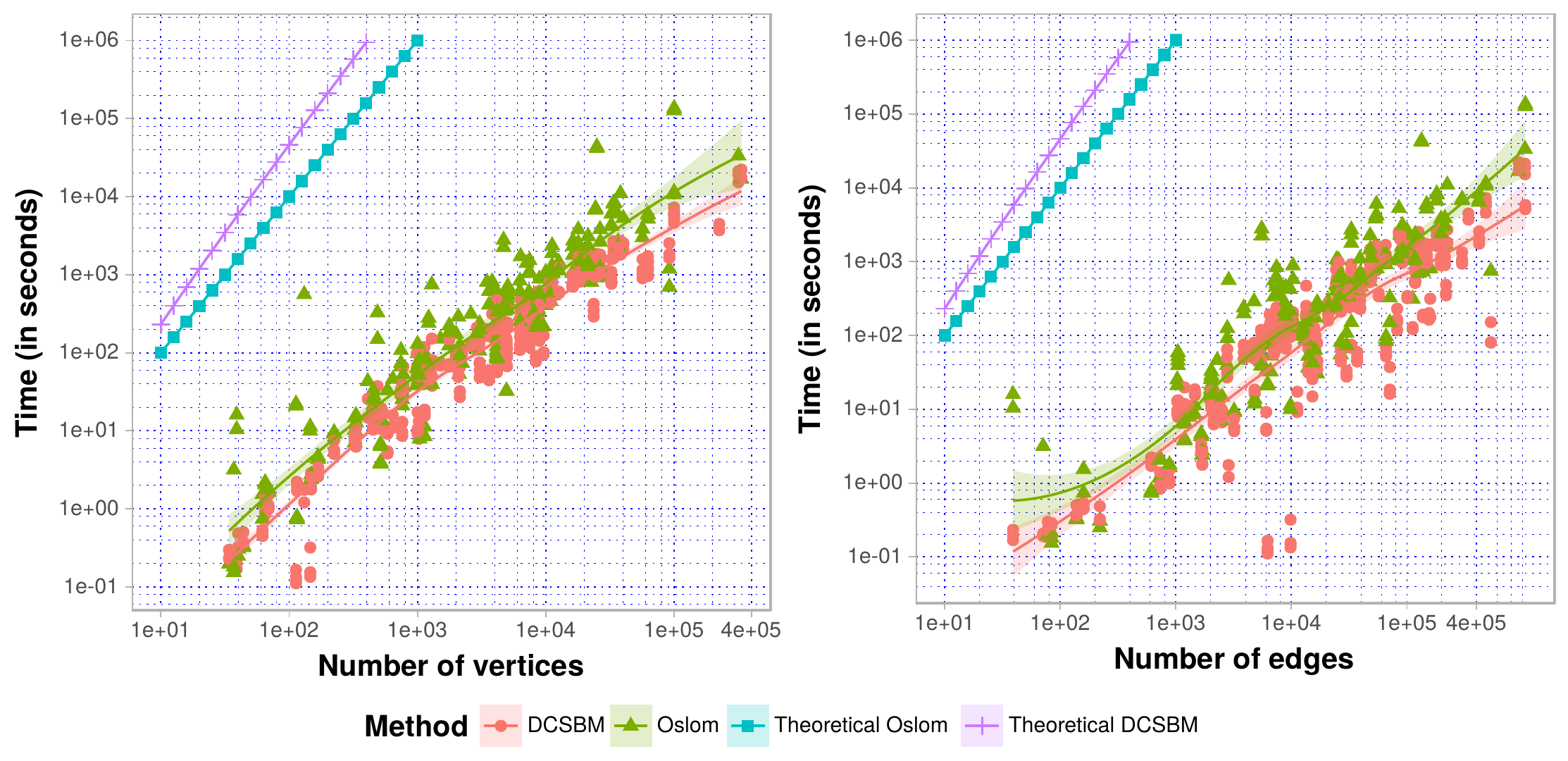}
  \caption{The execution time needed by DCSBM and Oslom methods to identify community structures on networks of the dataset.}
  \label{fig:time-gr4}
\end{figure}

The same analyses for methods in the two final groups are shown in Figure~\ref{fig:time-gr4} and Figure~\ref{fig:time-gr5}. We can easily see that \textit{DCSBM} and \textit{Oslom} have practically identical performances in terms of time consumption, \textit{DCSBM} being slightly better. In the last group, the results are quite different between different methods. The label propagation method \textit{LPA} shows a clear distinctive curve indicating its out-performance compared with the other methods. Besides, \textit{SLPA} works quite well, but not as fast as \textit{LPA} although it employs some additional techniques to reduce the number of necessary calculations \cite{xie:2012a}. This difference in performance is due to the more complicated mechanism that \textit{SLPA} uses in comparison to \textit{LPA}. It is due to the fact that \textit{SLPA} manages dedicated memories for all the nodes to stock the membership information and update it regularly. Therefore, despite the improvement of 5 to 10 times in the label update strategy, the global performance cannot surpass that of \textit{LPA} method. In terms of scalability, \textit{LPA} and \textit{SLPA} seem to exhibit the same behavior which is almost linear for small and medium graphs but increases in large graphs. The spin glass model \textit{RB} manifests a better than expected presentation with an undeviating linear augmentation. The only unexpected behavior is spotted in \textit{Conclude} method, as when the size of input graphs exceeds some thousands, the required time has been inflated by a factor of $n$, making it very demanding for large graphs.
\begin{figure}
 \centering
 \includegraphics[width=.9\linewidth]{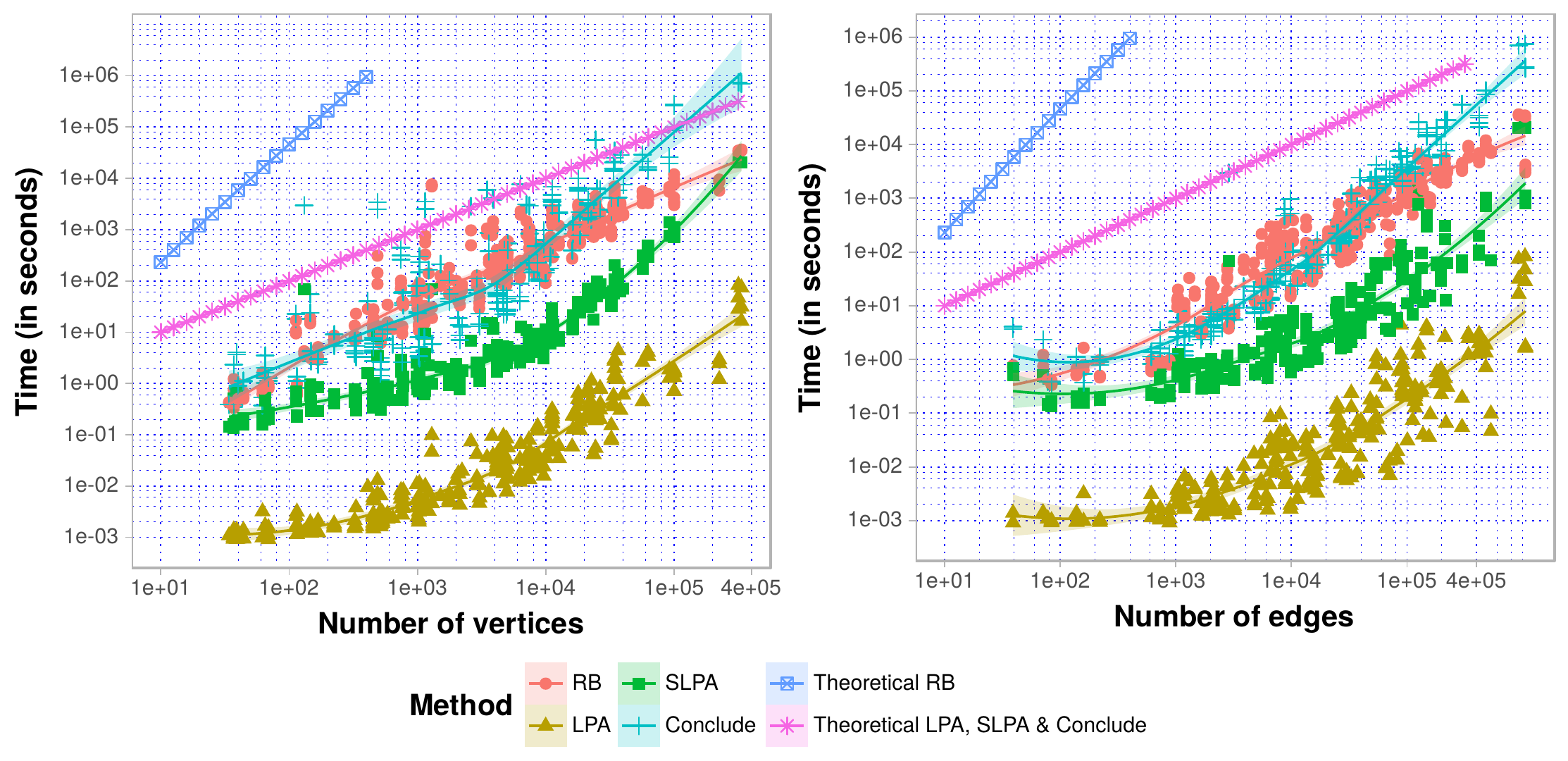}
  \caption{The execution time needed by RB, LPA, SLPA and Conclude methods to identify community structures on networks of the dataset.}
  \label{fig:time-gr5}
\end{figure}

\begin{figure}[ht]
 \centering
 \includegraphics[width=0.9\linewidth]{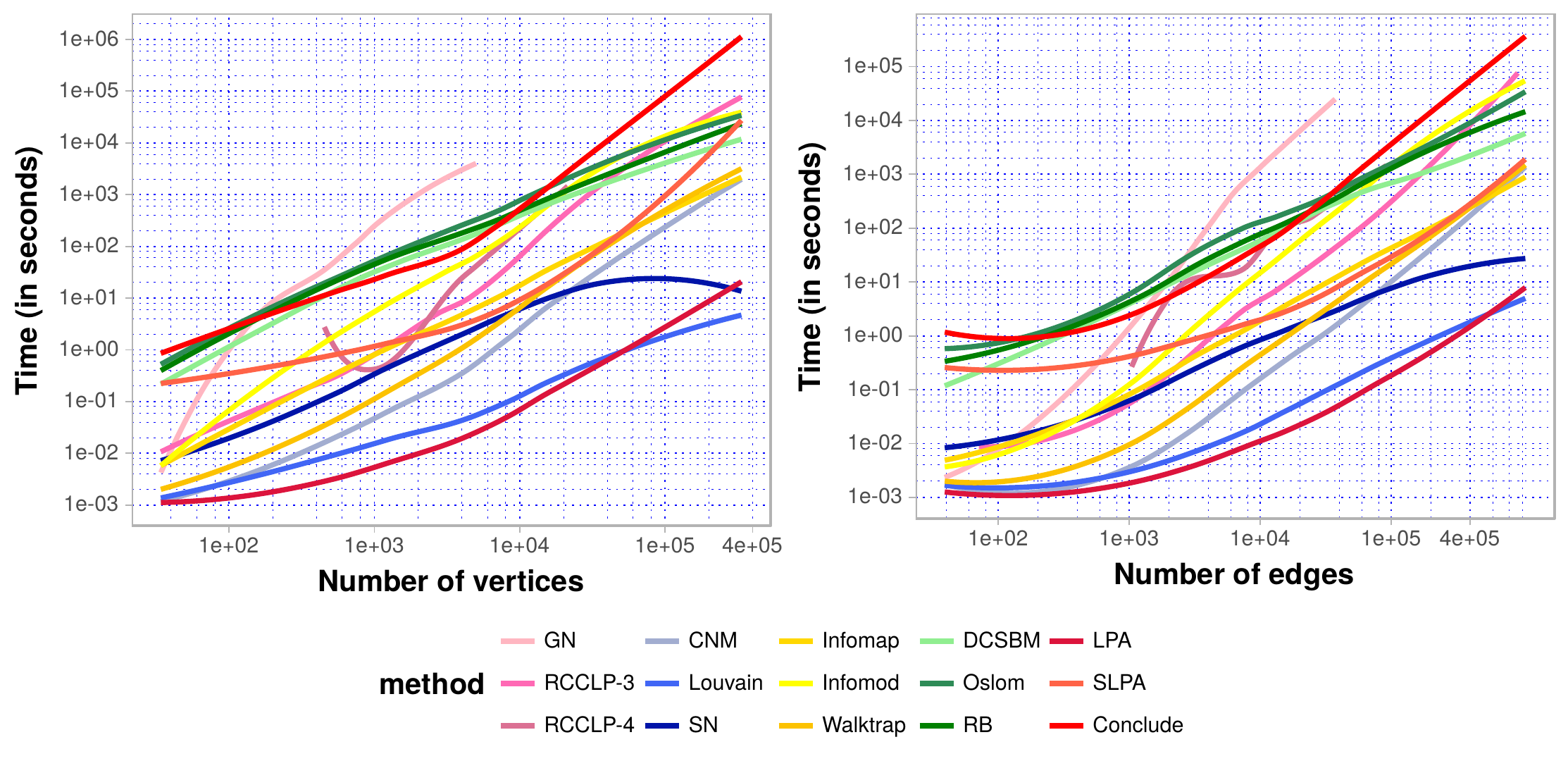}
  \caption[The estimated execution time needed for each method to identify community structures on networks of the dataset using a local regression model.]{The estimated execution time needed for each method to identify community structures on networks of the dataset using a local regression model. Methods of the same theoretical family (in the same group) are represented by similar colors.}
  \label{fig:time-all}
\end{figure}

Finally, we aggregate all the results into a common illustration as shown in Figure \ref{fig:time-all}. At the same time, for more convenient observation, we remove all the points corresponding to each experiment and keep only the regression curves, which are the execution time estimates as a function of the number of vertices on the left side and similarly on the right side for the number of edges. At first sight, it is easy to see that except for \textit{GN}, the necessary execution time for all other methods is limited in a range that increases polynomially with the network size, which accurately reflects theoretical estimates. This range is upper-bounded by \textit{Conclude}/\textit{Oslom} and lower-bounded by \textit{LPA}, corresponding to the worst and the best tested method(s) respectively. Another important fact which can be deduced from this figure is that, for most real world networks in the range up to 1 million edges, choosing a fast detection method could economize an order of $10^3$ times to $10^5$ times the calculation effort. This is an important element to be considered where time consumption is a serious problem. 

\begin{table}
 \caption{Ranking of analyzed methods according to the amount of time consumed, to identify community structures on networks of the dataset.}
\label{tab:timeranking}
 \begin{tabular}[1\textwidth]{l l l l}
  Method label \hspace{10pt} & Rank by average \hspace{10pt} & Rank by median \hspace{10pt} & Scalability \hspace{10pt}\\
  \hline\hline
%\rowcolor{lightred}
RCCLP-3 & 9 & 8 & Low \\
  \noalign{\vspace {.5cm}}
%\rowcolor{lightyellow}
CNM & 5 & 3 & Medium \\
%\rowcolor{lightgreen}
Louvain & 1 & 2 & High \\
%\rowcolor{lightgreen}
SN & 3 & 5 & High \\
  \noalign{\vspace {.5cm}}
%\rowcolor{lightgreen}
Walktrap & 4 & 4 & High \\
%\rowcolor{lightred}
Infomod & 12 & 9 & Low \\
%\rowcolor{lightyellow}
Infomap & 6 & 7 & Medium \\
  \noalign{\vspace {.5cm}}
%\rowcolor{lightred}
Oslom & 11 & 14 & Low\\
%\rowcolor{lightred}
DCSBM & 8 & 12 & Low \\
  \noalign{\vspace {.5cm}}
%\rowcolor{lightred}
RB & 10 & 13 & Low\\
%\rowcolor{lightgreen}
LPA & 2 & 1 & High\\
%\rowcolor{lightyellow}
SLPA & 7 & 6 & Medium\\
%\rowcolor{lightred}
Concude & 13 & 11 & Low \\
\hline\hline
 \end{tabular}
\end{table}

We demonstrate in Table \ref{tab:timeranking} the ranking of these methods according to our tests for reference purposes. \textit{GN} and \textit{RCCLP-4} are not involved in this ranking since they failed to accomplish their tasks in large graphs, which also means they are the most time-consuming methods within the methods that we analyzed. We show both the rankings according to the average and the median of time. Since the average-time ranking is heavily affected by the measurements on large graphs, the methods that were successful in discovering communities on very large graphs are ranked lower than methods that were not able to do so. In these cases, the median ranking is more accurate and reflects therefore the relative performance on small and medium graphs. For large graphs, we recommend using the average ranking.  

\subsection{Analysis on community size distribution}
\label{sec:fittinganalysis}
After these performance considerations, we focus on the nature of the results produced, i.e. the communities themselves. The number of latent communities that should be induced from a given network is one of the major questions in community detection context \cite{fortunato:2016a}, \cite{riolo:2017a}. It is equivalent to the subject of the expected number of clusters in a classical clustering problem. Observing the number of communities reveals useful information about the mesoscopic structure of a network. The variation of the number of communities in a network involves different levels of resolutions. An analogous way to describe the concept of resolution is the distance from an object that we prefer in order to contemplate it. The closer you get to an object, the greater the detail of its micro-structures that can be perceived, while, at the same time, information about the global organization tends to be less clear. Although several multi-resolution approaches \cite{lambiotte:2010a,pons:2011a} incorporate resolution parameters into their solutions providing more flexible mechanisms and different modular scales of networks, it is not always obvious to regulate these parameters appropriately without ad-hoc cases. The inclusion of multi-resolution parameters, of course, widens the possibility of understanding networks, but at the expense of automation convenience, that is sometimes required in clustering problems. 

In this section, we compare, once again, the previously mentioned methods but this time according to their resolution abilities. We use the same dataset collection and again we keep all default configurations of the implementations unchanged to ensure the consistency of future results. From the previous analyses, some modifications will be applied to our testing process as follows:

\begin{enumerate}
 \item From the observation of the network size distribution in Figure \ref{fig:network_dataset}(a), as well as the previous computation time analyses, the linear relation between number of vertices and number of edges of networks in our corpus becomes clear. As a consequence, it will be redundant to address the relation of dependent variables in respect of these two latter predictors. Therefore, only analyses according to the number of vertices will be rendered.
 
 \item In the case of the community detection problem, showing only the numbers of communities discovered would not always be enough. If we assume that community size in an arbitrary network exhibiting a negative power law distribution (as shown for example in \cite{clauset:2004a}), it means that the number of communities depends heavily on the number of tiny communities. Therefore, we propose to observe the distribution of community size to discern the differences between methods which could not be recognized by seeing solely the number of blocks.
 
\item Due to the huge number of calculations required and a limited hardware resource, discovering processes in the last section were interrupted unless they could be finished in a few hours. Here, some more efforts have been made, when a method is supposed to be finished in a reasonable amount of time (a few days).   
\end{enumerate}

For a given network in the dataset, we applied all of the methods presented, to identify the set of communities predicted by each one and then measured their volumes. Similarly, to the previous section, for simplicity of observation, we group methods by different families depending on their approaches. We illustrate the results in Figures \ref{fig:fitting-gr1} to \ref{fig:fitting-gr5} by using some conventions as follows:

\begin{ProblemSpecBox}{Conventions for Figures \ref{fig:fitting-gr1} to Figure \ref{fig:fitting-gr5}}
  \label{box:conventionfittingfigures}
  \begin{enumerate}
  \item A figure (denoted \textit{a}) at the top contains three following sub-figures:
  \begin{enumerate}
    \item The central figure (a1) shows a scatter plot of the distribution of community size. The solid lines in the figure represent the estimated average community size using a local regression model \cite{cleveland:1979a}. Dark colored backgrounds around the lines are $95\%$ confidence intervals of the estimates. 
    \item The top figure (a2) exhibits marginal density distributions of communities found in each range of network sizes. They are rendered by a Gaussian kernel estimator. 
    \item The right hand figure(a3) illustrates another type of marginal density distribution of communities, as a function of their sizes. They are also rendered by a Gaussian kernel estimator.
    \item The axes of marginal figures are the same as the axes of the corresponding central figure. We simply omit them for ease of representation. 
  \end{enumerate}
  \item A figure on the bottom (denoted \textit{b}) presents the number of communities as a function of network size (i.e. their number of vertices) as well as the estimated relation between these variables using the regression model stated above.  Dark colored backgrounds around the lines show $95\%$ confidence intervals of the estimate relations.
  \end{enumerate}
\end{ProblemSpecBox}

\subsubsection{Edge removal approach: GN, RCCLP-3 and RCCLP-4}
\begin{figure}[ht]
 \begin{subfigure}{0.8\textwidth}
  \centering
  \includegraphics[width=0.91\linewidth]{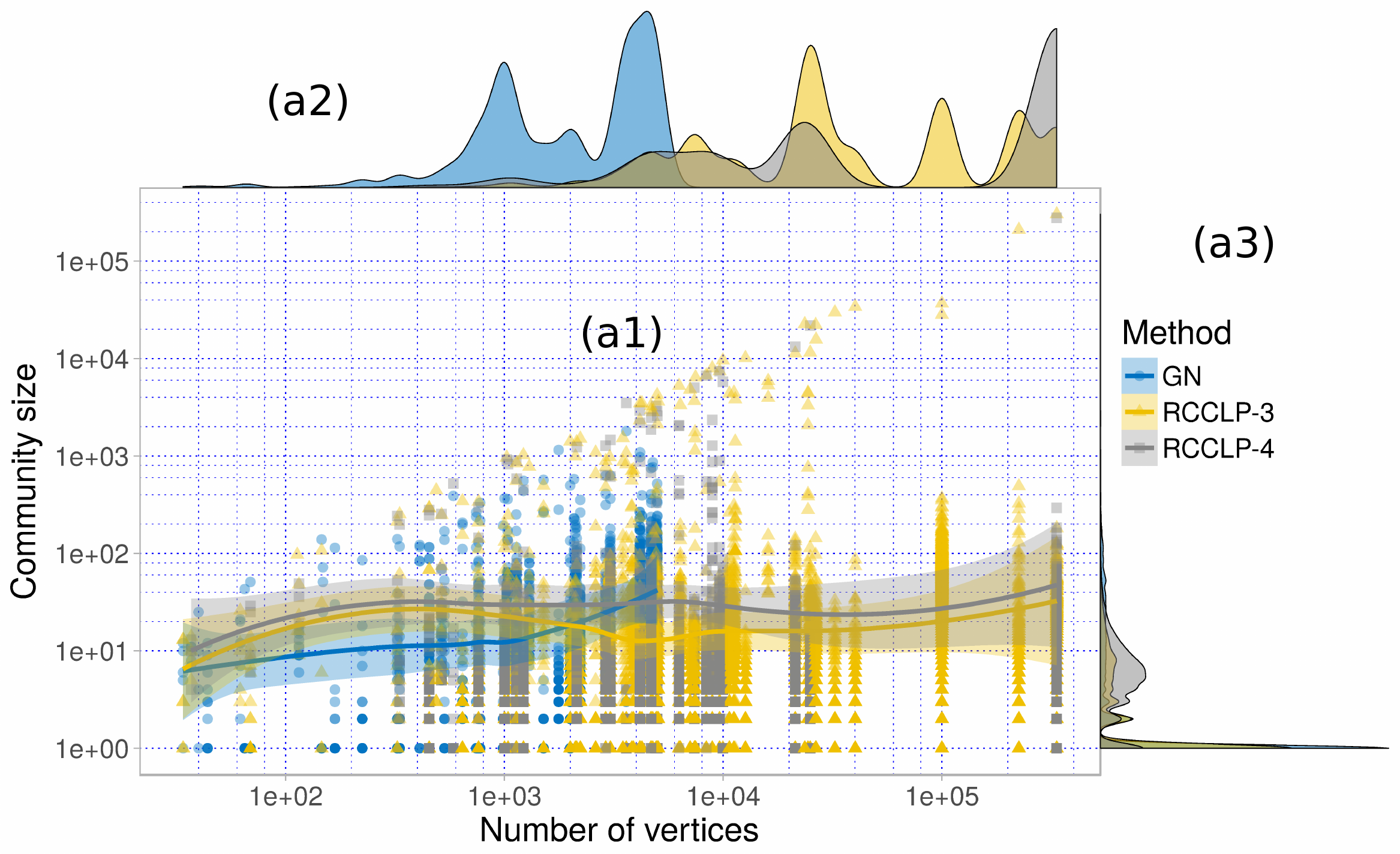}
  \caption{}
  \label{fig:size-gr1}
\end{subfigure}

\begin{subfigure}{0.8\textwidth}
  \hspace{12pt} \includegraphics[width=0.72\linewidth]{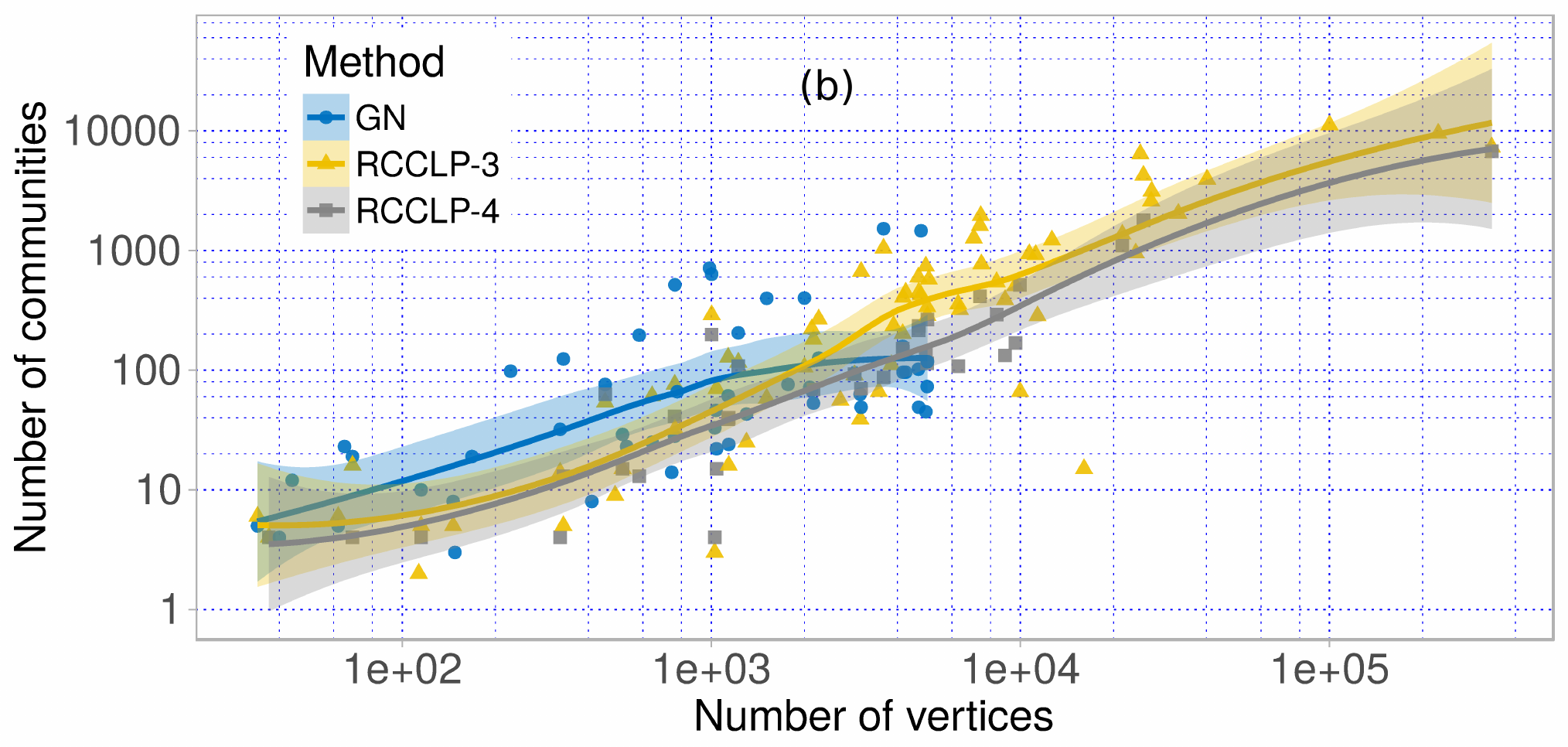}
    \caption{}
  \label{fig:nbcluster-gr1}
\end{subfigure}
\caption{Fitting quality of GN, RCCLP-3 and RCCLP-4 methods: number of communities and community size}
\label{fig:fitting-gr1}
\end{figure}
From Figure \ref{fig:fitting-gr1}, we can notice again that the \textit{GN} method is only able to function on small and medium networks, due to its high complexity, which is quite obvious from theoretical analysis. \textit{RCCLP-3} and \textit{RCCLP-4} can detect up to the largest networks in our corpus. By observing the right marginal density distribution, all of these methods identify a surprisingly high number of singleton communities. The average number of singleton communities is around $24\%$ but the number can reach $60\%$ in some cases. The reason for this aberrant phenomenon is that in some dense, small networks, there exist too many high and equivalent central vertices and edges. The separating mechanism employed here keeps removing central nodes or edges until a large number of vertices are isolated, creating singletons or very small communities. Since \textit{GN} only works on small graphs, it is highly impacted by this phenomenon in our experiment. Besides, from a global observation, we can see in the top figure that the majority of communities detected by these methods are very small for the same reason. From Figure \ref{fig:fitting-gr1}(a), we can see that a large number of communities have less than 10 vertices, even in very large networks. This makes the number of communities increase rapidly, as illustrated in Figure \ref{fig:fitting-gr1}(b). Bear in mind that the distributions of community size have right-skewed shapes, meaning that the majority of communities are small, and most are found under the lines of average community size. Therefore, the three methods of this family have very high resolutions. Notwithstanding, this result needs to be interpreted with caution, for the following reasons:

\begin{enumerate}
 \item The density function in Figure \ref{fig:fitting-gr1}(a1) reveals that the successful rates for discovering community structures of the three methods are fundamentally very different. In fact, due to the high complexity of time and memory, many networks are not successfully resolved, which significantly degrades the comparison quality. 
 \item As a consequence of the first reason, there is a high fluctuation in the dependent variables which makes the confidence intervals quite large. A deeper investigation of the quality on small and medium networks could partially palliate this problem. 
\end{enumerate}

Despite the previously mentioned issues, within this class of methods there is a large consensus as regards the discovery of the highest number of communities.

\subsubsection{Modularity optimization approach: CNM, Louvain and SN}
\begin{figure}[ht]
 \begin{subfigure}{0.85\textwidth}
  \centering
  \includegraphics[width=0.85\linewidth]{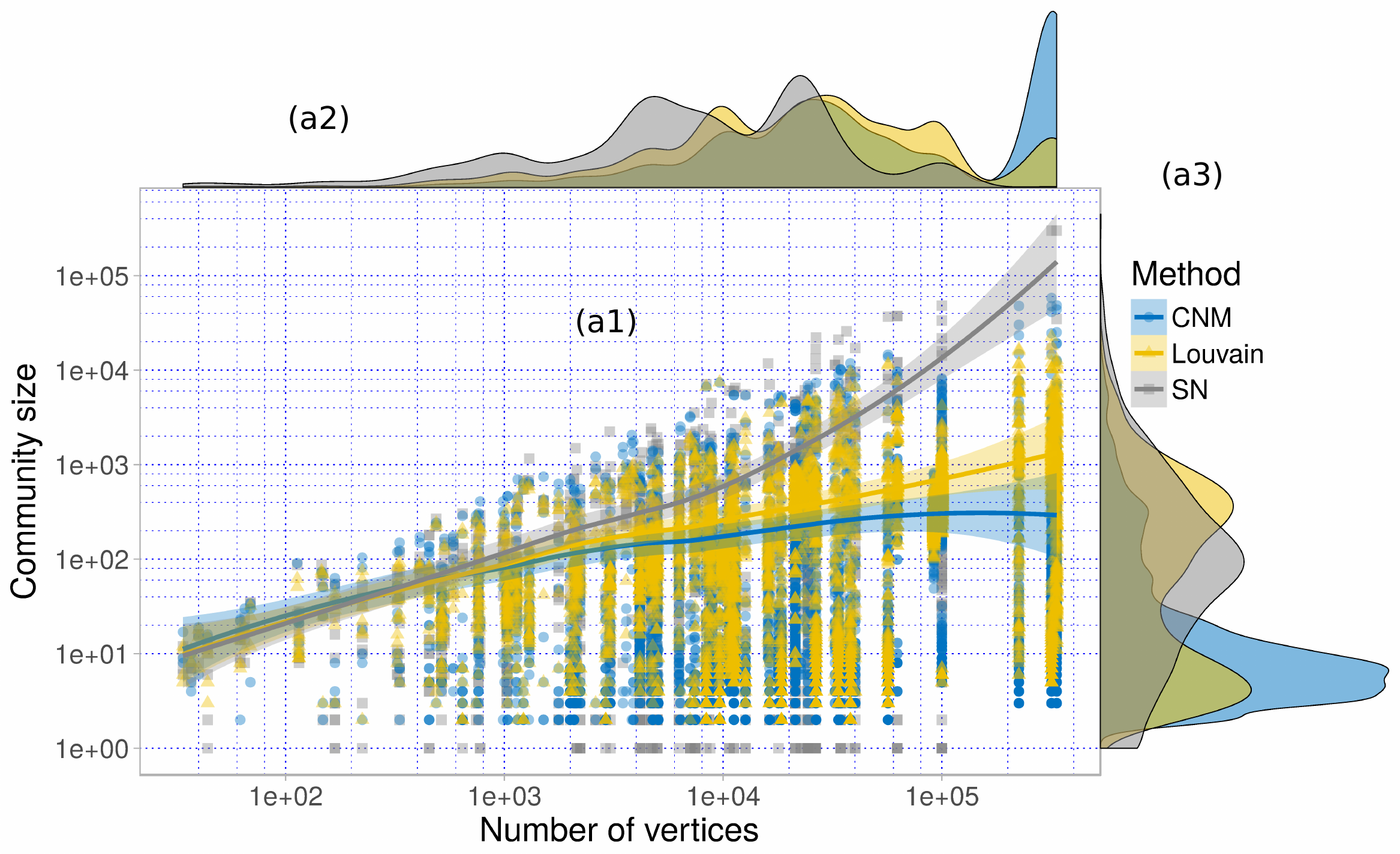}
  \caption{}
  \label{fig:size-gr2}
\end{subfigure}

\begin{subfigure}{0.78\textwidth}
  \hspace{12pt} \includegraphics[width=0.74\linewidth]{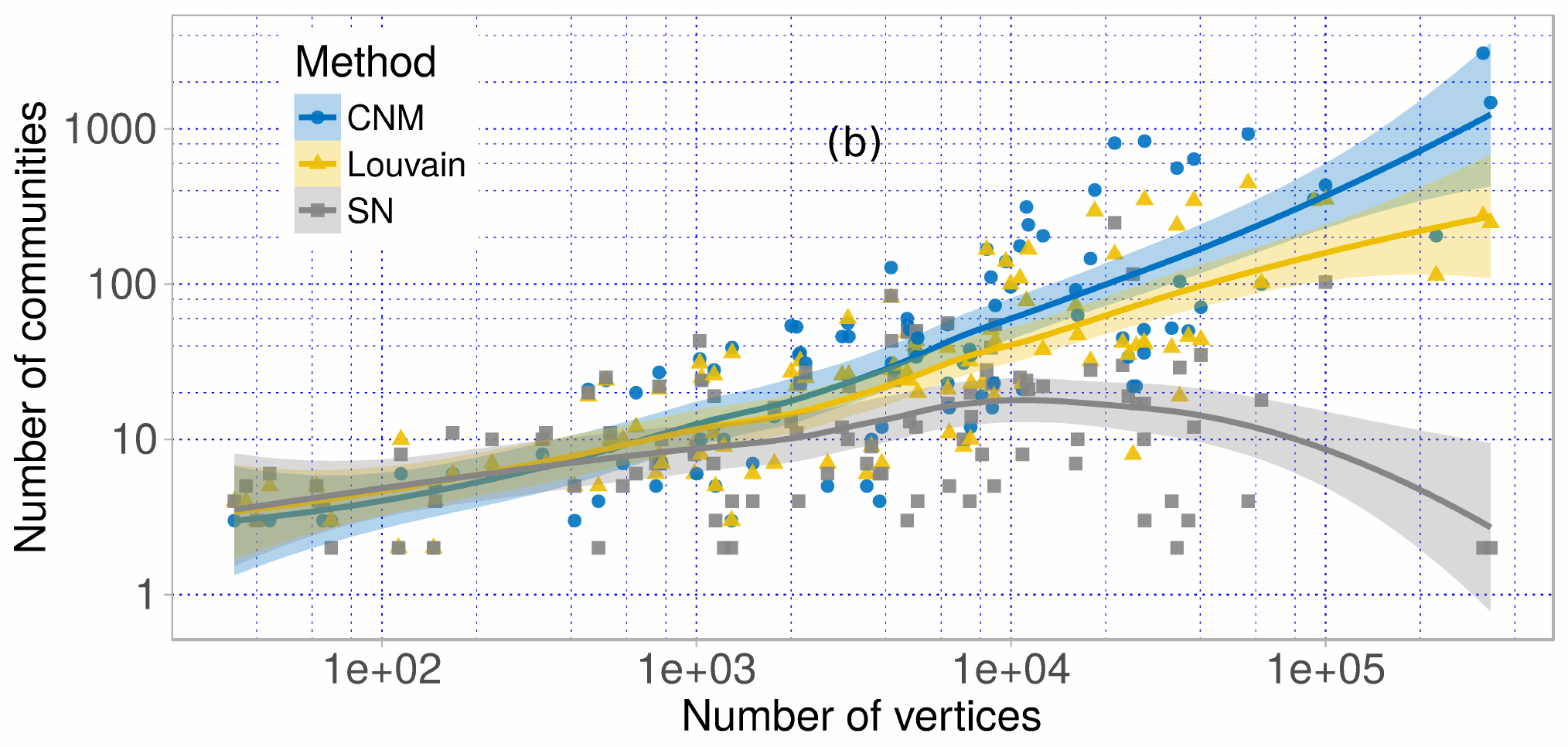}
    \caption{}
  \label{fig:nbcluster-gr2}
\end{subfigure}
\caption{Fitting quality of CNM, Louvain and SN methods: number of communities and community size}
\label{fig:fitting-gr2}
\end{figure}

In this second group, our measurements are more complete since all three methods successfully resolved large networks. From Figure \ref{fig:fitting-gr2}(a2), it can be seen that there is a regularity between the distributions of communities over the whole range of networks, except for the range of very large networks. Actually, in this range, the behavior is very different in the three methods. While \textit{CNM} determines a very large number of medium and small communities, \textit{Louvain} identifies fewer small communities and more medium and large communities. On the other hand, \textit{SN} only proposes a partition of two giant communities. For instance, if we take the $Amazon$ network \cite{snapnets}, while \textit{CNM} detected $1480$ clusters, the number is $249$ for \textit{Louvain}, and only two for \textit{SN}. The same phenomenon is also identified for another example, the $DBLP$ network \cite{snapnets}. The corresponding numbers are $3077$, $275$ and $2$ in the same order for the three methods. This fact can also be remarked in smaller networks as can be seen in Figure \ref{fig:fitting-gr2}(b). However the gap between the number of communities reduces gradually from the right to the left of the figure. But in general, the order remains unaltered as experienced in our observations, i.e. the average number of communities detected by \textit{CNM} is larger than that of \textit{Louvain} which is in turn larger than that of \textit{SN}. Consequently, the order of community sizes is inversed, since the sizes of graphs are fixed, as can be seen in Figure~\ref{fig:fitting-gr2}(a1). Another fact can be extracted from Figure \ref{fig:fitting-gr2}(a3) concerning the diversity of community size: while \textit{CNM} and \textit{SN} consistently move towards small and medium communities respectively, \textit{Louvain} on the other hand tends to propose both small and medium size communities.

\subsubsection{Dynamic process approach: Infomap, Infomod and Walktrap}
\begin{figure}[ht]
 \begin{subfigure}{0.85\textwidth}
  \centering
  \includegraphics[width=0.85\linewidth]{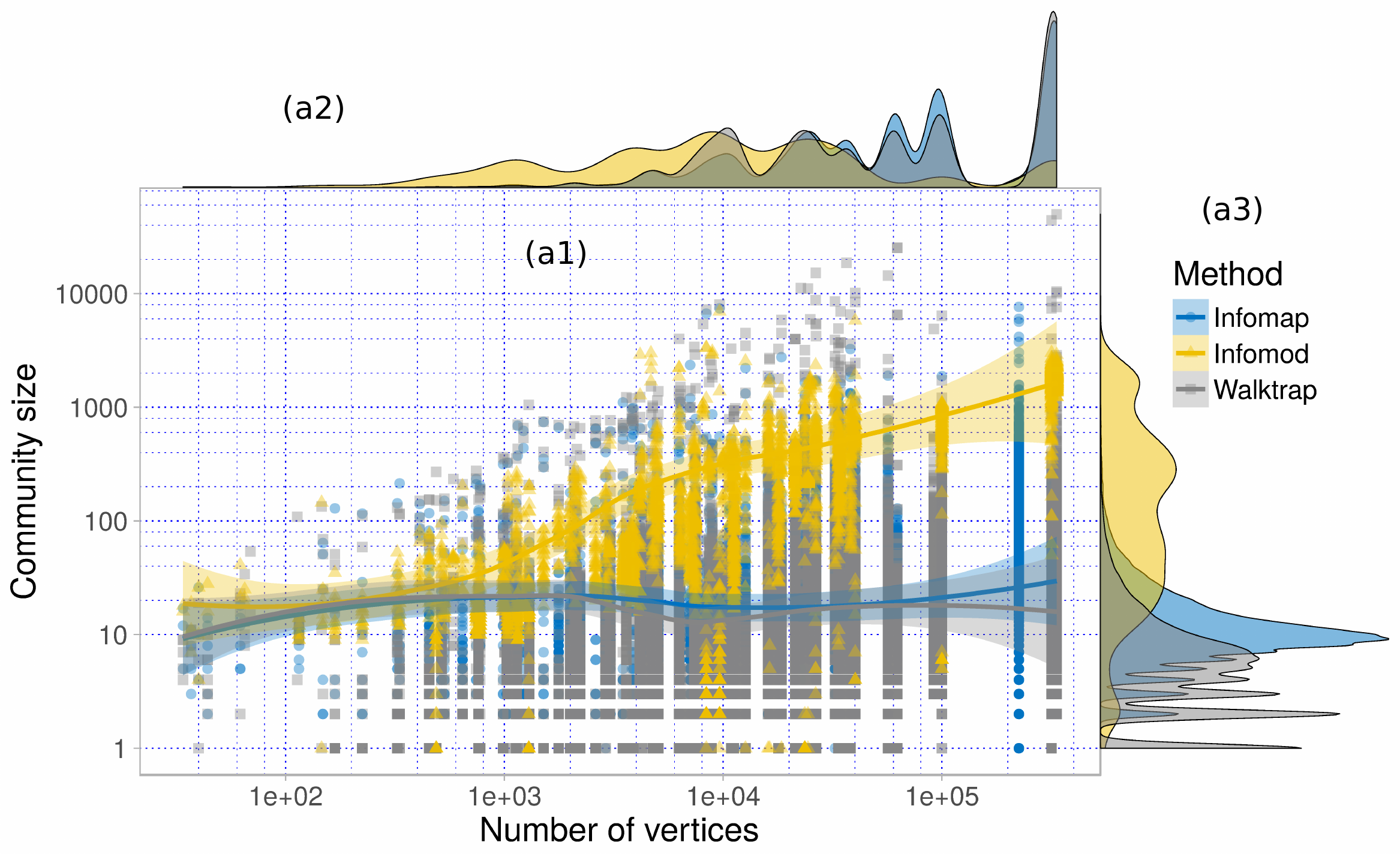}
  \caption{}
  \label{fig:size-gr3}
\end{subfigure}
\begin{subfigure}{0.8\textwidth}
  \hspace{12pt} \includegraphics[width=0.72\linewidth]{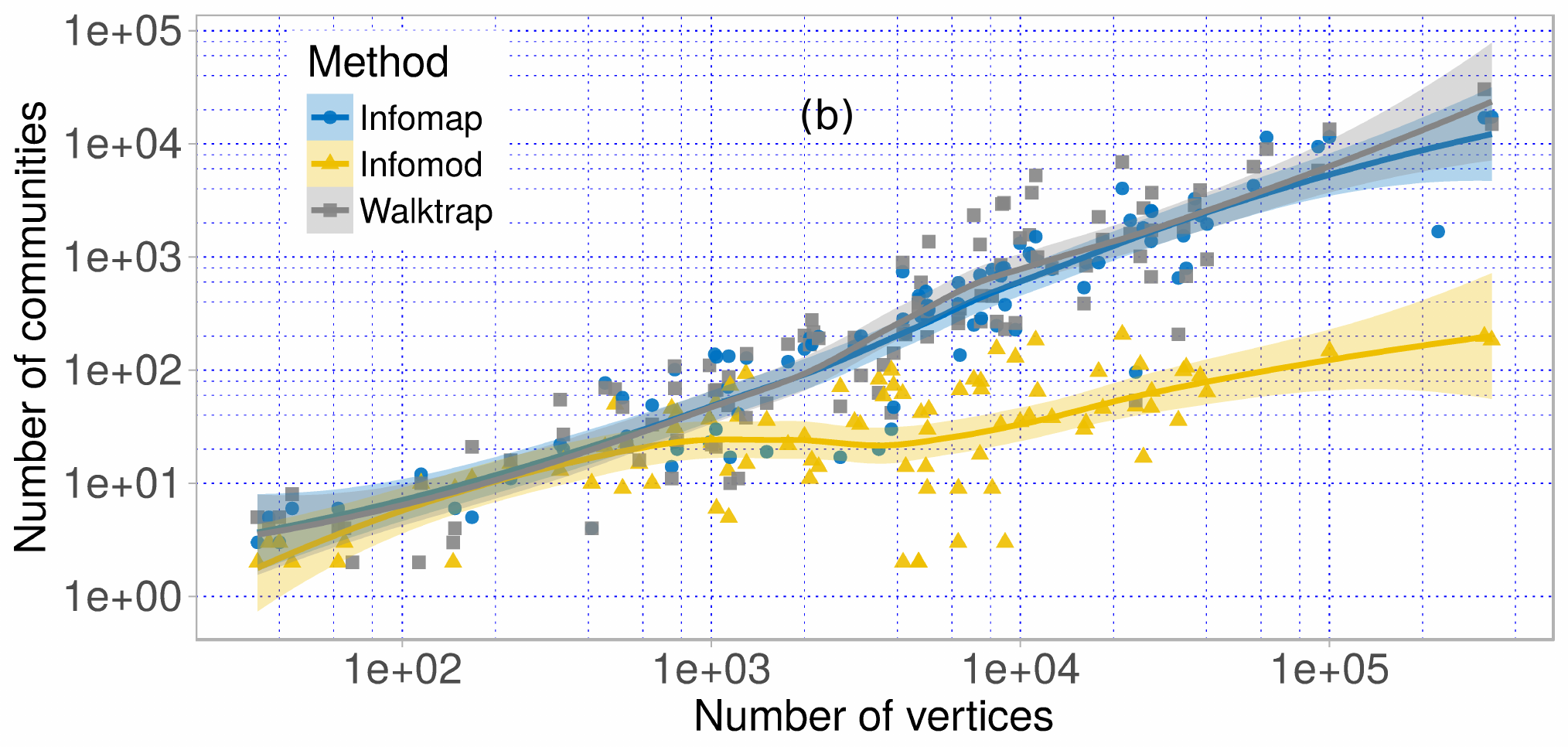}
    \caption{}
  \label{fig:nbcluster-gr3}
\end{subfigure}
\caption{Fitting quality of Infomap, Infomod and Walktrap methods: number of communities and community size.}
\label{fig:fitting-gr3}
\end{figure}
At first glance, we can see a clear separation within the three methods. While \textit{Infomap} and \textit{Walktrap} display quite a comparable evolution of average community size, depicted by Figure  \ref{fig:fitting-gr3}(a1), as well as marginal distribution, as depicted by Figure \ref{fig:fitting-gr3}(a2-a3), \textit{Infomod} is driven distinctly apart. After close examination, we notice that in \textit{Infomod}, there is a relatively uniform partitioning of communities which is upper-bounded by the largest community containing $6948$ vertices. Unlike many other methods including \textit{Infomap} and \textit{Walktrap}, the number of medium and large communities discovered by \textit{Infomod} does not outnumber the number of small communities, as stipulated by heavy-tailed distributions. As a consequence, the total number of communities observed remains low and increases at a slow, constant pace.

\textit{Infomap} and \textit{Walktrap} tend to keep their average community size limited to around $10$ to $30$, over the whole range of networks. This phenomenon keeps them apart from the resolution limit issue. In both methods, the most popular community size can be found around $10$ nodes or smaller. Our more specific experiment on the median community size shows almost similar results for \textit{Infomap} while the number decreases slightly for \textit{Walktrap}. Above these values, the number of communities decreases significantly. The biggest difference between these two methods can be easily observed at the spurious region on the marginal distribution of Figure \ref{fig:fitting-gr3}(a3). In fact, unlike \textit{Infomap}, which produces moderately small communities, \textit{Walktrap} identifies a huge number of isolated nodes (around $10\%$ according to the statistics), similarly to \textit{RCCLP-3} and \textit{RCCLP-4}, as indicated in the previeus section. This problem may be due to the agglomerative hierarchical clustering employed by \textit{Walktrap} to detect communities, which engenders orphaned peripheral vertices. This behavior has been identified in particular by Newman and Girvan, cf. Figure 3 in \cite{newman:2004a}. This problem, however, is quite simple to palliate since these peripheral vertices could be assigned to their closest neighbor's community. By removing this issue, we obtained quite a similar result for \textit{Infomap} and \textit{Walktrap}.     

In terms of average number of communities, \textit{Infomap} and \textit{Walktrap} show practically the same behavior. The evolutions coincide across almost all networks with small confidence intervals, particularly in the mid-range networks. For medium and large networks, as seen in Figure \ref{fig:fitting-gr3}(b), it is very likely that \textit{Infomod} identifies a much smaller number of communities. In fact, more than $75\%$ of \textit{Infomod}'s partitions have fewer communities than those of the other two methods.

\subsubsection{Statistical inference approach: SBM, DCSBM and Oslom}
\begin{figure}[ht]
\centering
 \begin{subfigure}{0.85\textwidth}
  \centering
  \includegraphics[width=0.9\linewidth]{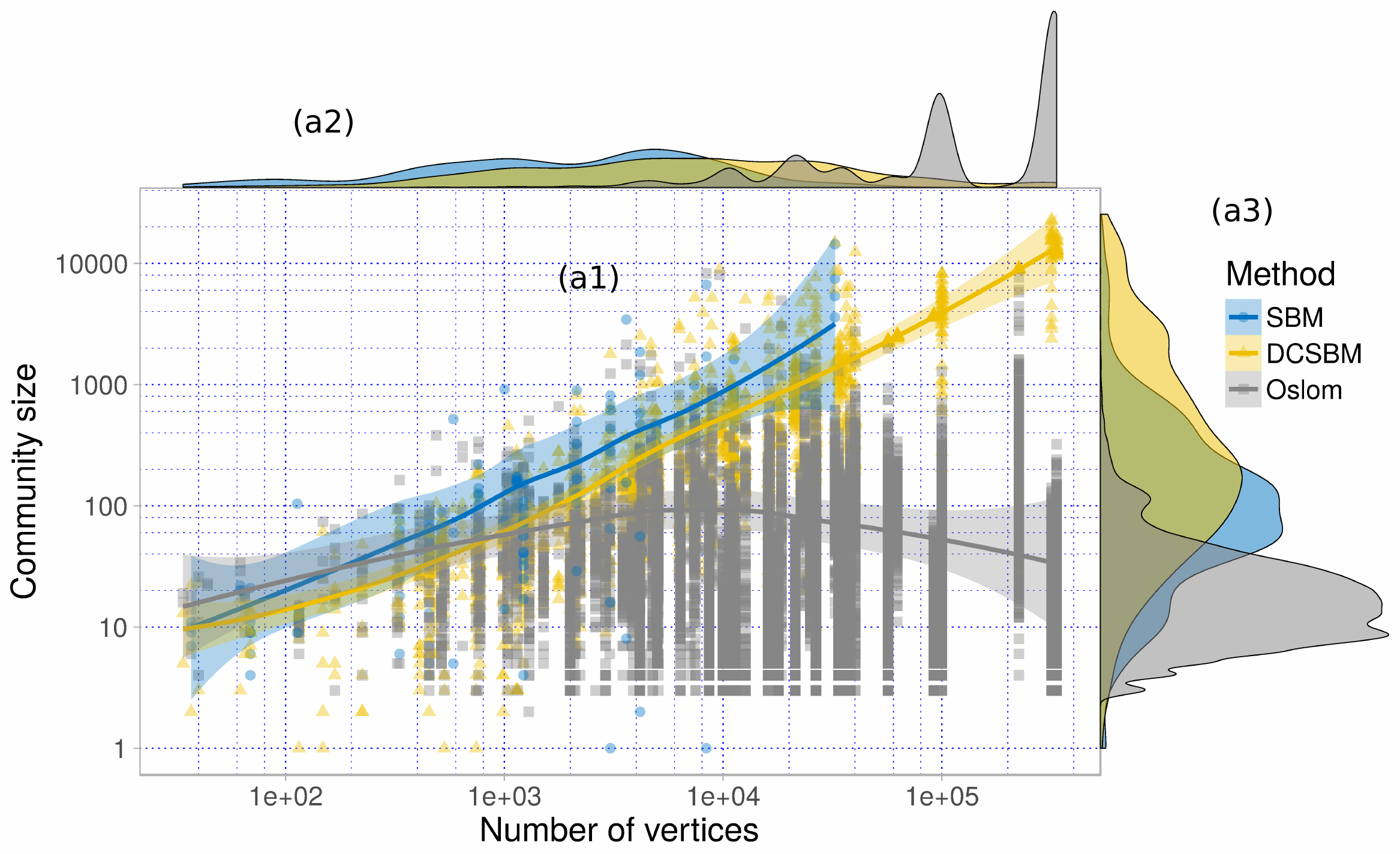}
  \caption{}
  \label{fig:size-gr4}
\end{subfigure}
\begin{subfigure}{0.85\textwidth}
  \hspace{12pt} \includegraphics[width=0.72\linewidth]{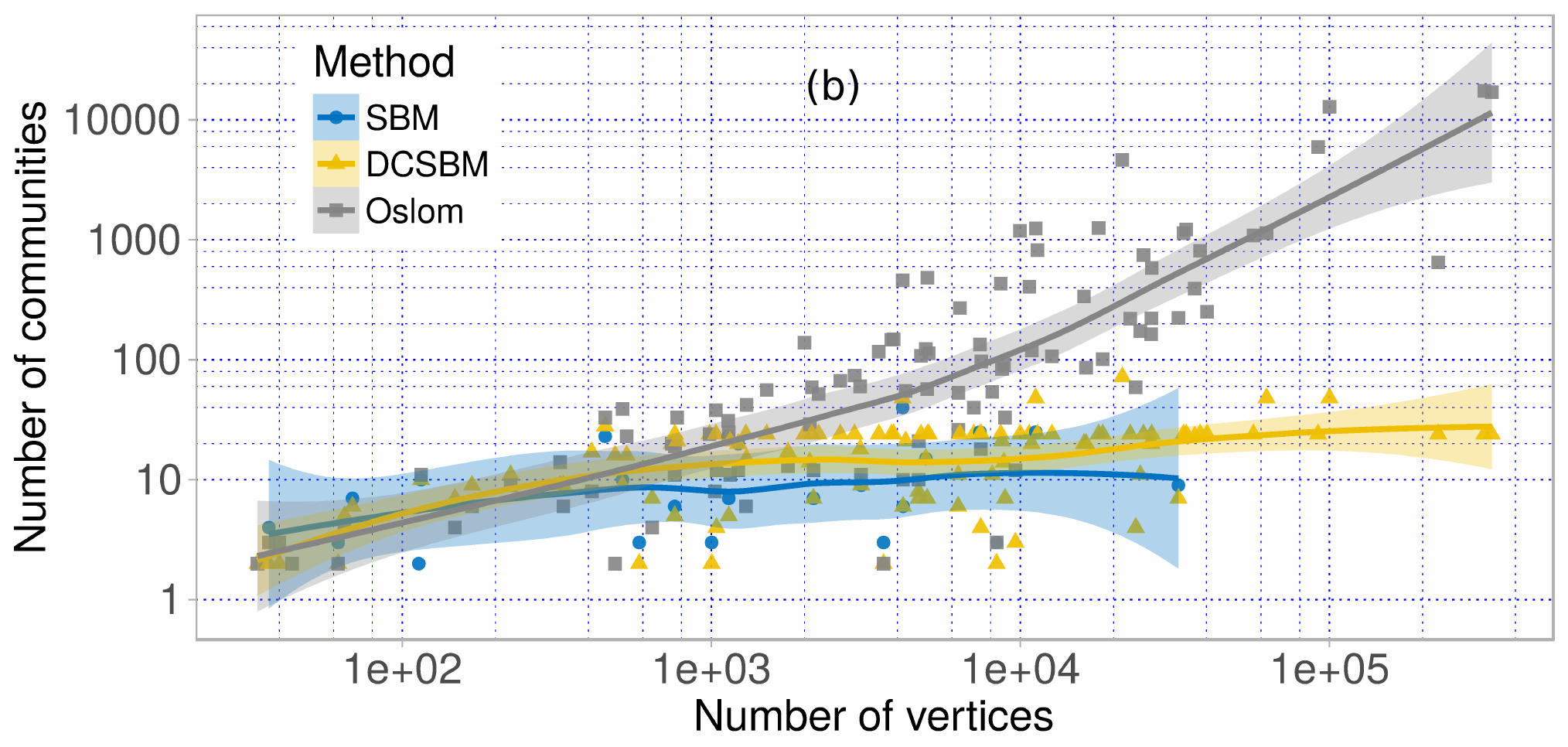}
    \caption{}
  \label{fig:nbcluster-gr4}
\end{subfigure}
\caption{Fitting quality of SBM, DCSBM and Oslom methods: number of communities and community size.}
\label{fig:fitting-gr4}
\end{figure}

In the case of statistical inference, we see quite a similar phenomenon ro that previously experienced in the dynamic approach. Specifically, the distributions of community size of the two implementations \textit{SBM} and \textit{DCSBM} nearly coincide with a slightly higher average community size for the former. In fact, in this Bayesian block model it is necessary that the prior distribution of number of blocks be given. %According to different block model variants, one could assume various hypotheses about underlying mechanisms that create an observed network under the corresponding regulations of block structures and define a prior probability. 
In the implementation that has been employed, the authors initialize the community discovering process by assigning nodes randomly to groups, according to a queuing-type mechanism and then use a Monte Carlo sampling process to maximize the posteriori probability. However, the calculation becomes extremely time-consuming when the maximum number of communities is too large \cite{riolo:2017a}. Hence, by default, the maximum number of communities is configured at $25$, as proposed by the authors, which leads to an underestimation of medium and large graphs, as shown in Figure \ref{fig:fitting-gr4}(b), and also noticed by the authors. One can see the impact of this regulation as the number of communities approaches asymptotically $25$, independently with the network size on the right hand side of the figure. Discovering community structures using this method in large networks (more than $25^2$ nodes, for example) would have to be done recursively to avoid resolution limits. In other words, one could apply community detection again on very large communities (larger than the square root of the number of nodes).   

By observing the distribution of community size in Figure \ref{fig:fitting-gr4}(a1), it is understandable that the average block size of \textit{SBM} and \textit{DCSBM} increases linearly according to the number of vertices. As the number of communities remains constant, the average community size must increase proportionately. Furthermore,  Figure \ref{fig:fitting-gr4}(a3) also reveals that community sizes are well distributed around their mean values, which makes the marginal distribution quite symmetric for both \textit{SBM} and \textit{DCSBM}. There is almost no particular inclination towards small communities, as acknowledged in some previous methods.

For the case of \textit{Oslom}, the separation is quite clear. It uncovers many more communities, making their sizes very small. Figure \ref{fig:fitting-gr4}(a1) shows that the majority of \textit{Oslom} communities are found under the average values of the associated partitions of \textit{SBM} and \textit{DCSBM}. Our demonstrations show that there is indeed a significant difference in the partitioning strategies of these methods.

\subsubsection{RB, LPA, SLPA and Conclude methods}
\begin{figure}[ht]
 \begin{subfigure}{0.85\textwidth}
  \centering
  \includegraphics[width=0.9\linewidth]{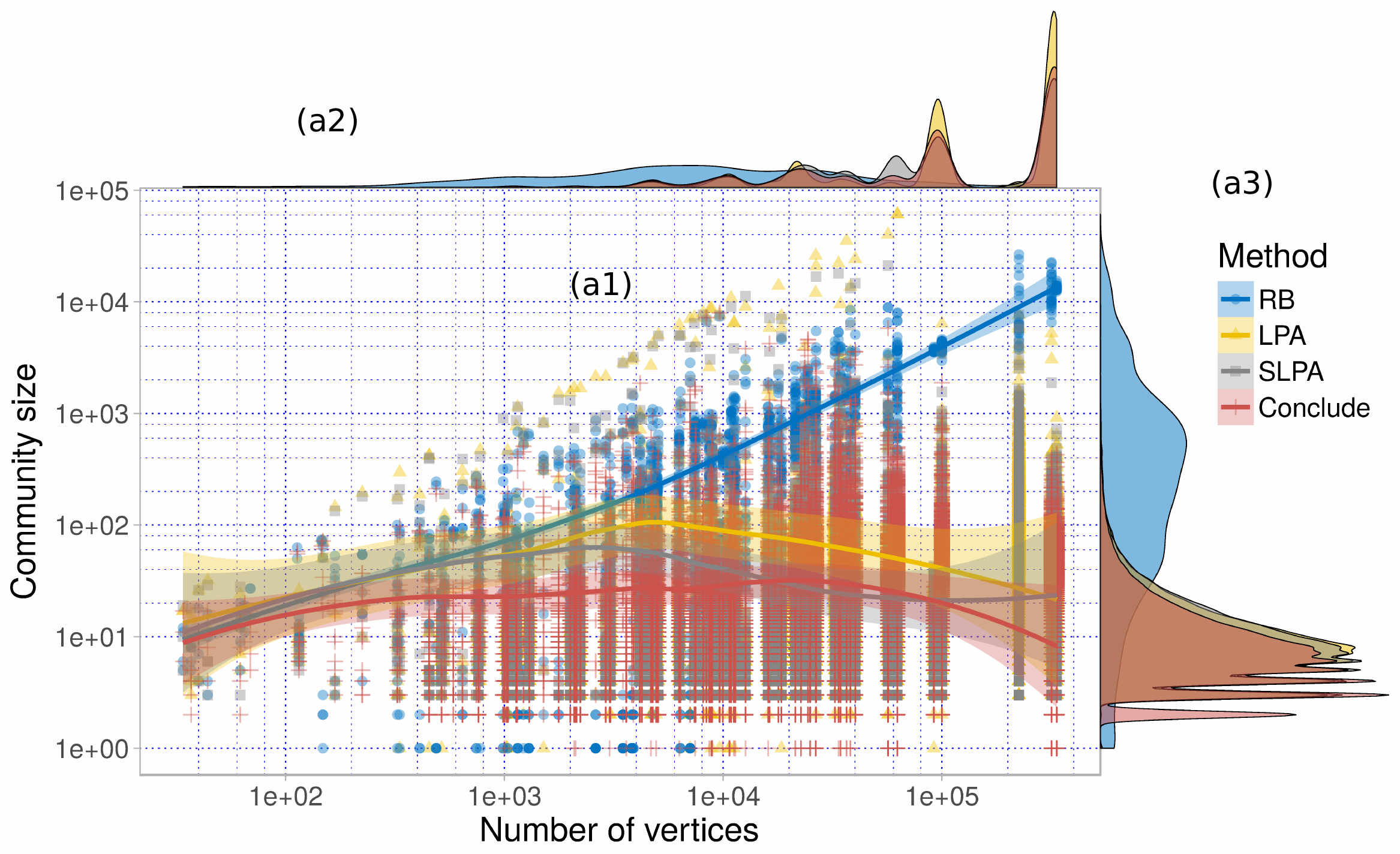}
  \caption{}
  \label{fig:size-gr5}
\end{subfigure}
\begin{subfigure}{0.85\textwidth}
  \hspace{12pt} \includegraphics[width=0.73\linewidth]{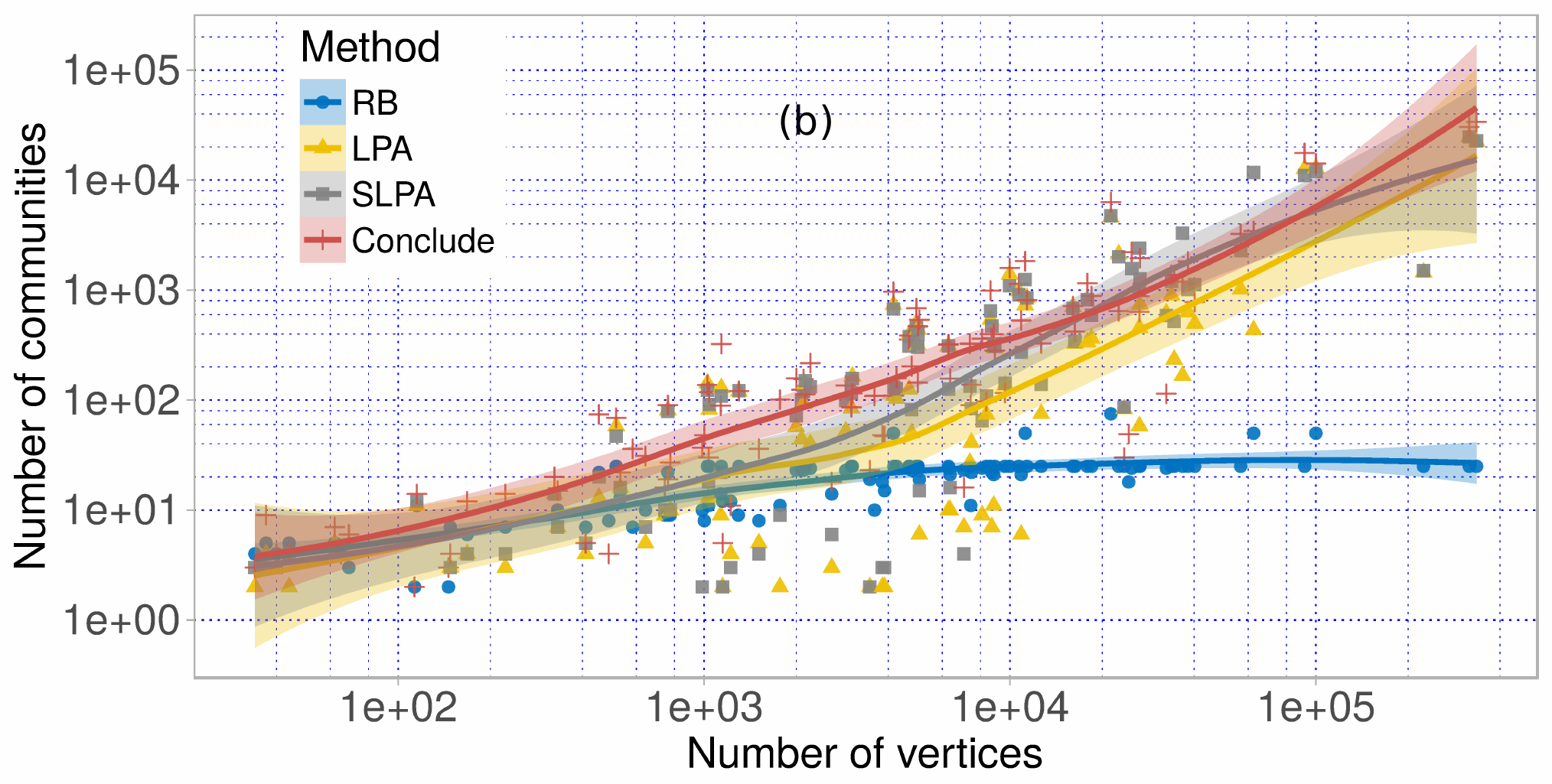}
    \caption{}
  \label{fig:nbcluster-gr5}
\end{subfigure}
\caption{Fitting quality of RB, LPA, SLPA and Conclude methods: number of communities and community size.}
\label{fig:fitting-gr5}
\end{figure}

In the last group, we discover that there is a remarkable coincidence in all distributions of the three methods \textit{LPA}, \textit{SLPA} and \textit{Conclude}. In fact, the difference between them is almost indistinguishable on the marginal measures. There is only a small discrepancy in the number of detected communities in very large networks, as can be seen in Figure \ref{fig:fitting-gr5}(a2), such that \textit{LPA} detected slightly more communities than \textit{SLPA} and \textit{Conclude}. From Figure \ref{fig:fitting-gr5}(a3), one can see that the majority of communities are quite small in these three methods. Similarly to \textit{CNM}, \textit{Infomap} or \textit{Walktrap}, the majority of communities are small, i.e. have less than 10 nodes.

\begin{figure}[ht]
 \begin{subfigure}{0.95\textwidth}
  \centering
  \includegraphics[width=0.73\linewidth]{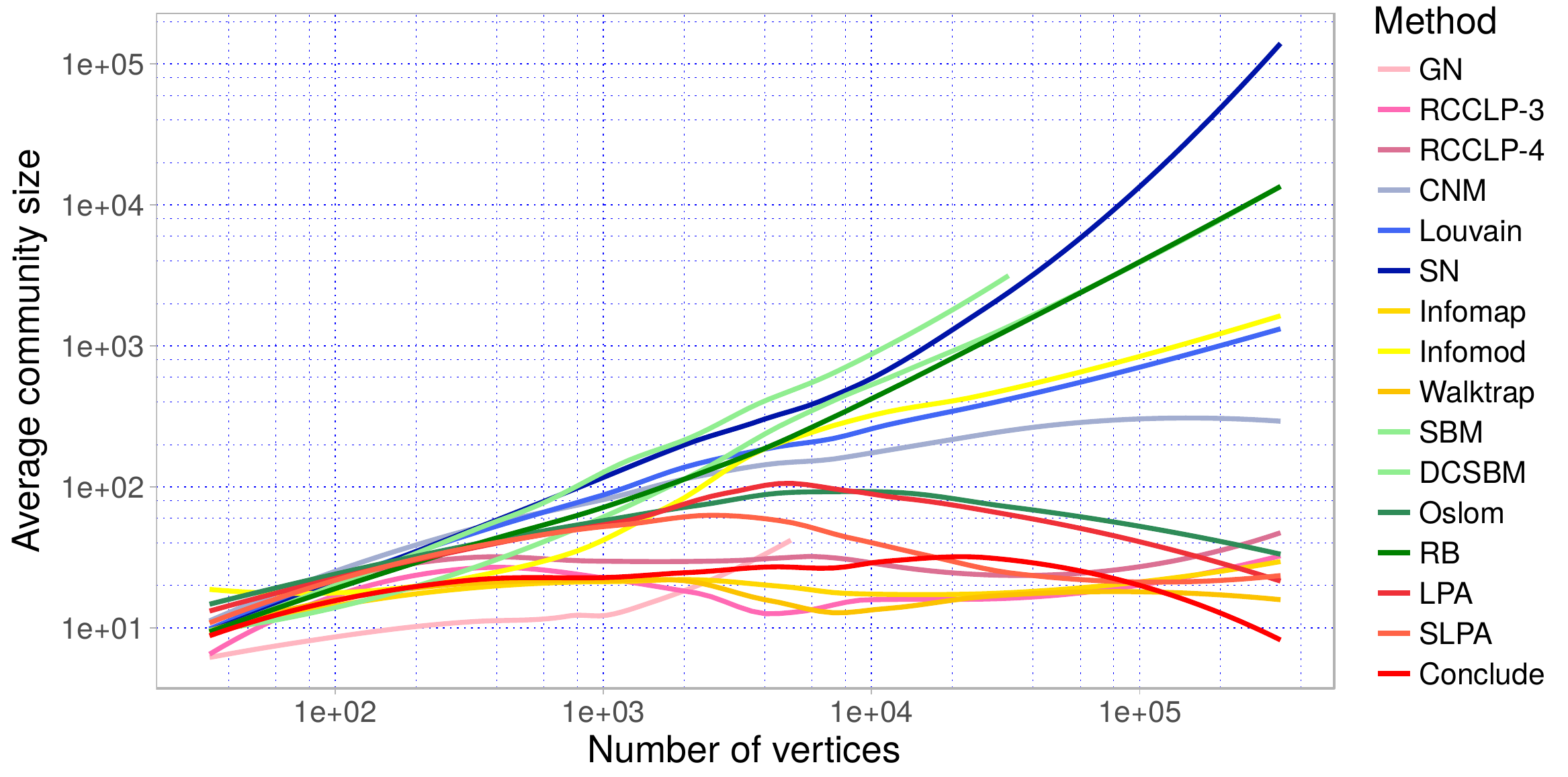}
  \caption{}
\end{subfigure}
\begin{subfigure}{0.95\textwidth}
\centering
  \includegraphics[width=0.75\linewidth]{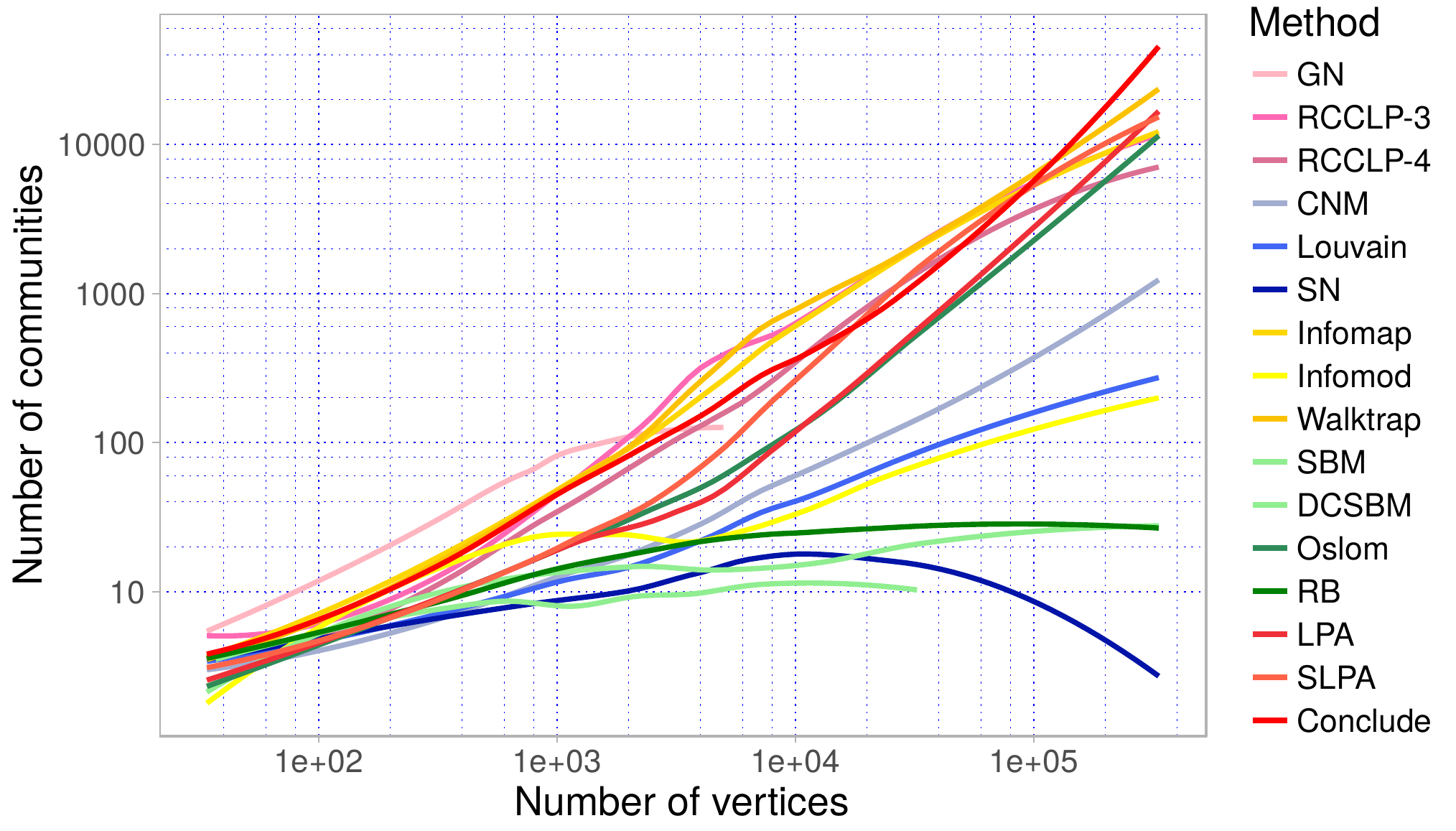}
    \caption{}
\end{subfigure}
\caption{A summary of community size estimation}
\label{fig:fitting-all}
\end{figure}

In the three methods, one can see that the variation of the data is significantly large, which also produces a large variation in our estimates. Since the associated prediction intervals for the estimates are likely to be larger, predictions related to community size distribution are not expected to be accurate.

On the other hand, $RB$ method shows a solid consistency with far fewer variations in our examination. Average community size increases regularly and the number of communities becomes saturated from medium size networks. The behavior of $RB$ method closely resembles that of \textit{DCSBM} observed in Figure \ref{fig:fitting-gr3}. Consequently, it is supposed to suffer the resolution limit for large networks. Nevertheless, since $RB$ is provided with a resolution tune parameter, the method may avoid this effect if the parameter is correctly chosen.

\subsubsection{Summary}
For the final step in this section, in the same manner as the previously presented time computational analysis, we aggregate, for all methods, the estimates of average community size and the number of detected communities, as a function of number of vertices in the network in Figures \ref{fig:fitting-all}(a) and \ref{fig:fitting-all}(b), respectively. One can see that there exist several partitioning strategies hidden in these methods. If we use the preference of a theoretical number of recoverable communities in a $k$-planted partition model \cite{ames:2013a}, being $O(\sqrt{n})$, the  methods studied could be considered to over-fit (create more than $k$ clusters) or under-fit (create less than $k$ clusters) as presented in Table \ref{tab:fittingsummary}, in the third column.

\begin{table}[ht]
 \centering 
 \caption{Ranking of analyzed methods according to their number of detected communities. A method is considered to over-fit if it detects asymptotically more than $\sqrt{n}$ clusters. The group numbers exhibit the estimated similarity based on fitting quality.}
\label{tab:fittingsummary}
 \begin{tabular}[1\textwidth]{l l l}
  Method label \hspace{10pt} & Size wrt. k-planted model \hspace{10pt} & Fitting \hspace{10pt} \\
  \hline\hline
GN & Bigger & Over-fit \\
RCCLP-3 & Bigger & Over-fit \\  
RCCLP-4 & Bigger & Over-fit\\
  \noalign{\vspace {.5cm}}
CNM & Close & Over-fit\\
Louvain & Close & Under-fit\\
SN & Smaller & Under-fit\\
  \noalign{\vspace {.5cm}}
Walktrap & Bigger & Over-fit\\
Infomod & Close & Under-fit\\
Infomap & Bigger & Over-fit\\
  \noalign{\vspace {.5cm}}
Oslom & Smaller & Under-fit\\
SBM & Smaller & Under-fit\\
DCSBM & Smaller& Under-fit\\
  \noalign{\vspace {.5cm}}
RB  & Smaller & Under-fit\\
LPA & Bigger & Over-fit\\
SLPA & Bigger & Over-fit\\
Concude & Bigger & Over-fit\\
\hline\hline
 \end{tabular}
\end{table}

With an overview of the second and third columns of Table \ref{tab:fittingsummary}, methods belonging to the same theoretical class, which shares a common assumption about the definition of community, have a tendency to show the same fitting quality. This has also been identified by \cite{ghasemian:2018a}. However, although being useful to help practitioners to presume the expected number of clusters which a method would detect with respect to the theoretical experience, it is still very challenging to decide which method to use, since the reference is based on a hypothesis about an underlying model. This also means that if the hypothesis about the partition model changes (another model than $k$-planted model), the expected number of communities will be diversified, and hence the indicated fitting quality preference becomes disproved. As a consequence, in the next section, we propose a novel technique to estimate the similarity of community detection methods based on community size distributions. 

%Certainly, this is only one of the interesting quality aspects that differentiate one method from the others. Nonetheless, we will demonstrate that it also allows to gain more insight into the difference in terms of partitioning strategy.

\subsection{Similarity based on community size distribution}
\label{sec:sizesimilarity}
A very naive but efficient approach to evaluate the similarity of two methods is to inquire into the \textit{``closeness''} of the two corresponding community
size distributions \cite{dao:2018b}. Two methods could be supposed to be similar, if their corresponding density distributions expose a large intersection area, as shown in Figure
\ref{fig:overlapsimilar}(a). From this notice, we can define our new similarity function as follows:

\begin{figure}[ht]
 \centering
 \includegraphics[width=1.0\linewidth]{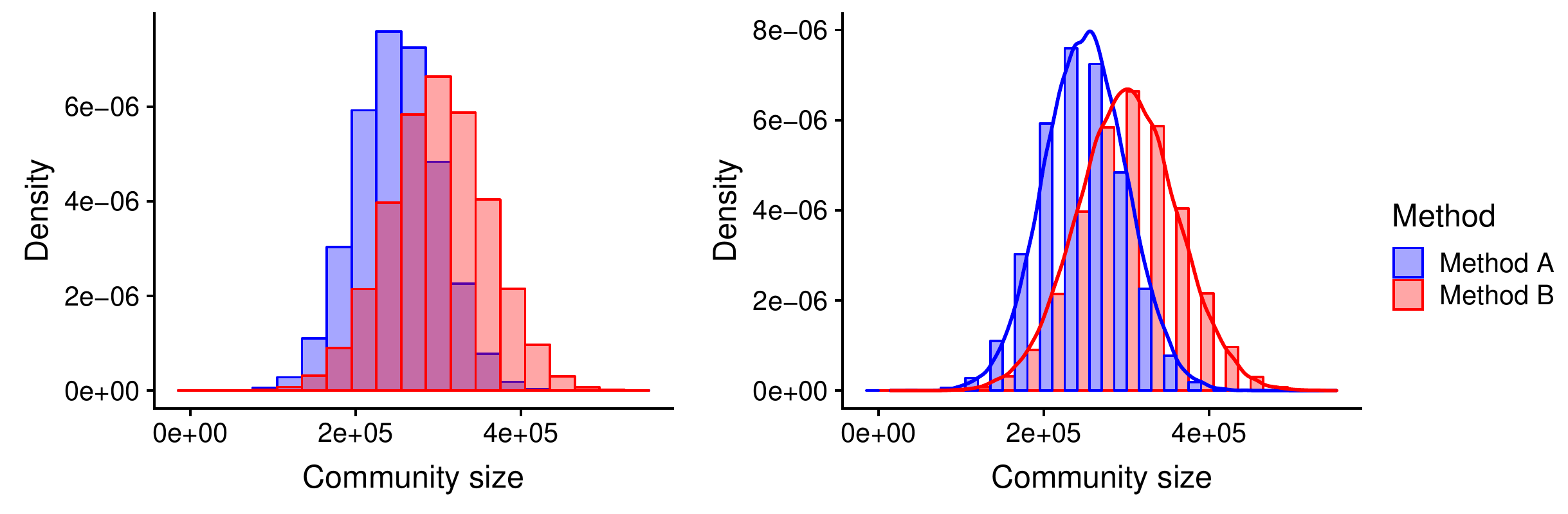}
  \caption[The distribution of sizes of communities detected by two different methods.]{The distribution of sizes of communities detected by two different methods. On the left (a) overlap fraction using a histogram, on the right (b) when community sizes interlace, the similarity is better estimated using a kernel density estimator. } 
  \label{fig:overlapsimilar}
\end{figure}

First, we denote two 2-tuples $(\mathcal{A},n^a)$ and $(\mathcal{B},n^b)$ being the multisets representing all communities detected on a set of networks $\mathcal{G} = \{G\}$ by method $A$ and method $B$ respectively, where $\mathcal{A} = \{x_1^a,x_2^a,...,x_r^a\}$ and $\mathcal{B} = \{x_1^b,x_2^b,...,x_s^b\}$ being the ascending ordered sets of sizes of communities: $1 \leq x_1^{a} < x_2^{a} < ... < x_r^{a}$ and $1 \leq x_1^{b} < x_2^{b} < ... < x_s^{b}$. The multiplicity functions $n^a: \mathcal{A} \rightarrow \mathbb{N}_{\geq 1}$ and $n^b: \mathcal{B} \rightarrow \mathbb{N}_{\geq 1}$ measure the number of communities of sizes $x_i^a$ and $x_i^b$ respectively. Let $N^a = \sum_{i=1}^{r} n^a(x_i^a)$ and $N^b = \sum_{i=1}^{s} n^b(x_i^b)$ being the total number of communities of all sizes detected by each method, we define a similarity function describing the closeness of $A$ and $B$ on $\mathcal{G}$ as:

\begin{equation}
S_{\mathcal{G}}(A,B) = \frac{1}{2} \sum\limits_{i=1}^{r} \sum\limits_{j=1}^{s} \min \left\{ \frac{n^a(x_i^a)}{N^a}, \frac{n^b(x_j^b)}{N^b} \right\} \delta(x_i^{a},x_j^{b}), 
\label{eqn:sim1}
\end{equation}

where $\delta(x_i^a,x_j^b) = 1$ if $x_i^a = x_j^b$ and $0$ otherwise. Equation (\ref{eqn:sim1}) is simply the common fraction of same-size communities detected on $\mathcal{G}$ by both $A$ and $B$: $0 \leq S_{\mathcal{G}}(A,B) \leq 1$. This definition seems to be intuitive but does not work well in practice. As illustrated in Figure \ref{fig:overlapsimilar}(b), when the sizes interlace with each other, a low score will be produced, although the level of similarity in this case is that of Figure \ref{fig:overlapsimilar}(a). Choosing an appropriate binning interval would mitigate the problem. This solution is quite inflexible, however.%: it is sensitive to the characteristic of data as well as to the functionality of the methods in use. 
A straightforward alternative can be envisioned by using a kernel density estimator to uncover the probability density function as shown by the solid lines in Figure \ref{fig:overlapsimilar}(b). In this way, we approximate the common fraction of same-size communities of Equation (\ref{eqn:sim1}) by the overlapping area of two corresponding continuous distributions. The premise behind this estimation is that two similar methods do not always produce a large portion of exact same-size communities but rather a large portion of comparable-size ones. Hence, we consider the following estimator to take into account local information of community size $x_0$:

\begin{equation}
\widehat{f}(x_0) = \frac{1}{hn} \sum\limits_{i} K\left( \frac{x_i - x_0}{h} \right),
\label{eqn:estimator1}
\end{equation}
where $h$ is the bandwidth controlling the neighborhood interval around $x_0$ and $K$ is the kernel function controlling the weight given to the observations $\{x_i\}$, chosen as Gaussian in our analysis. One would wonder why we use a Gaussian, whereas Power law may be a better fit, as some papers (eg. \cite{clauset:2004a}) show that community size in many real world networks exhibits a power law distribution. We would like to recall that expecting such groundtruth-like partitions does not mean that community detection methods impose this property into their mechanism and produce Power law distribution.
For example, on the Zachary network, although we try to discover two equivalent-size groups, community detection methods identify two to four similar-size communities. Some methods indeed fit a power law, but as a matter of fact, many others have a tendency to partition network nodes into quite balanced groups (spectral methods, modularity optimization methods, SBM, etc.) as seen in the previous section. This led us to use a Gaussian kernel, as it fits well in most of our cases on our large dataset and our panel of methods. In a less generic and exhaustive context, when the resulting community size varies significantly, following a Power law distribution, we, of course, recommend using the appropriate law.

Using our estimator, we rewrite the similarity function defined in Equation (\ref{eqn:sim1}) as follows:

\begin{equation}
S_{\mathcal{G}}(A,B) = \int \min\{\widehat{f}^{(a)}(x),\widehat{f}^{(b)}(x)\} dx,
\label{eqn:sim2}
\end{equation}
where 
\begin{equation}
\widehat{f}^{(u)}(x) = \frac{1}{hN^u} \sum\limits_{i}^{N^u} \left[ n^u(x_i^u) K\left( \frac{x_i^{u} - x}{h} \right)\right],
\label{eqn:estimator2}
\end{equation}
with $u \in \{a,b\}$. In the estimations of this paper, the bandwidth $h$ is selected based on the normal reference rule \cite{silverman:1986a} to minimize the mean integrated squared error.

Using equations (\ref{eqn:sim2}) and (\ref{eqn:estimator2}) to estimate the similarity between pairs of detection methods on a large dataset will help us to discover different behaviors of community detection methods. Since the accuracy of the estimator depends on the networks of the dataset that we analyze, the result will obviously have to be relativized. However, our large and representative corpus helps to reduce the dependency impact.

\begin{figure}[ht]
 \centering
 \includegraphics[width=0.9\linewidth]{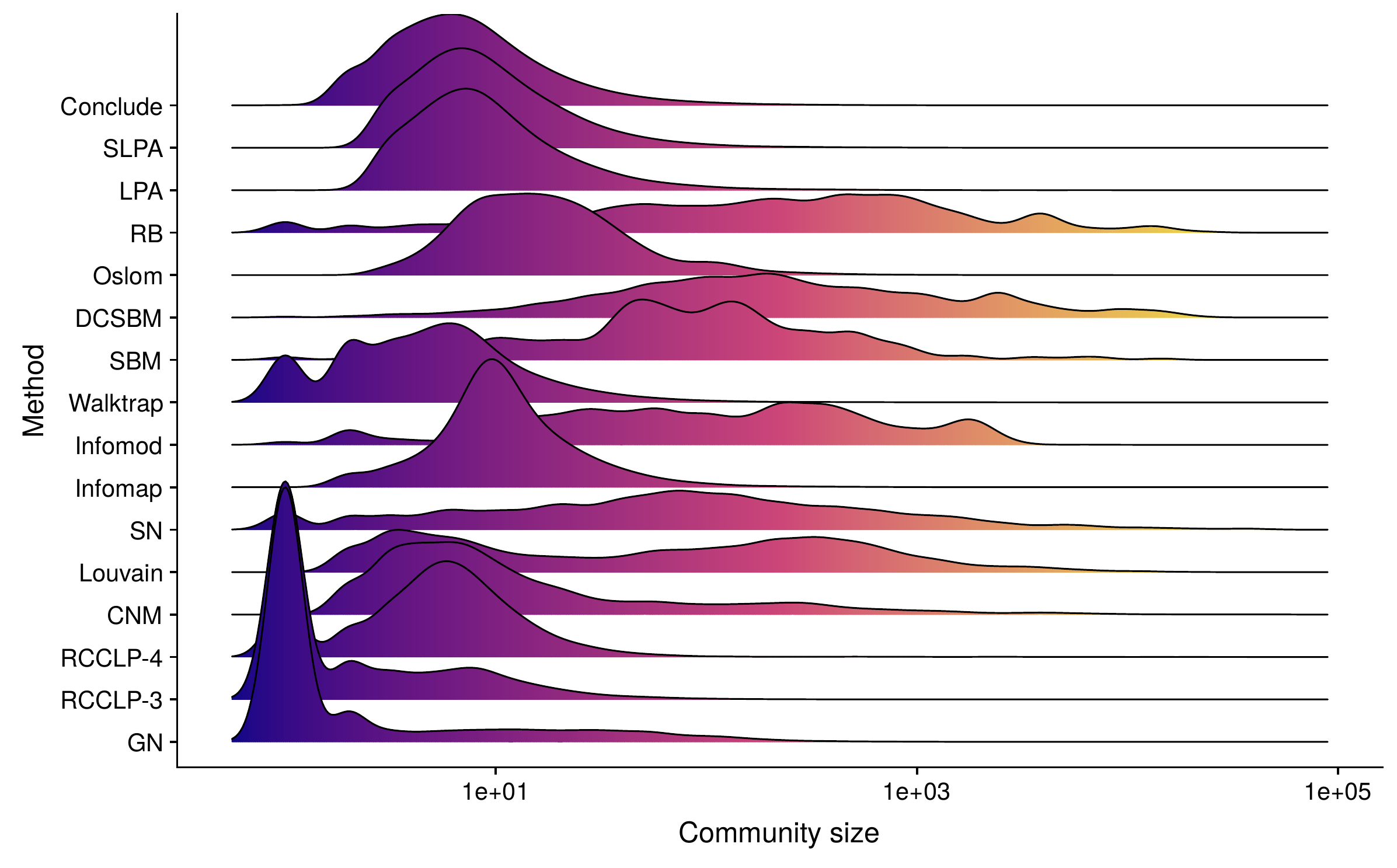}
  \caption[Community size distributions, for all the communities from the partitions detected on all the networks. The distributions are smooth using a Gaussian kernel estimator.]{Community size distributions, for all the communities from the partitions detected on all the networks. The distributions are smooth using a Gaussian kernel estimator. The gradient color is used only for ease of observation.}
  \label{fig:size-distribution-all}
\end{figure}

\subsubsection{Experimental results}
From the communities identified in the previous section, we proceed to measure the volumes of communities detected by each method to determine the elements of the corresponding 2-tuples. Finally, we use the similarity function defined by Equation (\ref{eqn:sim2}) to estimate the closeness between each pair of methods. Due to the huge number of experiments, only processes having a reasonable theoretical estimated time and memory consumption are maintained (less than a few days and at most 30 to 40 GBytes of memory). The outcome distributions are illustrated in Figure \ref{fig:size-distribution-all}. 

As we can see, there is a clear difference in the densities of community size, showing that these methods have various partitioning strategies. Knowing that methods belonging to the same theoretical group (as shown in Table \ref{tab:implementation}) are placed next to each other, we can see some agreements between the theoretical families, with practical outcomes as follows:

\begin{itemize}
  \item[] \textbf{Edge removal}: \textit{GN} and \textit{RCCLP-3} have very similar distributions where a large number of communities are very small. This is due to the fact that in some highly local centralized networks, having star-like structures \cite{dao:2018a}, they have a tendency to remove edges connecting hub and peripheral nodes and to create singletons (single node community). This phenomenon is less distinguishable on \textit{RCCLP-4}, since there are far fewer quadrangular than triangular connections in networks.
  
  \item[] \textbf{Modularity optimization}: Modularity is known to suffer from resolution limit phenomenon \cite{fortunato:2006a}, which often aggregates small communities in large scale networks. We can see from Figure \ref{fig:size-distribution-all} that \textit{Louvain} and \textit{SN} found very large communities, as predicted. At the same time, there are also a comparable number of small communities which are found on small graphs. However, the behavior is a little different on \textit{CNM} method, which is an agglomerative clustering algorithm based on modularity optimization.
  
  \item[] \textbf{Dynamic process}: Methods in this family show very discernible distributions although all are based on dynamic processes. In fact, they make different assumptions about community structure and searching mechanisms. Therefore, belonging to the same theoretical family does not lead to a similarity in practical results.
  
  \item[] \textbf{Statistical inference}: the Bayesian \textit{SBM}  and \textit{DCSBM} use the Monte Carlo sampling process, which is very time-consuming, in order to sweep the solution space. This makes the method unfeasible, if the maximum number of clusters is not limited. Indeed, in the default version, the maximum number of communities is limited to 25, meaning that the $(DC)SBM$ methods find very large communities in large networks. On the other hand, the Oslom method uses an agglomerative discovery mechanism and identifies globally smaller communities.
  
  \item[] \textbf{Other methods}: In this group, \textit{LPA}, $SPLA$ (both based on label propagation) and \textit{Conclude} display almost identical distributions. $RB$ method, being based on a very close concept with modularity (with a tuning parameter), exhibits a similarity with modularity optimization based methods.   
  
\end{itemize}

\begin{figure}[ht]
 \centering
 \includegraphics[width=0.70\linewidth]{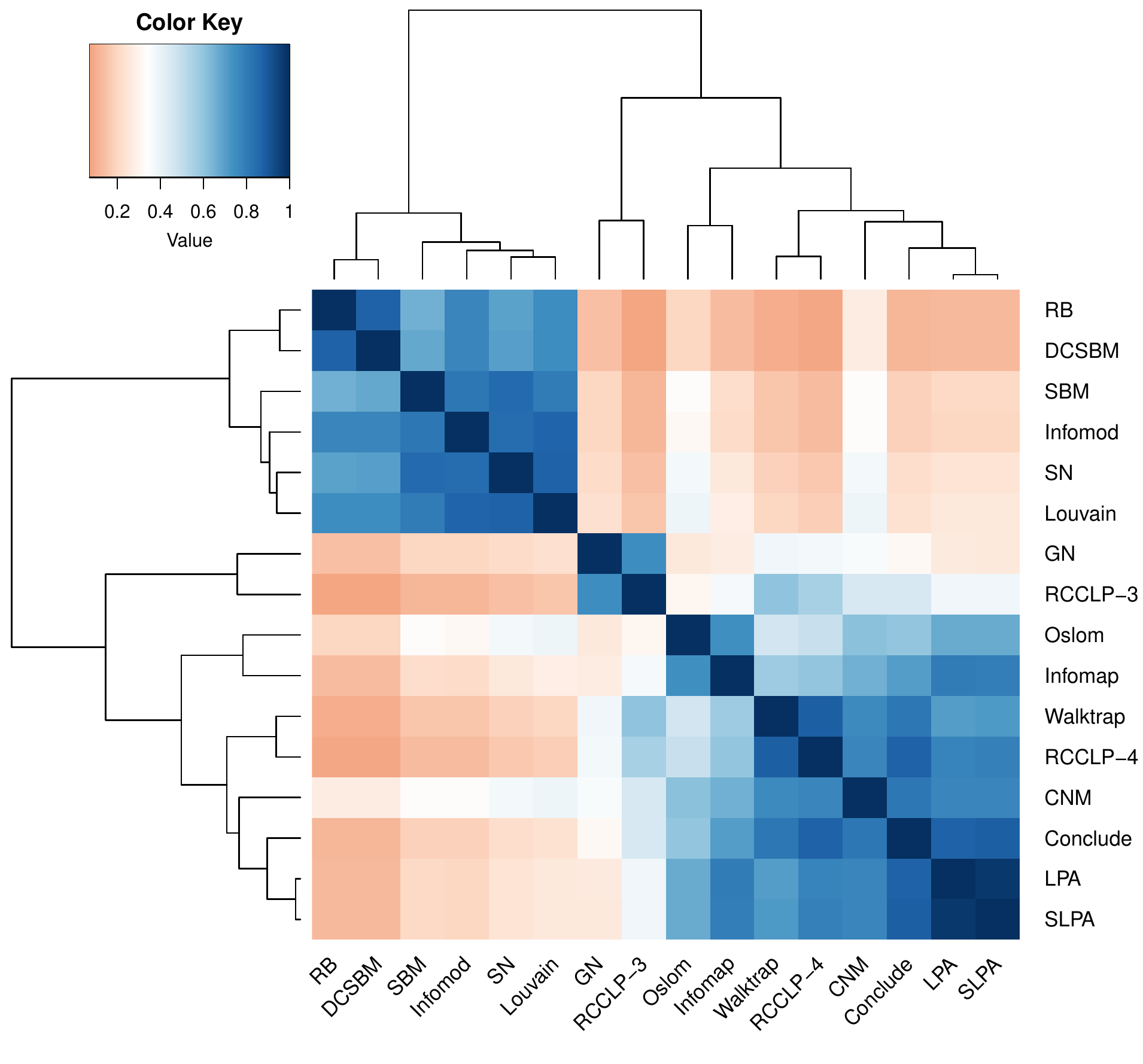}
  \caption[The similarity between community detection methods in terms of size fitting quality. Two methods are considered to be similar, if they share a large fraction of same-size communities. Methods are ordered using hierarchical clustering.]{The similarity between community detection methods in term of size fitting quality. Two methods are considered to be similar if they share a large fraction of same-size communities. Methods are ordered using hierarchical clustering \cite{joe:1963a}. The dendrogram proposes a hierarchical structure of the fitting closeness. Blue colors mean high similarity.}
  \label{fig:fit-sim-matrix}
\end{figure}

Quantitatively, applying the estimator presented in Equation (\ref{eqn:estimator2}), to compute pairwise similarities between the methods, leads us to the results demonstrated in Figure~\ref{fig:fit-sim-matrix}. 

A well-known method for comparing two distributions is the Kolmogorov-Smirnov test (KS). Under the null hypothesis, the two distributions are identical. The test generates the cumulative distributions of the distributions, reports the maximum difference between them and calculates a p-value. If the Kolmogorov-Smirnov statistic (distance) is small or the $p$ value is high (above the significance level, e.g. $0.05$), the assumption that the distributions of the two samples are identical cannot be rejected. Conversely, we can reject the null hypothesis, if the p-value is low. In our case, $p-value < 0.05$ for all the pairs, which means that the distributions do not fit the same model, except for a single pair $(Conclude,LPA)$ with a value $p = 0.07$. 
A deeper exploration shows that these two methods do not fit a normal law, but they fit better a power law distribution than other heavy-tailed distributions (with a non-significant difference, however). 
In practice, the KS distance is useful, if one is mostly interested in how similar two datasets are. %But as this is a local distance (the maximum difference at some point between two curves), differences are not very well rendered. However, 
As shown on the Table~\ref{tab:KS-distance}, the smallest distances occur with couples (SLPA, LPA), (Conclude, LPA), (DCSBM, RB), (RCCLP-4, Walktrap), (GN, RCCLP-3), (Infomod, SN), (Louvain, SN), which is consistent with our results.

\begin{table}[ht]
	\centering
	\caption{Extract of the smallest Kolmogorov-Smirnov distances between community size distributions}
	\label{tab:KS-distance}
	\begin{minipage}{\textwidth}
		\begin{tabular}{llcc}
			\hline \hline
			Method & Method & KS-Distance\hspace{4pt} & p-value \\
			\hline
			SLPA & LPA & 0,007 & 0,017 \\
			Conclude & LPA & 0,067 & 0,075 \\
			DCSBM & RB & 0,070 & 0,000 \\
			RCCLP-4 \hspace{4pt} & Walktrap & 0,080 & 0,000 \\
			GN & RCCLP-3 & 0,098 & 0,000 \\
			Conclude & SLPA & 0,118 & 0,000 \\
			Infomod & SBM & 0,118 & 0,000 \\
			Infomod & SN & 0,126 & 0,000 \\
			DCSBM & Infomod & 0,127 & 0,000 \\
			Louvain & SN & 0,127 & 0,000 \\
			\hline \hline
		\end{tabular}
	\end{minipage}
\end{table}

Thus, according to the community size criterion, the methods can be classified into different classes of partitioning strategy. Similar results are also discovered by \cite{ghasemian:2018a}, as they found that methodological similar methods usually lead to a similar outcome. The separations are clearly shaped, showing that the distinction is very clear between groups. Therefore, we choose to characterize these methods by 3 (possibly 4) principle groups as follows:

\begin{enumerate}
 \item \textbf{Group 1} - $RB$, \textit{DCSBM}, \textit{SBM}, \textit{Infomod}, \textit{SN}, \textit{Louvain}: Methods in this group discover communities whose sizes vary in wide range of spectra, from very small to very large communities. The characterized community size distribution is quite flat, meaning all sizes are nearly equally considered.
 
 \item \textbf{Group 2} - \textit{GN} and \textit{RCCLP-3}: These two methods identify a huge number of very small communities including singletons regardless of network size. As a consequence, there are few variations in community volume.
 
 \item \textbf{Group 3} - the others: These methods produce communities whose sizes approach bell-shaped distribution. The strategy can be translated as: neither  left nor right, i.e. communities that are neither too small and nor too big. 
\end{enumerate}

This characterization could help us to identify appropriate groups of community detection methods, according to different community size fitting strategies. Also, it helps to avoid brute-force attempts when a method does not succeed in proposing desired partitions by proposing substitute solutions. Moreover, by combining with the previous time-computation analysis in Section \ref{sec:timeanalysis}, one could also choose a group of methods corresponding to size distinction criteria and then select the fastest method that leads to a desired outcome.

Community distribution (or number of communities) is just one possible quality dimension, albeit possibly one of the most intuitive and important pieces of information when choosing a clustering method. In the next part, we demonstrate some techniques that can be used to define other similarity aspects. We show that these notions of similarity can be combined to accentuate the distinction between different community detection methods.

\section{Quality profiling of community detection methods}
\label{sec:goodnesssimilarity}
\subsection{Community quality metrics}
A popular way to evaluate the structure of communities is to design quality metrics, in order to measure different expected characteristics from subgraphs that we want to obtain. In practice, metrics using network generative models are sometimes preferable, as they reflect different assumptions about the underlying mechanisms that create community structure. One of the most widely used metrics, that quantifies the quality of community structure, is the \textit{modularity} function. The idea here is to reveal how the quality of an identified community structure is different from that which would be expected. Although some unexpected phenomena, known as resolution limit \cite{fortunato:2006a,traag:2011a} have been exposed, when the scale of community size is too small, modularity remains the standard measure of quality.  

The advantage of this approach is that one can \textit{``embed"} the assumption of community structure inside quality functions. Hence, they provide better performance in some cases. However, community structure is quite an open question, such that, according to different mechanisms that render the structure of networks, there will be models that are more suitable than others. %Modeling networks, therefore, contribute greatly to the evaluation of network structure as well as community structure.

We present some quality metrics (also called ``community scoring functions" or less often ``goodness metrics") to evaluate community structure. Many of them are initially or gradually employed as objective functions in some community detection methods, since they expose good performance in searching processes. Firstly, we recall some notations that will be used to describe structural characteristics of communities. A graph $\mathcal{G} = (\mathcal{V}, \mathcal{E})$ consisting of $n = |\mathcal{V}|$ vertices and $m = |\mathcal{E}|$ edges can be represented by an associated adjacency matrix $A$. Given a community $C$ of $n_C$ vertices being a subgraph of $\mathcal{G}$ in an arbitrary partition $P$, a function $f(C)$ or $f(P)$ quantifies a structural goodness feature of community $C$ or the whole partition $P$, according to a particular expectation of community structure. Let $m_c$ be the number of edges inside community $C$, $m_C = |(i,j) \in \mathcal{E} : i \in C, j \in C|$, $l_C$ be the number of edges that connect $C$ to other vertices outside of $C$, $l_C = |(i,j) \in \mathcal{E} : i \in C, j \not\in C|$. Any vertex $i$ belonging to community $C$ has an \textit{internal} degree $k_{iC}^{int}$ and an external degree $k_{iC}^{ext}$ satisfying $k_{iC}^{int} + k_{iC}^{ext} = k_i$, where $k_i$ is the total degree of vertex $i$. The internal and external degrees can be expressed via the adjacency matrix $A$ as: $k_{iC}^{int} = \sum_{j \in C} A_{ij}$ and $k_{iC}^{ext} = \sum_{j \notin C} A_{ij}$. If vertex $i$ in community $C$ has $k_{iC}^{ext} > 0$ and $k_{iC}^{int} \geq 0$ , $i$ is called \textit{boundary vertex} since $i$ has neighbor(s) outside of $C$. Otherwise, if $k_{iC}^{ext} = 0$ and $k_{iC}^{ext} > 0$, $i$ is called \textit{internal vertex} which only has connections with other vertices in the same community. In this paper, we employ the following functions to evaluate community structure:

\subsubsection{Newman-Girvan Modularity}
The standard version of modularity \cite{newman:2004a} reflects the difference for the fraction of intra community edges of a partition, with the expected number of such edges, if distributed according to a \textit{null model}. In the standard version of modularity, the null model preserves the expected degree sequence of the graph under consideration. In other words, the modularity compares the real network structure with a corresponding one where nodes are connected without any preference about their neighbors. There are several ways to mathematically express modularity. In order to compare standard modularity with other variants, it is convenient to consider modularity as a sum of contributions from pairs of vertices of the same community:

\begin{equation}
\label{eqn:standardmodularitymetric}
Q_{NG}(P) = \frac{1}{m}\sum\limits_{c \in P} \left[m_c-\frac{(2m_c + l_c)^2}{4m}\right]
\end{equation}

Despite some existing issues such as the widely discussed resolution limits \cite{fortunato:2006a} or the indistinguishability structures of two partitions having similar modularity scores, the Newman-Girvan modularity is still widely used in the community as a proof of quality.   

\subsubsection{Erd\H{o}s-R\'{e}nyi Modularity}
The Newman-Girvan modularity has attracted much attention in the research literature. Many alternative derivations have been proposed to adapt to different contexts. Some use different null models to quantify the modular structure of partitions. For example, one could assume that vertices in a network are connected randomly with a constant probability, $p$, as formulated in the Erd\H{o}s-R\'{e}nyi (ER) model \cite{erdos:1959a}. The connection probability is calculated as $p = \frac{2m}{n(n-1)}$, being the number of presented edges over the total number of edges that could be established. The expected number of edges in a community of size $n_c$ becomes $\langle m_c\rangle = p{n_c\choose 2}$. This null model leads us to the ER Modularity:

\begin{equation}
\label{eqn:ermodularitymetric}
Q_{ER}(P) = \frac{1}{m}\sum\limits_{c \in P} \left[m_c- \frac{mn_c(n_c-1)}{n(n-1)} \right]
\end{equation}

\subsubsection{Modularity Density (D-value or D-modularity)}
Standard modularity is found to be impacted by resolution limits \cite{fortunato:2006a}, i.e., it is claimed that the sizes of detected modules depend on the size of the whole network, such that optimizing standard modularity can not identify communities having a small number of vertices. The expected number of intra-community edges is highly sensitive to the total number of edges in the whole network \cite{rosvall:2007a}, as can be observed in the second term of Equation (\ref{eqn:standardmodularitymetric}). Modularity density \cite{li:2008a} is one of several propositions envisioned to palliate this issue. The idea of this metric is to include the information about community size in the expected density of a community to avoid the negligence of small and dense communities. For each community $C$ in partition $P$, it uses the \textit{average modularity degree} calculated by $d(C) = d^{int}(C) - d^{ext}(C)$ where $d^{int}(C)$ and $d^{ext}(C)$ are the average internal and external degrees of $C$ respectively to evaluate the fitness of $C$ in its network. Finally, the modularity density can be calculated as follows:

\begin{equation}
\label{eqn:modularitydensity}
Q_{D}(P) = 
\sum\limits_{c \in P} 
\frac{1}{n_c} \left( \sum_{i \in c} k^{int}_{ic} - \sum_{i \in c} k^{ext}_{ic} 
\right) 
\end{equation}

\subsubsection{Z-modularity}
Z-modularity is another variant of the standard modularity that is proposed to avoid the resolution limit \cite{miyauchi:2016a}. The founding idea of this version is based on the observation that the difference between the fraction of edges inside communities and the expected number of such edges in a null model should not be considered as the only contribution to the final quality of community structure. Specifically, the authors recommend that the statistical rareness of a community should also be taken into consideration. In other words, the rarer the structure of a commnunity, the greater is its contribution to the final modularity. Therefore, the variance of the probability distribution of the fraction of the number of edges within each community is included in the quality function through a standardization using Z-score. Following the null model of the standard modularity, the probability that an edge in placed inside community $C$ is $p = (\frac{D_C}{2m})^2$, where $D_C = 2m_C + l_C$ is the total degree of community $C$. The number of edges in each community follows a binomial distribution with the probability $p$ and its normalized value approaches a normal distribution when the number of edges is sufficiently large. The statistical rarity of partition $P$ in terms of the fraction of the number of intra-community edges using Z-score is thus translated into Z-modularity as follows:

\begin{equation}
\label{eqn:zmodularity}
Q_Z(P) = 
\left[\sum\limits_{c \in P} \frac{m_c}{m}- \sum\limits_{c \in P} \left( \frac{D_c}{2m} \right)^2 \right]
\left[\sum\limits_{c \in P} \left( \frac{D_c}{2m} \right)^2 
\left( 1 -  \sum\limits_{c \in P} \left( \frac{D_c}{2m} \right)^2 \right) \right]^\frac{-1}{2}
\end{equation}

\subsubsection{Surprise}
This statistical approach proposes a quality metric assuming that edges between vertices emerge randomly according to a hyper-geometric distribution \cite{aldecoa:2011a}. Specifically, for a graph of $n$ vertices and $m$ edges, there are $M = {n\choose 2}$ possible ways of drawing $m$ edges. For a particular partition, there are $M^{int} = \sum_{C \in P} {n_c\choose 2}$ possible ways of drawing an intra-community edge. Surprise metric computes the (minus logarithm of) probability of observing at least $m^{int} = \sum_{C \in P} \frac{k_{C}^{int}}{2}$ intra-community edges within $m$ draws without replacement from the population of $M$ possible choices in which there are precisely $M^{int}$ possible intra-community edges. This probability is formalized as follows:

\begin{equation}
\label{eqn:surprisemetric1}
S(P) = - \log \sum\limits_{k = m^{int}}^{\min(m,M^{int})} 
\frac{{M^{int}\choose k} {{M- M^{int}}\choose {m-k}}}{{M\choose m}}.
\end{equation}

However, this formulation is not easy to work with in large-scale networks due to numerical computational problems. Hence, \cite{traag:2015a} provides an asymptotic approximation for the metric, which is a good alternative. By assuming that the relative number of intra-community edges $q = \frac{m^{int}}{m}$ and the relative number of expected intra-community edges $\langle q \rangle = \frac{M^{int}}{M}$ remains fixed, Surprise metric is approximated at:

\begin{equation}
\label{eqn:surprisemetric2}
S(P) \approx m D(q || \langle q \rangle),
\end{equation}
where $D(q || \langle q \rangle)$ is the Kullback–Leibler divergence \cite{kullback:1951a}:

\begin{equation}
\label{eqn:kldivergence}
D(q || \langle q \rangle) = p \log \frac{p}{\langle q \rangle} + 
(1-q) \log \frac{1 - q}{1 - \langle q \rangle}.
\end{equation}

According to the Surprise metric, the higher the score of a partition, the less likely it is that it resulted from a random realization, and thus the better the quality of the community structure will be.  

\subsubsection{Significance}
This metric uses a similar approach to the Surprise metric. It also estimates how likely a partition of dense communities is to appear in a random graph, but in a different way \cite{traag:2015a}. While Surprise uses global quantities $q$ and $1 - \langle q \rangle$, Significance compares each community density $p_C = \frac{m_C}{{n_C\choose 2}}$ to the average graph density $p = \frac{m}{M}$. The asymptotic form of Significance can be written as:

\begin{equation}
\label{eqn:significancemetric1}
Z(P) = \sum\limits_{C \in P} {n_C\choose 2} D(p_C || p).
\end{equation}
Similarly, $D(x||y)$ is the Kullback–Leibler divergence defined in Equation (\ref{eqn:kldivergence}). Generally, if the number of communities is relatively large or the graph is relatively dense, Significance is more discriminative than Surprise. On the other hand, in the case where $\langle q \rangle > p$, Surprise can be better than Significance \cite{traag:2015a}. 

\subsection{Co-performance index}
We propose a new comparative approach using a matrix called community detection \textit{co-performance} matrix. The idea is that, given an expected quality function, one could investigate whether there exists a correlation in the efficiency of enhancing (or aggravating) its scores between different methods. The co-performance matrices reveal how understanding the performance of a method in optimizing a quality would allow us to predict the performance of other methods on the same quality. Therefore, an exhaustive analysis of co-performance matrices on many qualities allows us to profile the characteristics of community detection methods in a comparative way. The index could be calculated as follows:

Let methods $A$ and $B$ divide a graph $G_i = (V_i, E_i)$ of dataset $\mathcal{G} = \{G_i|i =1 .. N\}$ into $\alpha$ and $\beta$ communities described by partitions $P_{G_i}^a = \{C^a_{1G_i},C^a_{2G_i},...,C^a_{\alpha G_i}\} \in \mathcal{P}_{G_i}$ and  $P_{G_i}^b = \{C^b_{1G_i},C^b_{2G_i},...,C^b_{\beta G_i}\} \in \mathcal{P}_{G_i}$ respectively. We consider, solely, hard clustering methods, meaning $C_{uG_i}^a \cap C_{vG_i}^a = \varnothing : 1 \leq u < v \leq \alpha$ and $C_{uG_i}^b \cap C_{vG_i}^b = \varnothing : 1 \leq u < v \leq \beta$. A function $Q: \mathcal{P}_{G_i} \rightarrow \mathbb{R}$ quantifies a quality of a partition of graph $G_i$ according to a particular goodness aspect (or model). 

We define a \textit{co-performance index} of two methods $A$ and $B$ on $\mathcal{G}$ by their mutual capacity in discovering community structures showing a particular quality $Q$. In other words, each couple of methods should be assigned a high index according to a quality $Q$, if knowing the performance of one method significantly reveals  the information about the performance of the other. A straightforward solution for defining the index is using the Pearson correlation. In other words, it reflects the covariance of a quality score on two sets of partitions detected by two methods, indicating whether two methods \textit{agree} or \textit{disagree} about a quality criterion in the context of a network dataset.

Denoting $q^a_i = Q(P_{G_i}^a)$ and $q^b_i = Q(P_{G_i}^b)$, the co-performance index can be calculated as follows:

\begin{equation}
I_{\mathcal{G}}(A,B,Q) = \frac{N\sum q^a_i q^b_i - \sum q^a_i \sum q^b_i
}
{
[N \sum (q^a_i)^2 - (\sum q^a_i)^2]^{1/2}
[N \sum (q^b_i)^2 - (\sum q^b_i)^2]^{1/2}
},
\end{equation}
where $0 \leq I_{\mathcal{G}}(A,B,Q) \leq 1$. With a Pearson correlation, a high positive (resp. negative) score implies that two methods often find a strong consensus (resp. disagreement) in discovering communities with regards to a particular quality. Given a co-performance index, knowing the quality scores of one method could provide predictive information about the outcomes of the other method on the same dataset. This information could, in fact, be very useful in a context where alternative solutions must be deployed while maintaining an assumed quality, as expected.
We present in the following section the mutual performance of the presented detection methods by the previously presented quality functions. 
\begin{figure}[ht]
\centering
 \begin{subfigure}{0.40\textwidth}
  \centering
  \includegraphics[width=1\linewidth]{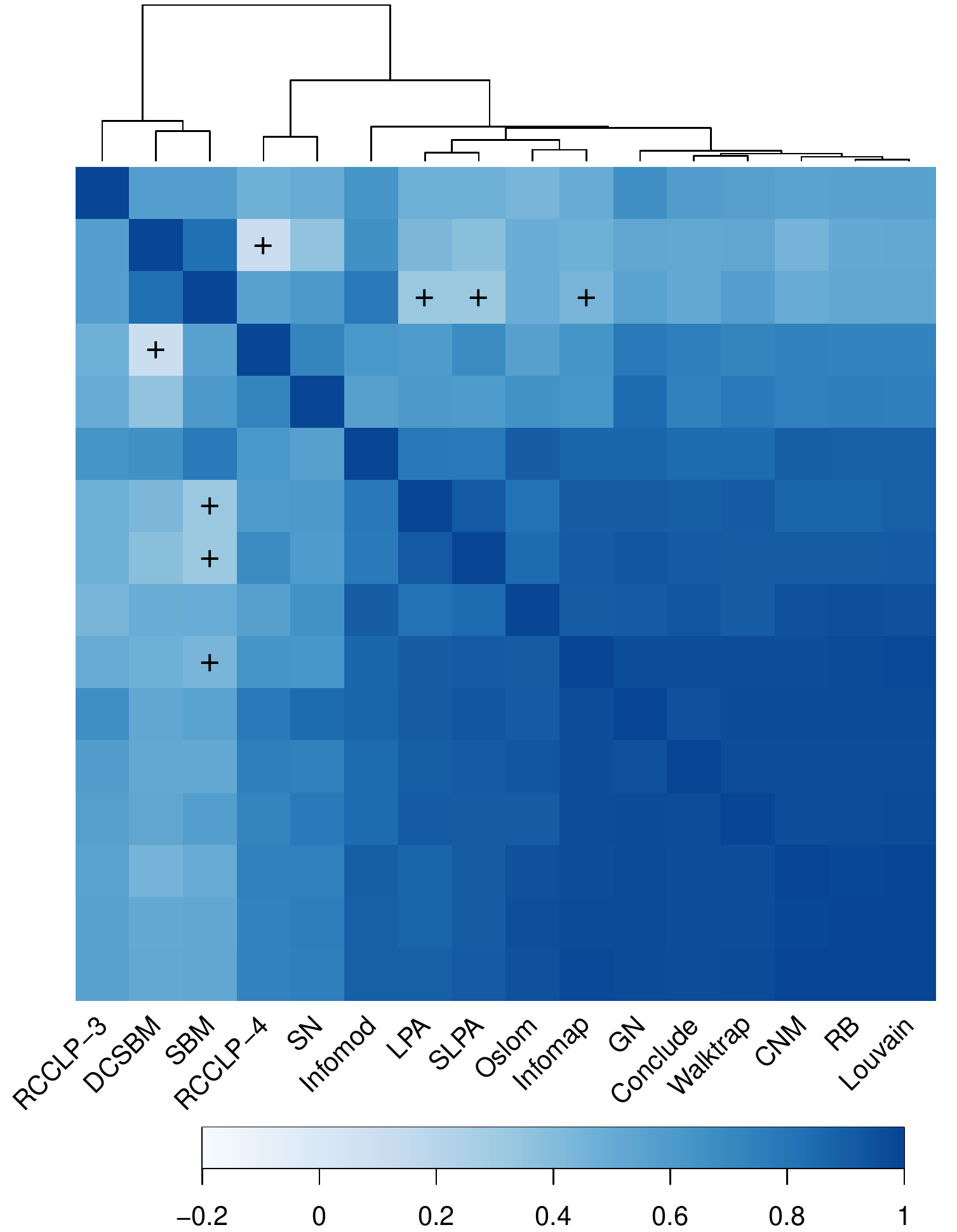}
  \caption{}
  \label{fig:qsd-corr-matrix}
\end{subfigure}
\begin{subfigure}{0.40\textwidth}
\includegraphics[width=1\linewidth]{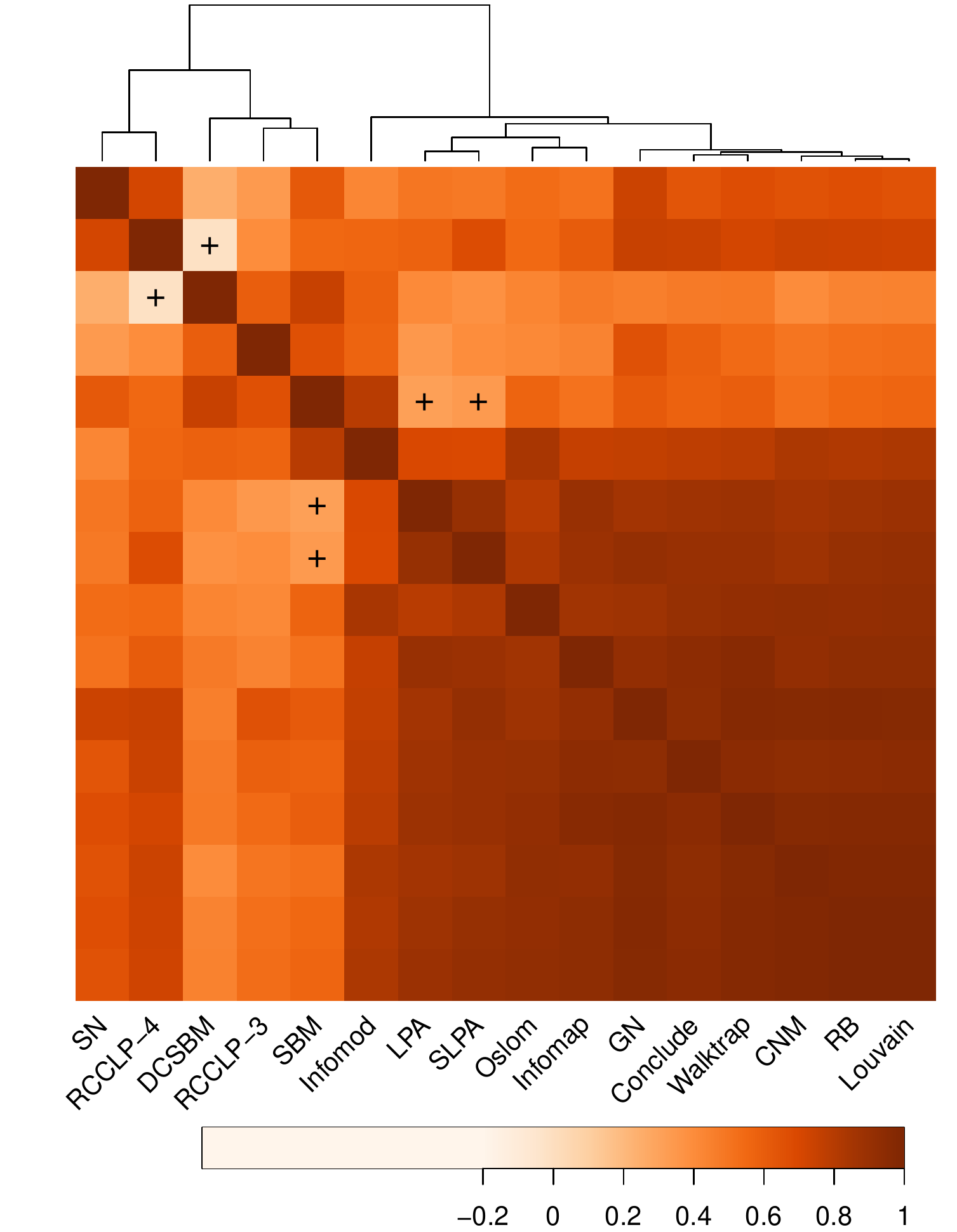}
    \caption{}
  \label{fig:qer-corr-matrix}
\end{subfigure}

 \begin{subfigure}{0.40\textwidth}
  \centering
  \includegraphics[width=1\linewidth]{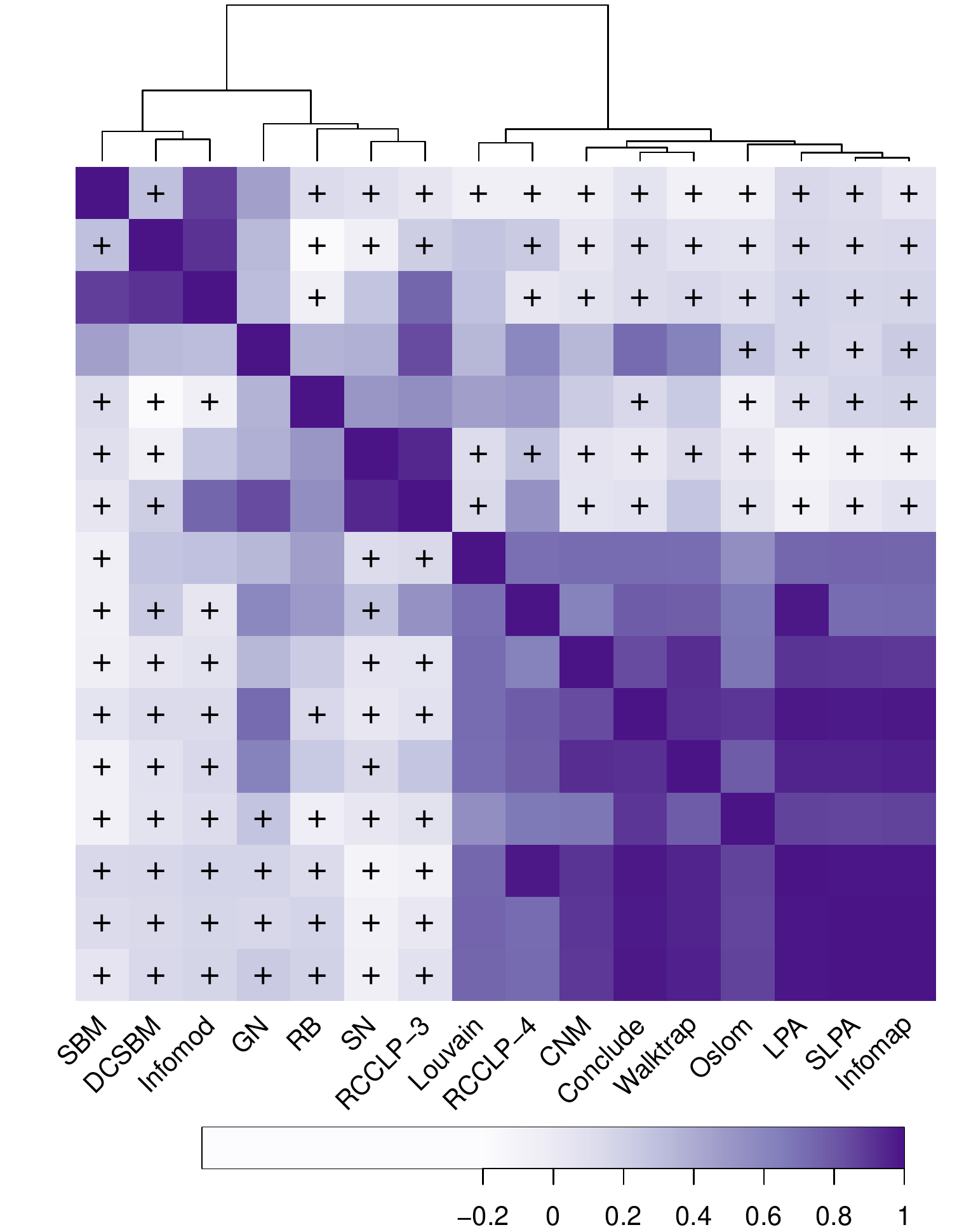}
  \caption{}
  \label{fig:qd-corr-matrix}
\end{subfigure}
\begin{subfigure}{0.40\textwidth}
\includegraphics[width=1\linewidth]{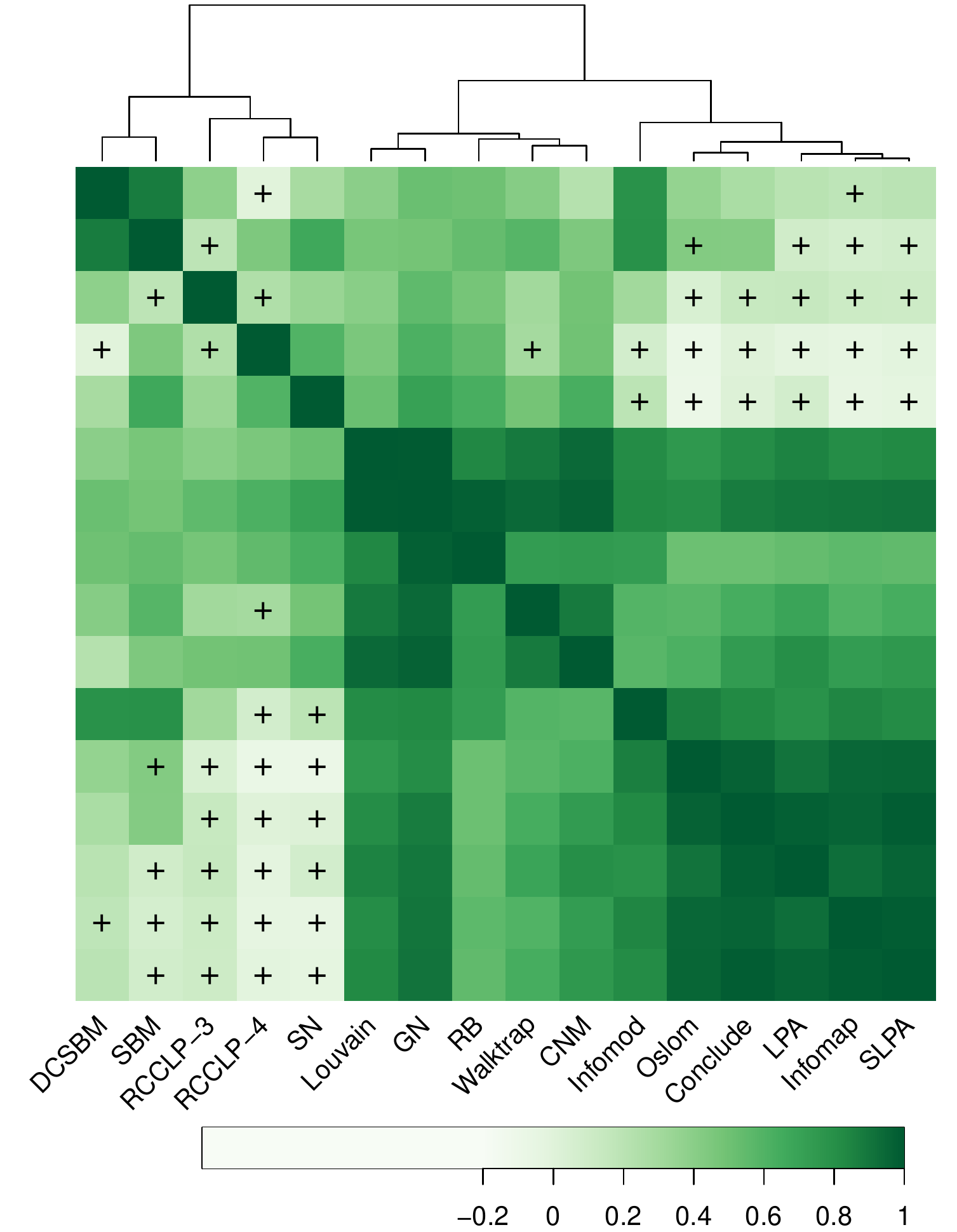}
    \caption{}
  \label{fig:qz-corr-matrix}
\end{subfigure}
\caption[The \textit{co-performance} matrices of different methods.]{The \textit{co-performance} matrices of different methods. The "+" marks indicate cases where p-values are larger than $0.05$. (a) Newman-Girvan modularity, (b) Erd\H{o}s-R\'{e}nyi modularity, (c) Density modularity and (d) Z modularity.}
\label{fig:q-corr-matrix}
\end{figure}

\begin{figure}[ht]\ContinuedFloat
\centering
 \begin{subfigure}{0.40\textwidth}
  \centering
  \includegraphics[width=1\linewidth]{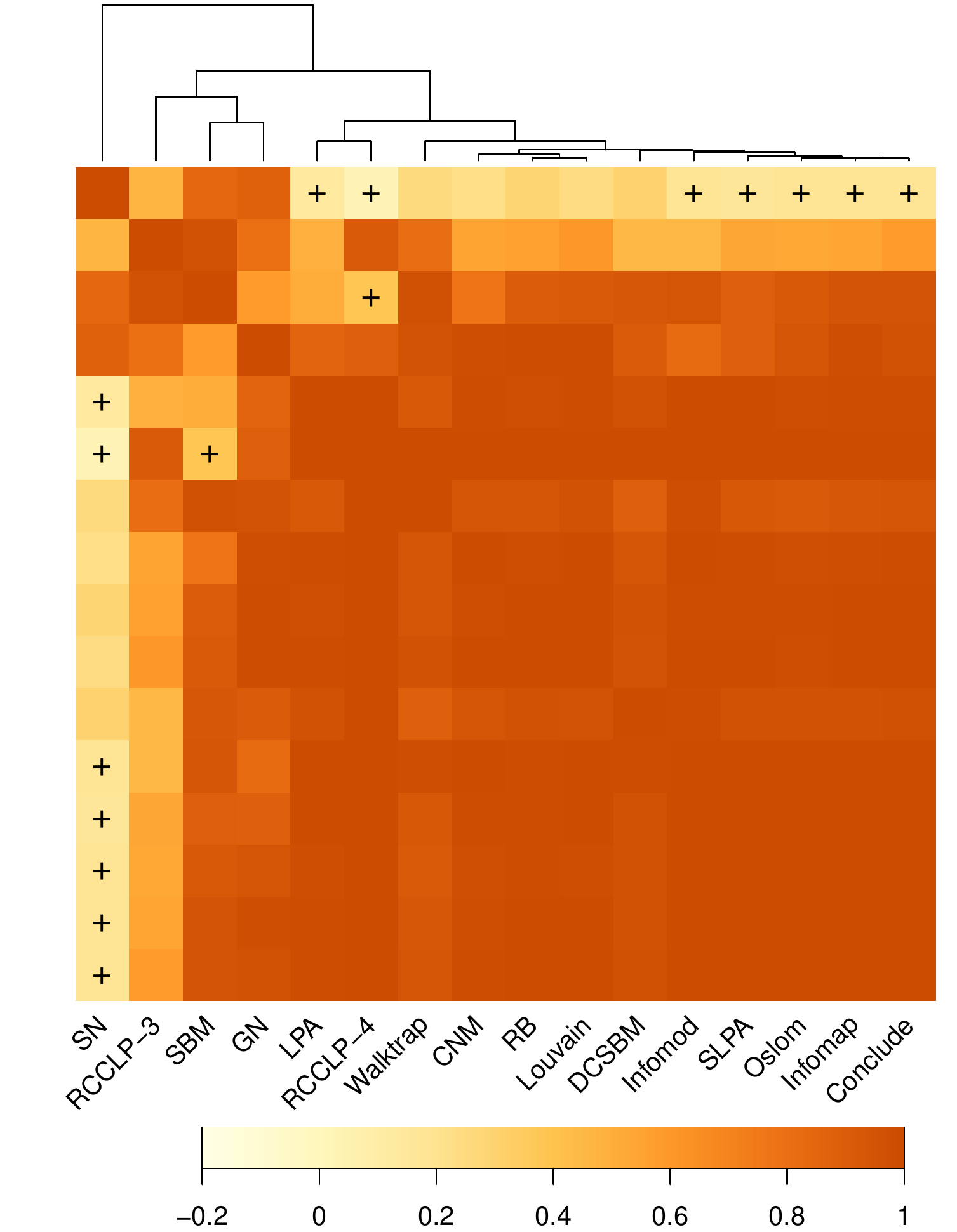}
  \caption{}
  \label{fig:sur-corr-matrix}
\end{subfigure}
\begin{subfigure}{0.40\textwidth}
  \includegraphics[width=1\linewidth]{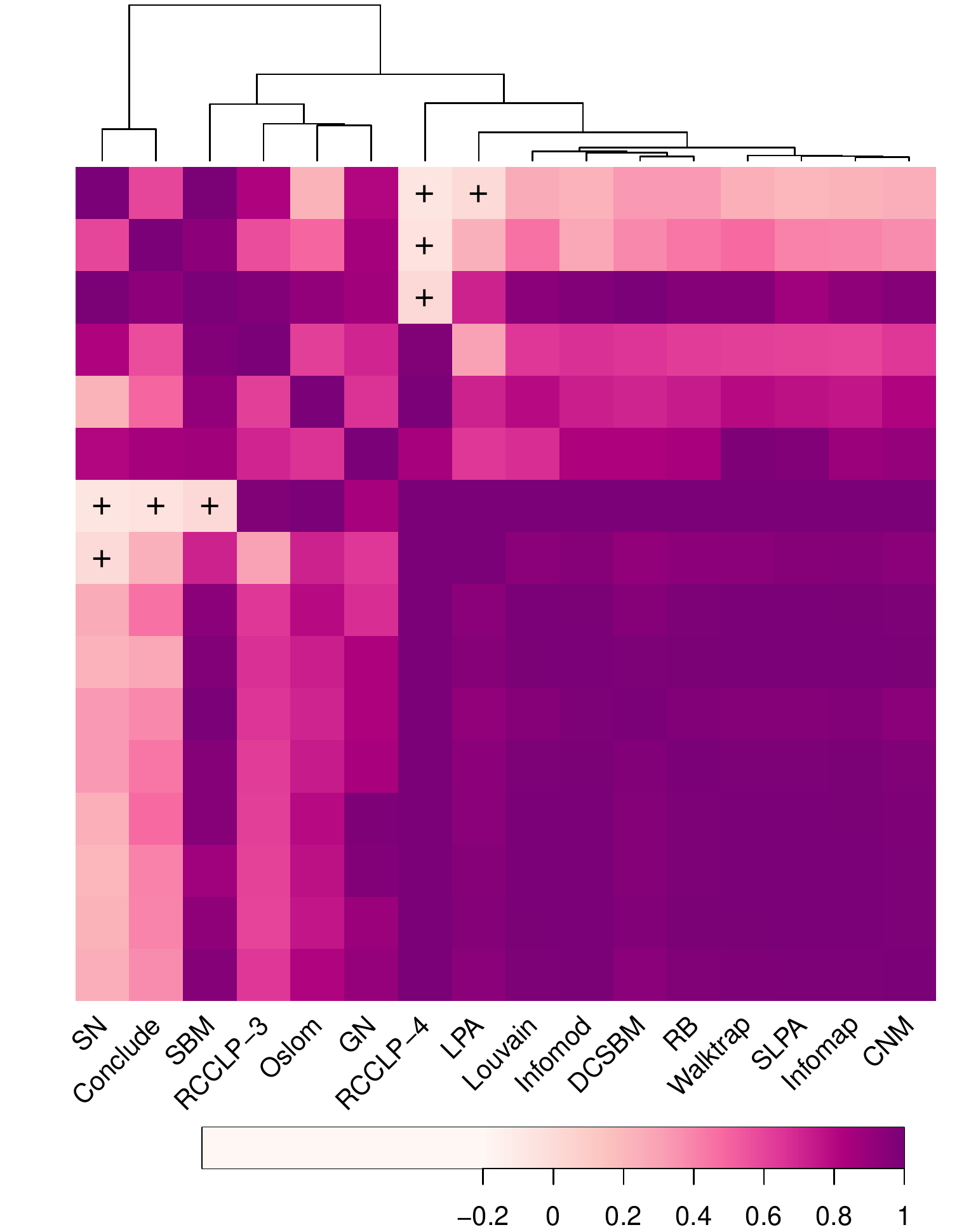}
    \caption{}
  \label{fig:sig-corr-matrix}  
\end{subfigure}
\caption[The \textit{co-performance} matrices of different methods (cont.)]{The \textit{co-performance} matrices of different methods (cont.) (E) Surprise and (F) Significance.}
\label{fig:corr-matrix}
\end{figure}

Figure \ref{fig:corr-matrix} illustrates the co-performance matrices, according to six different quality goodness criteria. Again, similarly to the previous section, goodness functions with a \textit{close} concept are placed together. For instance, \textit{NG} modularity and \textit{ER} modularity are both based on null models whose concept uses an expected fraction of intra-community edges. While the hypothesis of \textit{NG} version is to keep the expected degree sequence of the graph in question, the \textit{ER} version redistributes edges randomly with a constant average degree for every node. $D$-modularity and $Z$-modularity attempt to penalize large communities by including community sizes and significance level, respectively. One can notice a very slight similarity in the experimental results of the co-performance indexes between different quality functions. Also, it seems that the assumption about the quality of community structure has an impact on the co-performance outcome.

As shown in Figure~\ref{fig:corr-matrix}, there is a class of methods (\textit{Louvain, GN, CNM, RB, Infomod, Infomap, Walktrap, Oslom, LPA, SLPA, Conclude}) in which all methods show very consistent results, except for the case of $D$-modularity\footnote{In fact, density modularity is somewhat apart from other traditional ways of defining modularity, as it is not defined based on a null model but solely on edge density. The term $D$-modularity is misused in this sense}. Furthermore, there is also a strong relation between \textit{SBM} and \textit{DCSBM}. For the other methods (\textit{RCCLP} and \textit{SN}), no clear tendency could be observed from this experiment. The similarity of a large number of methods in many quality functions implies that, globally, if a method performs well on a given network, there is a signal that the others (from the same group) could also obtain good results. In other words, if the community structure in a network is clear, most methods will be able to detect it with more or less accuracy and conversely. However, as the co-performance indexes also vary significantly ($0.2$ to $0.3$) inside each group, there will be always a remarkable difference if one goes from one method to another. 

Within the case of density modularity shown in Figure \ref{fig:corr-matrix}(c), we discover that the sizes of detected communities have a great impact on co-performance. Since density is a measure that penalizes large size communities heavily, especially in sparse networks, $D$-modularity gives very small values for giant communities and very high values for small ones. Concretely, the methods \textit{SBM}, \textit{DCSBM}, \textit{Infomod}, \textit{RB}, \textit{SN} discover very large communities (as shown in Section \ref{sec:sizesimilarity}) and their co-performances in terms of $D$-modularity are very weak, showing that internal densities of communities detected by these methods are not linearly correlated. The reason is that the corresponding densities fluctuate unpredictably around zero. Similarly, \textit{GN} and \textit{RCCLP-3} found many tiny communities making the density either very high or zero (if the internal degree is equal to the external degree), consequently the co-performance index can not show significant information. On the other hand, we notice a consistency between the similarity of community size and the co-performance when methods identify medium-sized communities. Specifically, we find high co-performance indexes between \textit{CNM, Conclude, Oslom, Walktrap, LPA, SLPA, Infomap} methods in most of the cases of the six quality fitness functions. This finding exposes a global agreement with our categorization, determined by community size distributions.             

The co-performance matrices also disclose interesting information about quality functions. As we can see in Figure~\ref{fig:corr-matrix}(a,b,d), the matrices imply a similarity between \textit{NG} modularity, \textit{ER} modularity and $Z$-modularity in the assumptions of quality. In the same way, Surprise and Significance are quite close in practice, as illustrated in \ref{fig:corr-matrix}(d,e). This experiment gives more proof about the closeness between the theoretical assumptions of community structure and the practical outcome. Moreover, although being based on different aspects of goodness, the performance of many methods tends to reach agreement on the modular structure of networks in general. This is to say, methods in the same group identify roughly and globally comparable results, although there are always significant differences. In order to strengthen and validate our conclusion, we are interested in using other popular approaches in the literature to compare these community detection algorithms, which will be presented in the next section. 

\section{Partitioning strategy comparison}
\label{sec:validationproximity}
This section is dedicated to using conventional clustering validation metrics from the literature to verify the previous similarity analyses. We employ some popular metrics in the traditional clustering context (and also widely used in community detection context), which measure directly the similarity of partitions using their corresponding contingency tables. These metrics do not take into consideration the structural information of community structures, but only use the common numbers of nodes that are shared by pairs of communities in two partitions.

\subsection{Validation metrics}
\label{sec:similarityevaluation}
 The consensus of two partitions $P_1 = \{c_1^{(1)}, c_2^{(1)}, ..., c_R^{(1)}\}$ and $P_2 = \{c_1^{(2)}, c_2^{(2)}, ..., c_S^{(2)}\}$ can be more practically observed using a contingency table (sometimes called confusion matrix or association matrix), whose elements $n_{ij} = |c_{i}^{(1)} \cap c_{j}^{(2)}|$ correspond to the number of common vertices between the $i$-th community of $P_1$ and the $j$-th community of $P_2$ as shown in Table \ref{tab:contingencytable}.

\begin{table}[ht]
\centering
\caption{Contingency table of $P_1$ and $P_2$ on the same graph provides information about the similarity between the two partitions.}
\label{tab:contingencytable}
\begin{tabular}{cc|cccc|c}
\centering
\multirow{7}{*}{\vspace{-30pt} Partition $P_1$} &
\multicolumn{6}{c}{Partition $P_2$} \\
& \hspace{30pt} & \hspace{10pt} $c_1^{(2)}$ \hspace{10pt} & \hspace{10pt} $c_2^{(2)}$ \hspace{10pt} & \hspace{10pt} $\cdots$ & \hspace{10pt} $c_S^{(2)}$ \hspace{10pt} & \hspace{10pt} $\sum$ \hspace{10pt} \\
\cline{2-7}
& $c_1^{(1)}$ & $n_{11}$ & $n_{12}$ & $\cdots$ & $n_{1S}$ & $n_{1\boldsymbol{\cdot}}$ \\ 
& $c_2^{(1)}$ & $n_{21}$ & $n_{22}$ & $\cdots$ & $n_{2S}$ & $n_{2\boldsymbol{\cdot}}$ \\ 
& $\vdots$ & $\vdots$ & $\vdots$ & $\ddots$ & $\vdots$ & $\vdots$ \\
& $c_R^{(1)}$ & $n_{R1}$ & $n_{R2}$ & $\cdots$ & $n_{RS}$ & $n_{R\boldsymbol{\cdot}}$ \\ 
\cline{2-7}
& $\sum$  & $n_{\boldsymbol{\cdot}1}$ & $n_{\boldsymbol{\cdot}2}$ & $\cdots$ & $n_{{\boldsymbol{\cdot}}S}$ & $n$ \\ 
\end{tabular}
\end{table}

In the evaluation of community structure using a validation metric, some of the following validation metrics are often used in the context of community detection, to define the matching coefficient between two arbitrary partitions of a network.
\subsubsection{Rand Index (RI)}
The rand index is a pair-counting based measure, defined as the ratio of the number of vertex pairs correctly classified (either in the same community or in different communities) by the total number of pairs \cite{rand:1971a}. The RI penalizes both false positive and false negative decisions of the clusterings. When the false positive has to be neglected, we can refer to the \textit{Jaccard index} \cite{kuncheva:2004a}. The rand index value of two partitions can be calculated by the following:

\begin{equation}
RI(P_1,P_2) = \frac{{n\choose 2} + 2 \sum_i \sum_j {n_{ij}\choose 2} - \left[ \sum_i {n_{i\boldsymbol{\cdot}}\choose 2} + \sum_j {n_{\boldsymbol{\cdot}j}\choose 2}\right]}{{n\choose 2}}
\end{equation}

The value varies between $1$ (meaning the two partitions are identical) and $0$ (indicating that the two partitions do not agree on any pair of vertices). However, this value is only observed in the scenario when one partition consists of one community and the other consists of $n$ community of $1$ vertex, which has little practical value. Another shortcoming of the rand index is that its expected value for two randomly chosen partitions does not take a constant value, which is normally expected for a good matching index \cite{vinh:2010a}. Therefore, a modified version of RI has been suggested, taking into consideration the expected value of randomness \cite{hubert:1985a}, which is introduced in the following. 

\subsubsection{Adjusted Rand Index (ARI)}
The corrected version of rand index takes the form:
\begin{equation}
Adjust\_index = \frac{Index - Expected\_Index}{Max\_Index - Expected\_Index}
\end{equation}

It quickly becomes a replacement recommended for measuring agreement between two partitions in the analysis of clusterings. Its values ranges from $-1$ to $1$ indicating completely different and identical partitions, respectively. It is known to be less sensitive to the difference of the number of communities between two partitions. An ARI value of $0$ indicates that the similarity is equal to the expected value from randomly chosen partitions. It can be calculated as:

\begin{equation}
ARI(P_1,P_2) = \frac{\sum_{ij} {n_{ij}\choose 2} - \left[\sum_i {n_i\choose 2} \sum_j {n_j\choose 2}\right]/{n\choose 2}}{\frac{1}{2}\left[\sum_i {n_i\choose 2} + \sum_j {n_j\choose 2} \right] - \left[ \sum_i {n_i\choose 2} \sum_j {n_j\choose 2}\right]/{n\choose 2}}
\end{equation}

\subsubsection{Normalized Mutual Information (NMI)}
Information theoretic based metrics constitute another approach for validating community structure with a given reference partition. Using the same notations as previously presented, the Mutual Information (MI) between two partitions quantifying the mutual dependence is calculated as:

\begin{equation}
\label{eqn:mutualinformationpartition}
I(P_1,P_2) = \sum_{ij} p(c_{i}^{(1)},c_{j}^{(2)}) \log \frac{p(c_{i}^{(1)},c_{j}^{(2)})}{p(c_{i}^{(1)})p(c_{j}^{(2)})} = 
\sum_{ij} \frac{n_{ij}}{n} \log \frac{n_{ij}n}{n_{i\boldsymbol{\cdot}}n_{\boldsymbol{\cdot}j}}
\end{equation}

It measures how much knowing a partitioning of vertices in one way would reduce the uncertainty about the other way. In order words, it could be considered as an indicator of information \textit{closeness} expressed by the joint distribution between two variables. Therefore, the mutual information can be used as a similarity metric between two partitions. However, it needs to be normalized to reflect a consistency between different measures. The normalization uses the entropy of each partition defined by:

\begin{equation}
\label{eqn:entropypartition}
H(P) = - \sum_k \frac{n_k}{n} \log \frac{n_k}{n}
\end{equation}
Several variants of normalization can be considered, for instance taking the average, the root or the maximum of entropy of the two partitions as the denominator \cite{ana:2003a}. In this document, we use the average version which is widely used in the context of community analysis~\cite{danon:2005a,chakraborty:2017a}. The closed form of NMI is hence defined from Equation (\ref{eqn:mutualinformationpartition}) and (\ref{eqn:entropypartition}), as follows:

\begin{equation}
\label{eqn:nmi} 
NMI(P_1,P_2) = \frac{2I(P_1,P_2)}{H(P_1)+H(P_2)} = \frac{-2 \sum_{ij} n_{ij} \log \left( \frac{n_{ij}n}{n_in_j}\right)}{\sum_i n_i \log \left( \frac{n_i}{n} \right) + \sum_j n_j \log \left( \frac{n_j}{n} \right)}
\end{equation}

Likewise, the NMI similarity between two partitions varies between $0$, corresponding to independent relation, and $1$ when two partitions are identical. NMI does not follow triangle inequality.

\subsubsection{Adjust Mutual Information (AMI)}
Similarly to the Rand Index, Mutual Information is also subject to the effect of randomness, i.e., there is not a constant baseline value between random partitions of a graph. This issue raises many difficulties in the comparison mechanism, since it is expected that a comparative index should preserve the relativity between different clusterings and enhance intuitiveness about the mutual agreement. For that reason, traditional Mutual Information is normalized with a supplementary correction for chance. This normalization method has recently received attention for comparing graph partitions. It is calculated as follows \cite{vinh:2010a}:  

\begin{equation}
\label{eqn:ami}
AMI(P_1,P_2) = \frac{I(P_1,P_2) - E\{I(M)|n_{i\boldsymbol{\cdot}}n_{\boldsymbol{\cdot}j}\}}{\frac{1}{2} (H(P_1)+H(P_2)) - E\{I(M)|n_{i\boldsymbol{\cdot}}n_{\boldsymbol{\cdot}j}\}},
\end{equation}
where $I(P_1,P_2)$ and $H(P)$ are introduced in equations (\ref{eqn:mutualinformationpartition}) and (\ref{eqn:entropypartition}) respectively. $E\{I(M)|n_{i\boldsymbol{\cdot}}n_{\boldsymbol{\cdot}j}\}$ is the expected mutual information value of all feasible contingency tables constructed from the actual table $M$ with the same marginals $n_{i\boldsymbol{\cdot}}$, $n_{\boldsymbol{\cdot}j}$.

\subsubsection{Normalized Variation of Information (NVI)}
Another popular metric that is often used in the context of comparing community partition similarity is the Variation of Information (VI) \cite{meila:2003a}, which is defined as:

\begin{equation}
\label{eqn:variationofinfomation}
VI(P_1,P_2) = H(P_1) + H(P_2) - 2I(P_1,P_2)
\end{equation}

The VI metric can be interpreted as an index of shared information distance between two partitions. Its lower bound is $0$ and it occurs when the two partitions are identical, whereas the upper bound $\log(n)$ happens when they are completely different. It is also preferable to use a normalized version, chance-corrected, to avoid the effect of randomness. Similarly to the construction of the Adjusted Mutual Information, with the same notation, the Normalized Variation of Information is calculated as follows:

\begin{equation}
\label{eqn:nvi}
NVI(P_1,P_2) = \frac{H(P_1) + H(P_2) - 2I(P_1,P_2)}{H(P_1) + H(P_2) - 2E\{I(M)|n_{i\boldsymbol{\cdot}}n_{\boldsymbol{\cdot}j}\}}.
\end{equation}
However, it turns out that $NVI$ discloses the same information with $AMI$, since from Equation (\ref{eqn:ami}) and (\ref{eqn:nvi}), one has $NVI(P_1,P_2) = 1 - AMI(P_1,P_2)$. Consequently, calculating $VI$ and $NVI$ is unnecessary. We have chosen the $RI$, $ARI$, $NMI$, and $AMI$ metrics in our experiment. A summary of these validation metrics are shown in Table \ref{tab:validationmetrics}. 

\begin{table}[ht]
\centering
\caption{Some popular validation metrics for comparing community partitions}
\label{tab:validationmetrics}
\begin{tabular}{p{1.3cm} p{1.7cm} p{9.5cm}}
Label & Range & Measure \\
\hline\hline
RI & $[0,1]$ & Fraction of commonly grouped and separated vertices in two partitions.\\
ARI & $[0,1]$ & Rand index with a chance correction, less sensitive to differences of community sizes.\\
NMI & $[0,1]$ & Information theoretic approach, indicates how much information in one partition will help to guess the other.\\
AMI & $[0,1]$ & Similar to NMI, with a chance correction to set a constant baseline for random partitions.\\
VI & $[0,\log (n)]$ & Shared information distance measures the amount of mutual information. The higher the value, the less resemblance between the two partitions.\\
NVI & $[0,1]$ & Normalized version of shared information distance with chance correction.\\
\hline\hline
\end{tabular}
\end{table}

Validation metrics are often used in the context of community structure evaluation to measure the difference between the partition identified by a method with an expected partition of the network under consideration (\textit{ground-truth}). The more similar the discovered partition is to the ground-truth, the better the performance of the method. However, in this section, validation metrics are exploited as a tool to compare community structures of different detection methods. They estimate the practical proximity of different algorithms through detected partitions, which constitutes a supplementary source of information for evaluating their performance in a comparative approach. 

\subsection{Empirical results}
\begin{figure}[ht]
\centering
\caption[The similarity between community detection methods quantified by different validation metrics based on partitions discovered on networks of the dataset: (a). NMI, (b). AMI, (c). RI, (d). ARI.]{The similarity between community detection methods quantified by different validation metrics based on partitions discovered on networks of the dataset. Rows and columns are ordered according a hierarchical clustering method \cite{joe:1963a}. In order, the average score of (a). NMI, (b). AMI, (c). RI, (d). ARI.}
\label{fig:validation-proximity-matrix}
 \begin{subfigure}{0.4\textwidth}
  \centering
  \includegraphics[width=1\linewidth]{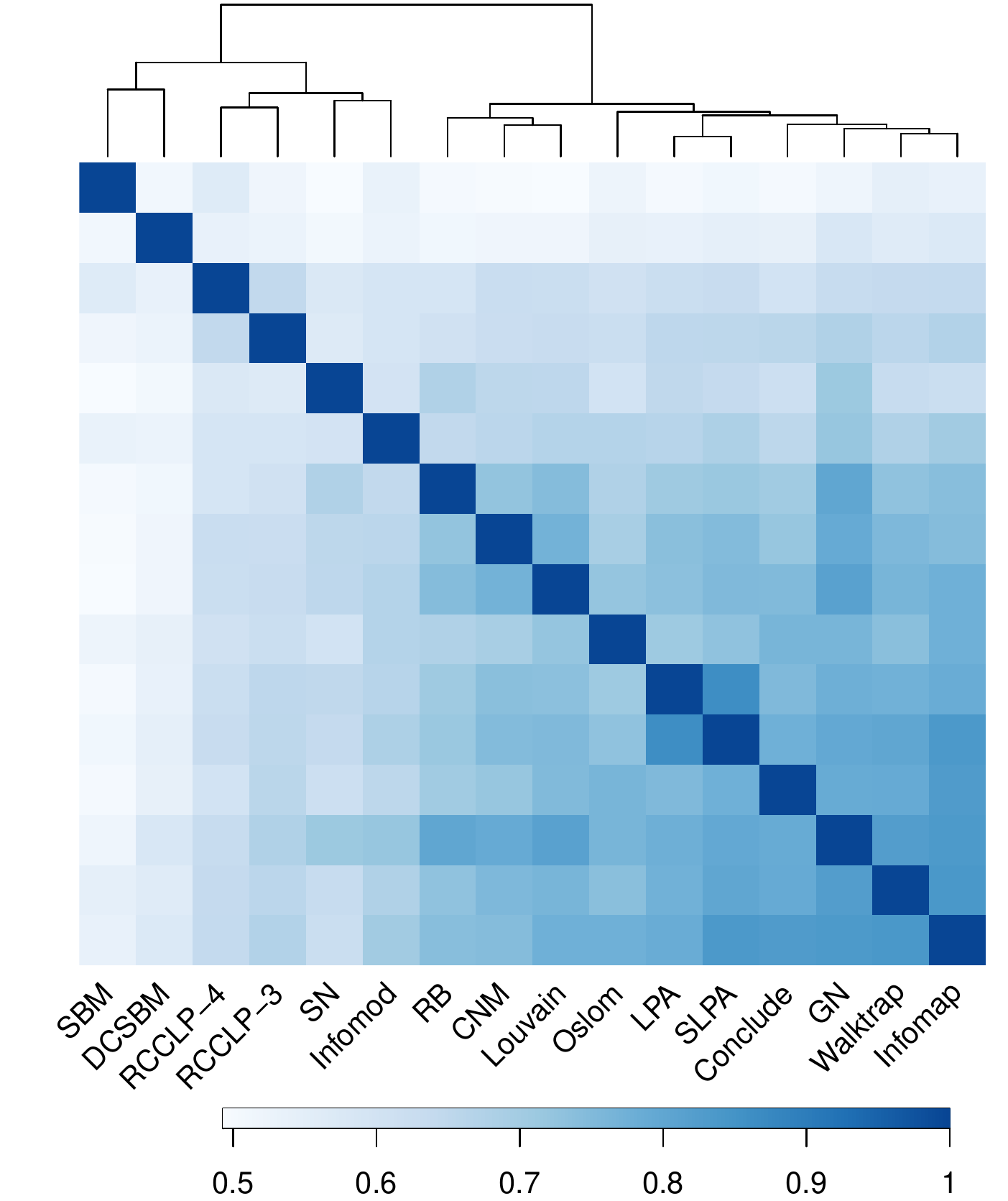}
  \caption{}
  \label{fig:nmi-matrix}
\end{subfigure}
\begin{subfigure}{0.4\textwidth}
\includegraphics[width=1\linewidth]{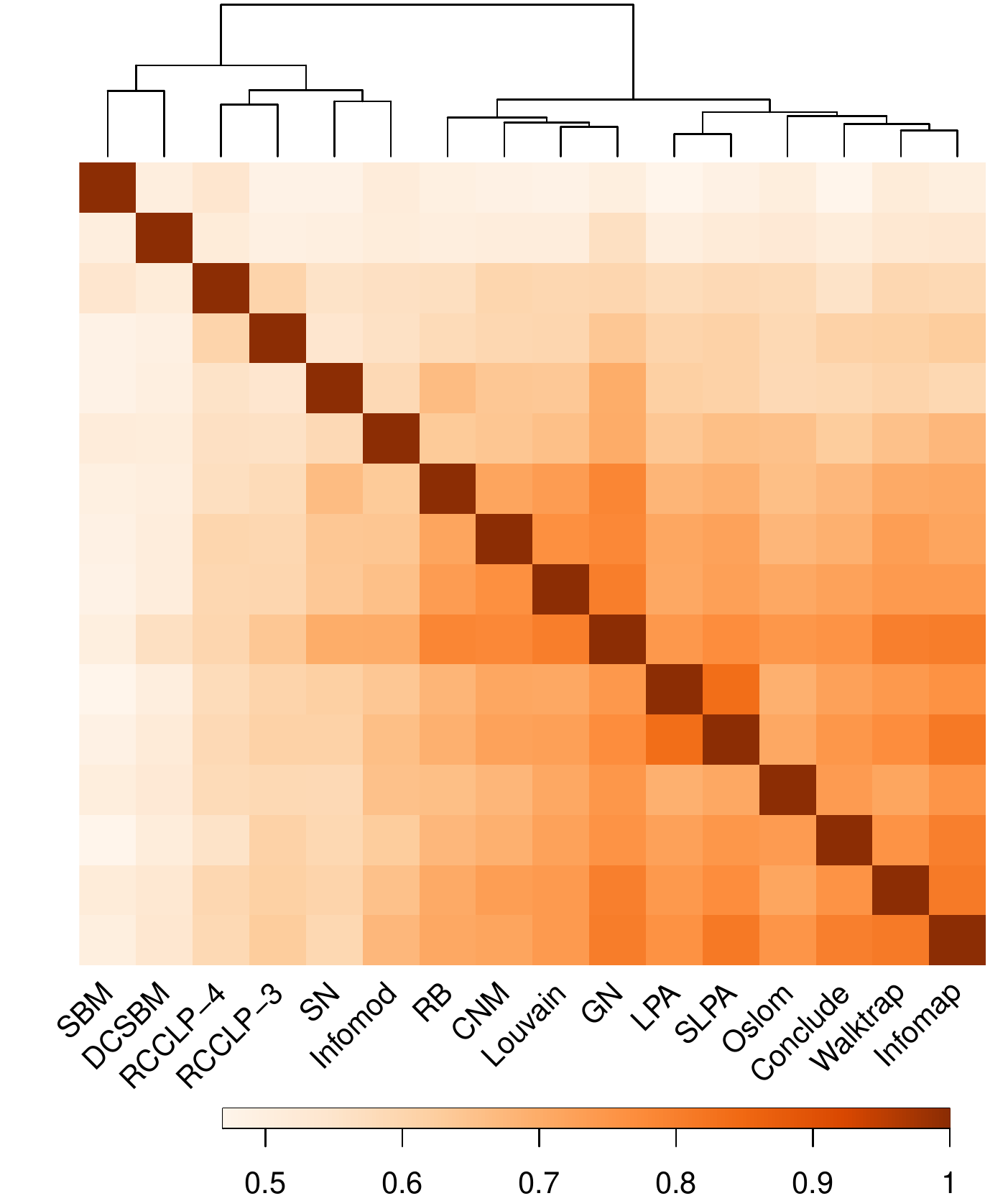}
    \caption{}
  \label{fig:ami-matrix}
\end{subfigure}

 \begin{subfigure}{0.4\textwidth}
  \centering
  \includegraphics[width=1\linewidth]{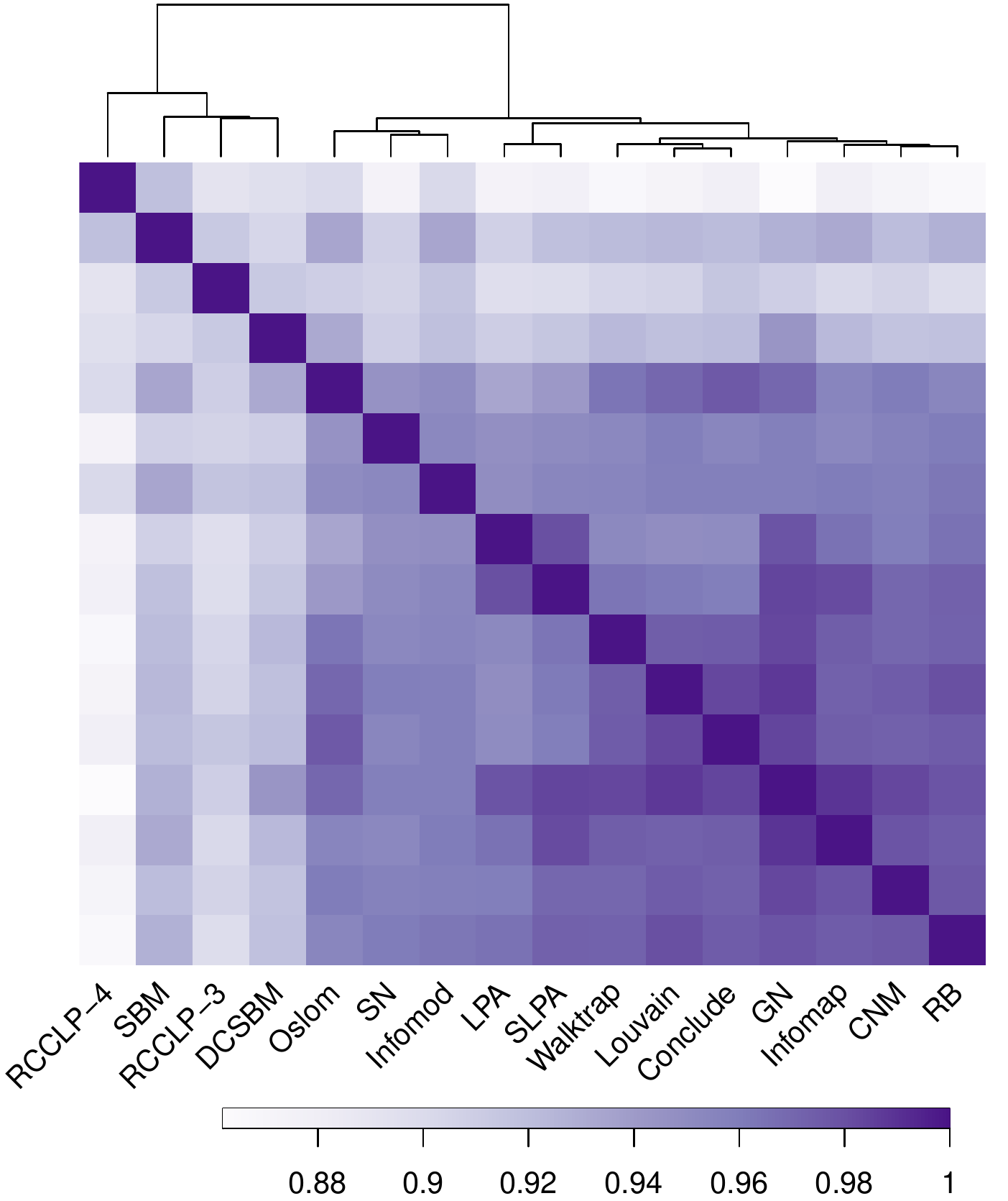}
  \caption{}
  \label{fig:ri-matrix}
\end{subfigure}
\begin{subfigure}{0.4\textwidth}
\includegraphics[width=1\linewidth]{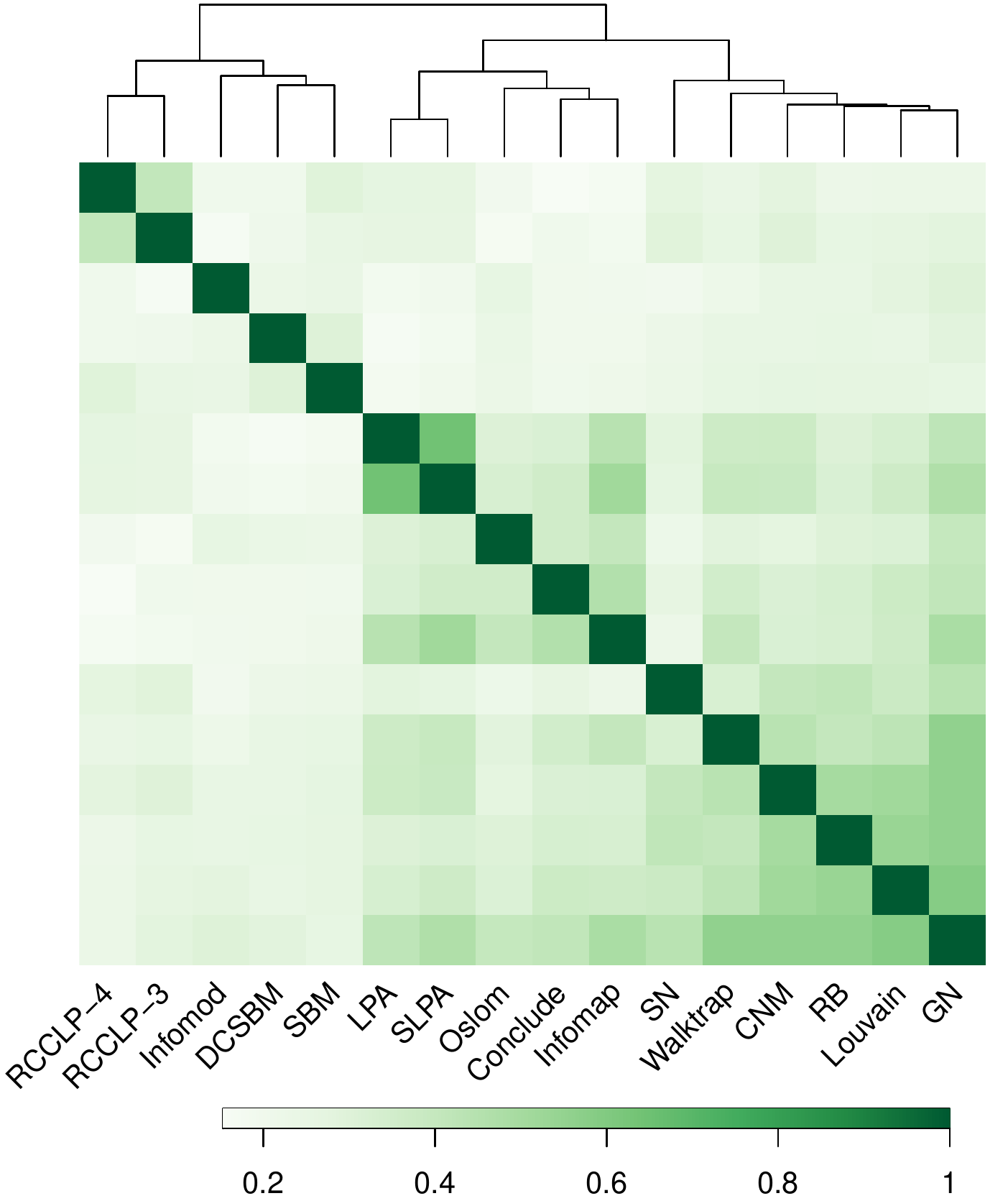}
    \caption{}
  \label{fig:ari-matrix}
\end{subfigure}
\end{figure}

Once again, the experimental process is the same as those of the previous sections. From the partitions detected by the methods on the dataset, we calculate pairwise scores, quantified by each validation metric on each network. Figure \ref{fig:validation-proximity-matrix} illustrates pairwise average scores of the 4 metrics over the networks of the dataset\footnote{Where the corresponding methods are able to finish using a reasonable amount of time and memory as mentioned in the previous experiments.}.

Again, by observing the dendrograms in Figure \ref{fig:validation-proximity-matrix}, one can see that all 4 metrics classify methods into two principle groups in a similar way to the co-performance matrices exposed in the previous section. The group of methods \textit{CNM, Conclude, Oslom, Walktrap, LPA, SLPA, Infomap} mentioned in the last section also show very strong similarities in this experiment. \textit{LPA} and \textit{SLPA}, especially, being based on label propagation mechanism show nearly identical results in many cases. In addition, one could discern another group including \textit{RB, CNM, GN} and \textit{Louvain} (modularity based), which shows high consistency in general. Additionally, even with weaker scores, \textit{SBM} and \textit{DCSBM} are often found in the same group, with \textit{RCCLP-3} and \textit{RCCLP-4} too. Globally, it seems that methods with a close theoretical approach tend to provide more similar results, which is also revealed in the previous sections.

Another fact that could be extracted from this experiment is that $RI$ should not be used as a validation metric for evaluating detection performance. Since its average values vary generally over a small range ($0.9$ to $1.0$), it is more difficult to see the difference between partitions. On the other hand, $NMI$ and $AMI$ show very close results in our experiment (between $0.5$ and $1.0$), meaning that structural communities detected by different methods are quite comparable, as concluded in the previous section. Finally, $ARI$ seems to magnify the differences between methods. However, there is no major difference in the similarity evaluation in comparison with the other metrics.

\section{Related work}
\label{sec:relatedwork}
Orman \textit{et al.} published a comparative evaluation of eight community detection algorithms \cite{orman:2012a}, most of which are also studied in this paper. Different validation metrics are used to evaluate the agreement between the discovered partitions and reference community structures (ground-truths). As in our work, they find that these metrics ($RI$, $ARI$, $NMI$) \textit{``agree with each other with small differences when considering the way they rank algorithms''}, as illustrated in Section \ref{sec:validationproximity}. Furthermore, the authors also focus on analyzing many topological aspects of community structure including transitivity, density,  community size, etc. These topological qualities are then used to inspect community structures detected by different algorithms. The analyses allow the authors to conclude that these two approaches (topological metrics and validation metrics), used to evaluate community structures, are \textit{``complementary and needed to perform a relevant and complete analysis of community detection results''}. They also note that the \textit{``traditional approach ($RI$, $ARI$, $NMI$) is much faster and easier to apply''}, and hence they advise using these metrics first. However, in practice, ground-truths are not usually available\footnote{In the context where a new algorithm is invented, one normally uses networks whose community structures are well known in order to validate the proposed method. In reality, since community detection is often employed to discover structures of new networks, it is therefore not likely that reference community structures always exist.}. Therefore, from the above observations, our analyses in this paper could be an important support, dispensing additional information about the closeness between methods, both in terms of the topological aspect and the partition-based aspect.      

Agreste \textit{et al.} evaluate different community detection algorithms in an empirical and comparative approach, especially for the context of web data analytic~\cite{agreste:2017a}. The authors find that \textit{``time complexity is a crucial factor in the selection of a community detection algorithm''} and find that the label propagation method (LPA) \textit{``has outstanding performance in scalability on both artificial and real graphs''}, which is also in  global agreement with our analysis in Section \ref{sec:timeanalysis} providing predictions about required time for each method, according to network size. They also conclude that \textit{``\textit{Infomap} algorithm showcased the best trade-off between accuracy and computational performance''} based on $NMI$ score. Such a conclusion is valid in some specific cases where the expected ground-truth community structure is well understood. Otherwise, some additional analyses should be done to determine whether \textit{ground-truth} information aligns the final objective of community detection algorithms \cite{peel:2017a}. Indeed, metadata information of nodes are often used in practice as a ground-truth community structure, whereas it has been found that metadata communities are sometimes very sparse \cite{dao:2017a}. One should be cautious as regards generalizing. Algorithms may have difficulty agreeing with some non-aligned metadata-based reference, but one should not assume this level of performance on other datasets.

Ghasemian \textit{et al.} present, in a recent publication, an evaluation of overfitting and underfitting of several community detection models~\cite{ghasemian:2018a}. The authors study the number of communities detected in practice by many methods, and the maximum number of detectable clusters, according to a theoretical model. Some conclusions are drawn about fitting qualities of methods in comparison to theoretical estimates. This study helps to choose an appropriate method according to a fitting quality. Community detection methods are also grouped in distinct families, based on their outputs on many real-world networks (similarly to our analysis in Section \ref{sec:similarityevaluation}) using $AMI$ metric. The authors also find that \textit{``what an algorithm finds in a network depends strongly on the assumptions it makes about what to look for''}, which is aligned with our results through several analyses.

Jebabli \textit{et al.} also propose a framework to assess the performance of community detection algorithms, based on the topological characteristics of the resulting community-graphs \cite{jebabli:2018a}. In their paper, the authors restrict their attention to networks with overlapping community structures, whereas our comparative evaluation concerns hard clustering. However, they discuss how their framework can be applied to networks with non-overlapping community structure. They present and evaluate an efficient alternative methodology, as compared to the classical quality and clustering measures. Similarly to our work, algorithms are compared in a decision-making scheme. They provide different rankings of the algorithms, according to different topological properties. They also introduce a Multiple-Criteria Decision-Making strategy in order to find the best compromise between these different rankings. A single ranking is produced. Such a decision making strategy is therefore a very interesting approach. Converting recommendations into a decision process is one of our perspective studies.

\section{Conclusion}
\label{sec:conclusion}
It is quite challenging to recommend which method is best for which scenario. It is at least as demanding as defining all possible scenarios. We carried out several experiments, demonstrating different aspects of community structure quality, which can be combined together in a flexible way to assist network analysts to find appropriate methods, according to the context. The following questions, for example, could be asked sequentially during decision making processes:

\begin{enumerate}
 \item What is the size of the network under consideration and what is the acceptable computation time for the task of community detection? 
 \item What are the expectations about the number of communities as well as the community size distribution?
 \item Is there any fitness function that should be optimized?
 \item In circumstances where the targeted method cannot be deployed, is there an alternative solution?
\end{enumerate}

The experiments and results in this paper could help to identify suitable method(s)  quickly, if one is able to answer the above questions. %Even still very far from being an exclusive analysis, our experiments cover a wide range of popular aspects of community structure that are studied in the state-of-the-art. Some primary conclusions that could be extracted from the analyses can be cited:

The consideration of computation time is indeed crucial in the process of choosing a community detection method for a problem at hand. Even if the theoretical estimate of time complexity is important and reveals the scalability of a community detection method, computation time is worth being studied in practice. Our estimates provide detailed information of practical time required by many popular community detection methods, according to network size. In particular the most scalable methods that we tested (\textit{Louvain}, \textit{LPA} and \textit{SN}) approximately reduce $10^4$ times the required computation time, compared with most of the other methods. This is not only significant but also crucial for a large network. Given a network size, our results help in filtering non suitable methods.
 
In addition, the expected number of communities to be obtained is another important criterion in choosing a community detection method. Depending on the context, one could prefer different granularity levels. Our study shows that there are globally three main strategies that community detection methods use to decompose a network. Specifically, some identify communities whose sizes vary regularly over a wide range of values, from very small to very large communities. Some others divide networks into a huge number of very small communities and very few large communities. And finally, the latter distribute nodes into similar medium-size communities (around 10 members). Therefore, knowing how a network should be broken down is very useful in order to end up with an appropriate community discovery method. 
 
In cases where (advanced) network analysts can determine a targeted objective function, designing new algorithms (or employing existing algorithms) that optimize the function would be the most effective. Since improving an objective function usually means expending more computation time, a compromise between getting higher fitness scores and using less time needs to be considered. However, finding a good method to optimize an objective function, satisfying a time constraint condition, is not straightforward and needs much investigation. Our approach, as presented in the co-performance analysis, provides network practitioners with a quick view of how different methods perform in improving some widely-used quality functions. This predictive information about the effect of using alternative methods in achieving good fitness scores would suggest multiple solutions for network analysts to reach a certain objective function. This scenario is specifically useful when the desired method is too expensive, in terms of computation time. Therefore, a combination of our empirical analyses about scalability and/or community size distribution with the co-performance index can identify eligible alternatives for specific cases.
 
Finally, we find that using some validation metrics to estimate the similarity between community detection methods could also provide interesting information that could help the decision process of network analysts. %However, the distinction between the performance of different methods is less significant than that of the previous analyses. 
In situations where one knows exactly (or has notions of) what should be found (ground-truth information), studying the way nodes are allocated to communities is important, as it provides useful information about how a method is able to reach the desired clusters. However, this scenario does not often occur, since community detection is in general used to discover the structures of networks when no a-priori information is available. When performing several methods is possible, validation metrics are used to compare their results and then identify different types of partitions. % and then verify the appropriateness of using a method. 
From our empirical study, we noticed significant differences in the way that nodes are distributed in communities. Methods such as \textit{SBM} or \textit{RCCLP-4}, especially, seem to detect partitions which are very discernible from those of the others. Hence, the use of these methods needs to be examined, and we shall recommend their use, along with other methods, as they may bring totally different, and probably complementary insights to the data.

\bibliography{./dao.bib}

\begin{thebibliography}{}

\bibitem[\protect\citename{Agreste {\em et~al.}\relax, }2017]{agreste:2017a}
Agreste, Santa, Meo, Pasquale~De, Fiumara, Giacomo, Piccione, Giuseppe,
  Piccolo, Sebastiano, Rosaci, Domenico, Sarne, Giuseppe M.~L., \& Vasilakos,
  Athanasios~V. (2017).
\newblock An empirical comparison of algorithms to find communities in directed
  graphs and their application in web data analytics.
\newblock {\em {IEEE} transactions on big data}, {\bf 3}(3), 289--306.

\bibitem[\protect\citename{Aldecoa \& Mar{\'{\i}}n, }2011]{aldecoa:2011a}
Aldecoa, Rodrigo, \& Mar{\'{\i}}n, Ignacio. (2011).
\newblock Deciphering network community structure by surprise.
\newblock {\em {PLoS} {ONE}}, {\bf 6}(9), e24195.

\bibitem[\protect\citename{Ames, }2013]{ames:2013a}
Ames, Brendan P.~W. (2013).
\newblock Guaranteed clustering and biclustering via semidefinite programming.
\newblock {\em Mathematical programming}, {\bf 147}(1-2), 429--465.

\bibitem[\protect\citename{Ana \& Jain, }2003]{ana:2003a}
Ana, L.N.F., \& Jain, A.K. (2003).
\newblock Robust data clustering.
\newblock {\em {IEEE} computer society conference on computer vision and
  pattern recognition. proceedings.}

\bibitem[\protect\citename{Arifin {\em et~al.}\relax, }2017]{arifin:2017a}
Arifin, S, Zulkardi, Putri, R I~I, Hartono, Y, \& Susanti, E. (2017).
\newblock Developing ill-defined problem-solving for the context of south
  sumatera.
\newblock {\em Journal of physics: Conference series}, {\bf 943}(dec), 012038.

\bibitem[\protect\citename{Blondel {\em et~al.}\relax, }2008]{blondel:2008a}
Blondel, Vincent~D, Guillaume, Jean-Loup, Lambiotte, Renaud, \& Lefebvre,
  Etienne. (2008).
\newblock Fast unfolding of communities in large networks.
\newblock {\em Journal of statistical mechanics: Theory and experiment}, {\bf
  2008}(10), P10008.

\bibitem[\protect\citename{Bohlin {\em et~al.}\relax, }2014]{bohlin:2014a}
Bohlin, Ludvig, Edler, Daniel, Lancichinetti, Andrea, \& Rosvall, Martin.
  (2014).
\newblock Community detection and visualization of networks with the map
  equation framework.
\newblock {\em Pages  3--34 of:} {\em Measuring scholarly impact}.
\newblock Springer International Publishing.

\bibitem[\protect\citename{Chakraborty {\em et~al.}\relax,
  }2017]{chakraborty:2017a}
Chakraborty, Tanmoy, Dalmia, Ayushi, Mukherjee, Animesh, \& Ganguly, Niloy.
  (2017).
\newblock Metrics for community analysis: A survey.
\newblock {\em Acm comput. surv.}, {\bf 50}(4), 1--37.

\bibitem[\protect\citename{Clauset {\em et~al.}\relax, }2004]{clauset:2004a}
Clauset, Aaron, Newman, M. E.~J., \& Moore, Cristopher. (2004).
\newblock Finding community structure in very large networks.
\newblock {\em Physical review e}, {\bf 70}(6).

\bibitem[\protect\citename{Cleveland, }1979]{cleveland:1979a}
Cleveland, William~S. (1979).
\newblock Robust locally weighted regression and smoothing scatterplots.
\newblock {\em Journal of the american statistical association}, {\bf 74}(368),
  829--836.

\bibitem[\protect\citename{Coscia {\em et~al.}\relax, }2011]{coscia:2011a}
Coscia, Michele, Giannotti, Fosca, \& Pedreschi, Dino. (2011).
\newblock A classification for community discovery methods in complex networks.
\newblock {\em Statistical analysis and data mining}, {\bf 4}(5), 512--546.

\bibitem[\protect\citename{Danon {\em et~al.}\relax, }2005]{danon:2005a}
Danon, Leon, D{\'{\i}}az-Guilera, Albert, Duch, Jordi, \& Arenas, Alex. (2005).
\newblock Comparing community structure identification.
\newblock {\em Journal of statistical mechanics: Theory and experiment}, {\bf
  2005}(09), P09008--P09008.

\bibitem[\protect\citename{Dao {\em et~al.}\relax, }2017]{dao:2017a}
Dao, Vinh~Loc, Bothorel, C{\'{e}}cile, \& Lenca, Philippe. (2017).
\newblock Community structures evaluation in complex networks: A descriptive
  approach.
\newblock {\em {The 3rd International Winter School and Conference on Network
  Science, NetSciX}},  11--19.

\bibitem[\protect\citename{Dao {\em et~al.}\relax, }2018a]{dao:2018a}
Dao, Vinh~Loc, Bothorel, C{\'{e}}cile, \& Lenca, Philippe. (2018a).
\newblock An empirical characterization of community structures in complex
  networks using a bivariate map of quality metrics.
\newblock {\em Arxiv e-prints}, June.

\bibitem[\protect\citename{Dao {\em et~al.}\relax, }2018b]{dao:2018b}
Dao, Vinh~Loc, Bothorel, C{\'{e}}cile, \& Lenca, Philippe. (2018b).
\newblock Estimating the similarity of community detection methods based on
  cluster size distribution.
\newblock  {\em The 7th international conference on complex networks and their
  applications}.
\newblock Springer International Publishing.

\bibitem[\protect\citename{Erd\H{o}s \& R\'{e}nyi, }1959]{erdos:1959a}
Erd\H{o}s, P., \& R\'{e}nyi, A. (1959).
\newblock {On random graphs, I}.
\newblock {\em Publicationes mathematicae (debrecen)}, {\bf 6}, 290--297.

\bibitem[\protect\citename{Fortunato \& Barthelemy, }2006]{fortunato:2006a}
Fortunato, S., \& Barthelemy, M. (2006).
\newblock Resolution limit in community detection.
\newblock {\em Proceedings of the national academy of sciences}, {\bf 104}(1),
  36--41.

\bibitem[\protect\citename{Fortunato, }2010]{fortunato:2010a}
Fortunato, Santo. (2010).
\newblock Community detection in graphs.
\newblock {\em Physics reports}, {\bf 486}(3-5), 75--174.

\bibitem[\protect\citename{Fortunato \& Hric, }2016]{fortunato:2016a}
Fortunato, Santo, \& Hric, Darko. (2016).
\newblock Community detection in networks: A user guide.
\newblock {\em Physics reports}, {\bf 659}(Nov.), 1--44.

\bibitem[\protect\citename{{Ghasemian} {\em et~al.}\relax,
  }2018]{ghasemian:2018a}
{Ghasemian}, A., {Hosseinmardi}, H., \& {Clauset}, A. (2018).
\newblock Evaluating overfit and underfit in models of network community
  structure.
\newblock {\em Arxiv e-prints}, Feb.

\bibitem[\protect\citename{Girvan \& Newman, }2002]{girvan:2002a}
Girvan, M., \& Newman, M. E.~J. (2002).
\newblock Community structure in social and biological networks.
\newblock {\em Proceedings of the national academy of sciences}, {\bf 99}(12),
  7821--7826.

\bibitem[\protect\citename{Holland {\em et~al.}\relax,
  }1983]{holland1983stochastic}
Holland, Paul~W, Laskey, Kathryn~Blackmond, \& Leinhardt, Samuel. (1983).
\newblock Stochastic blockmodels: First steps.
\newblock {\em Social networks}, {\bf 5}(2), 109--137.

\bibitem[\protect\citename{Hric {\em et~al.}\relax, }2014]{hric:2014a}
Hric, Darko, Darst, Richard~K., \& Fortunato, Santo. (2014).
\newblock Community detection in networks: Structural communities versus ground
  truth.
\newblock {\em Physical review e}, {\bf 90}(6).

\bibitem[\protect\citename{Hubert \& Arabie, }1985]{hubert:1985a}
Hubert, Lawrence, \& Arabie, Phipps. (1985).
\newblock Comparing partitions.
\newblock {\em Journal of classification}, {\bf 2}(1), 193--218.

\bibitem[\protect\citename{Jebabli {\em et~al.}\relax,
  }2015]{jebabli2015overlapping}
Jebabli, Malek, Cherifi, Hocine, Cherifi, Chantal, \& Hamouda, Atef. (2015).
\newblock Overlapping community detection versus ground-truth in amazon
  co-purchasing network.
\newblock {\em Pages  328--336 of:} {\em 2015 11th international conference on
  signal-image technology \& internet-based systems (sitis)}.
\newblock IEEE.

\bibitem[\protect\citename{Jebabli {\em et~al.}\relax, }2018]{jebabli:2018a}
Jebabli, Malek, Cherifi, Hocine, Cherifi, Chantal, \& Hamouda, Atef. (2018).
\newblock Community detection algorithm evaluation with ground-truth data.
\newblock {\em Physica a: Statistical mechanics and its applications}, {\bf
  492}, 651--706.

\bibitem[\protect\citename{Jerome, }2013]{jerome:2013a}
Jerome, Kunegis. (2013).
\newblock The koblenz network collection.
\newblock {\em Pages  1343--1350 of:} {\em Proceedings conference on world wide
  web companion}.

\bibitem[\protect\citename{Joe H.~Ward, }1963]{joe:1963a}
Joe H.~Ward, Jr. (1963).
\newblock Hierarchical grouping to optimize an objective function.
\newblock {\em Journal of the american statistical association}, {\bf 58}(301),
  236--244.

\bibitem[\protect\citename{Kullback \& Leibler, }1951]{kullback:1951a}
Kullback, S., \& Leibler, R.~A. (1951).
\newblock On information and sufficiency.
\newblock {\em The annals of mathematical statistics}, {\bf 22}(1), 79--86.

\bibitem[\protect\citename{Kuncheva \& Hadjitodorov, }2004]{kuncheva:2004a}
Kuncheva, L.I., \& Hadjitodorov, S.T. (2004).
\newblock Using diversity in cluster ensembles.
\newblock {\em Ieee international conference on systems, man and cybernetics}.

\bibitem[\protect\citename{Lambiotte, }2010]{lambiotte:2010a}
Lambiotte, R. (2010).
\newblock Multi-scale modularity in complex networks.
\newblock {\em 8th international symposium on modeling and optimization in
  mobile, ad hoc, and wireless networks}, May, 546--553.

\bibitem[\protect\citename{Lancichinetti {\em et~al.}\relax,
  }2010]{lancichinetti:2010a}
Lancichinetti, Andrea, Kivel\"{a}, Mikko, Saram\"{a}ki, Jari, \& Fortunato,
  Santo. (2010).
\newblock Characterizing the community structure of complex networks.
\newblock {\em {PLoS} {ONE}}, {\bf 5}(8), e11976.

\bibitem[\protect\citename{Lancichinetti {\em et~al.}\relax,
  }2011]{lancichinetti:2011a}
Lancichinetti, Andrea, Radicchi, Filippo, Ramasco, Jos{\'{e}}~J., \& Fortunato,
  Santo. (2011).
\newblock Finding statistically significant communities in networks.
\newblock {\em {PLoS} {ONE}}, {\bf 6}(4), e18961.

\bibitem[\protect\citename{Leskovec \& Krevl, }2014]{snapnets}
Leskovec, Jure, \& Krevl, Andrej. (2014).
\newblock {SNAP Datasets}: {Stanford} large network dataset collection.
\newblock June.

\bibitem[\protect\citename{Leskovec {\em et~al.}\relax, }2008]{leskovec:2008a}
Leskovec, Jure, Lang, Kevin~J., Dasgupta, Anirban, \& Mahoney, Michael~W.
  (2008).
\newblock Statistical properties of community structure in large social and
  information networks.
\newblock {\em Proceeding of the 17th international conference on world wide
  web - {WWW} 08}.

\bibitem[\protect\citename{Li {\em et~al.}\relax, }2008]{li:2008a}
Li, Zhenping, Zhang, Shihua, Wang, Rui-Sheng, Zhang, Xiang-Sun, \& Chen,
  Luonan. (2008).
\newblock Quantitative function for community detection.
\newblock {\em Physical review e}, {\bf 77}(3).

\bibitem[\protect\citename{Meil{\u{a}}, }2003]{meila:2003a}
Meil{\u{a}}, Marina. (2003).
\newblock Comparing clusterings by the variation of information.
\newblock {\em Learning theory and kernel machines},  173--187.

\bibitem[\protect\citename{Meo {\em et~al.}\relax, }2014]{demeo:2014a}
Meo, Pasquale~De, Ferrara, Emilio, Fiumara, Giacomo, \& Provetti, Alessandro.
  (2014).
\newblock Mixing local and global information for community detection in large
  networks.
\newblock {\em Journal of computer and system sciences}, {\bf 80}(1), 72--87.

\bibitem[\protect\citename{Miyauchi \& Kawase, }2016]{miyauchi:2016a}
Miyauchi, Atsushi, \& Kawase, Yasushi. (2016).
\newblock Z-score-based modularity for community detection in networks.
\newblock {\em {PLOS} {ONE}}, {\bf 11}(1), e0147805.

\bibitem[\protect\citename{Newman, }2006a]{newman:2006b}
Newman, M. E.~J. (2006a).
\newblock Finding community structure in networks using the eigenvectors of
  matrices.
\newblock {\em Phys. rev. e}, {\bf 74}(Sep), 036104.

\bibitem[\protect\citename{Newman, }2006b]{newman:2006a}
Newman, M. E.~J. (2006b).
\newblock Modularity and community structure in networks.
\newblock {\em Proceedings of the national academy of sciences}, {\bf 103}(23),
  8577--8582.

\bibitem[\protect\citename{{Newman}, }2010]{newman:2010a}
{Newman}, M.~E.~J. (2010).
\newblock {\em Networks: An introduction}.
\newblock Oxford University Press.

\bibitem[\protect\citename{Newman \& Girvan, }2004]{newman:2004a}
Newman, M. E.~J., \& Girvan, M. (2004).
\newblock Finding and evaluating community structure in networks.
\newblock {\em Physical review e}, {\bf 69}(2).

\bibitem[\protect\citename{Orman {\em et~al.}\relax, }2012]{orman:2012a}
Orman, G\"{u}nce~Keziban, Labatut, Vincent, \& Cherifi, Hocine. (2012).
\newblock Comparative evaluation of community detection algorithms: a
  topological approach.
\newblock {\em Journal of statistical mechanics: Theory and experiment}, {\bf
  2012}(08), P08001.

\bibitem[\protect\citename{Papadopoulos {\em et~al.}\relax,
  }2011]{papadopoulos:2011a}
Papadopoulos, Symeon, Kompatsiaris, Yiannis, Vakali, Athena, \& Spyridonos,
  Ploutarchos. (2011).
\newblock Community detection in social media.
\newblock {\em Data mining and knowledge discovery}, {\bf 24}(3), 515--554.

\bibitem[\protect\citename{Peel {\em et~al.}\relax, }2017]{peel:2017a}
Peel, Leto, Larremore, Daniel~B., \& Clauset, Aaron. (2017).
\newblock The ground truth about metadata and community detection in networks.
\newblock {\em Science advances}, {\bf 3}(5), e1602548.

\bibitem[\protect\citename{Pons \& Latapy, }2005]{pons:2005a}
Pons, Pascal, \& Latapy, Matthieu. (2005).
\newblock Computing communities in large networks using random walks.
\newblock {\em Pages  284--293 of:} Yolum, pInar, G{\"u}ng{\"o}r, Tunga,
  G{\"u}rgen, Fikret, \& {\"O}zturan, Can (eds), {\em Computer and information
  sciences - iscis 2005}.
\newblock Springer Berlin Heidelberg.

\bibitem[\protect\citename{Pons \& Latapy, }2011]{pons:2011a}
Pons, Pascal, \& Latapy, Matthieu. (2011).
\newblock Post-processing hierarchical community structures: Quality
  improvements and multi-scale view.
\newblock {\em Theoretical computer science}, {\bf 412}(8-10), 892--900.

\bibitem[\protect\citename{{Porter} {\em et~al.}\relax, }2009]{porter:2009a}
{Porter}, M.~A., {Onnela}, J.-P., \& {Mucha}, P.~J. (2009).
\newblock {Communities in Networks}.
\newblock {\em Notices of the american mathematical society}, Feb.

\bibitem[\protect\citename{Radicchi {\em et~al.}\relax, }2004]{radicchi:2004a}
Radicchi, F., Castellano, C., Cecconi, F., Loreto, V., \& Parisi, D. (2004).
\newblock Defining and identifying communities in networks.
\newblock {\em Proceedings of the national academy of sciences}, {\bf 101}(9),
  2658--2663.

\bibitem[\protect\citename{Raghavan {\em et~al.}\relax, }2007]{raghavan:2007a}
Raghavan, Usha~Nandini, Albert, R{\'{e}}ka, \& Kumara, Soundar. (2007).
\newblock Near linear time algorithm to detect community structures in
  large-scale networks.
\newblock {\em Physical review e}, {\bf 76}(3).

\bibitem[\protect\citename{Rand, }1971]{rand:1971a}
Rand, William~M. (1971).
\newblock Objective criteria for the evaluation of clustering methods.
\newblock {\em Journal of the american statistical association}, {\bf 66}(336),
  846--850.

\bibitem[\protect\citename{Reichardt \& Bornholdt, }2006]{reichardt:2006a}
Reichardt, J\"{o}rg, \& Bornholdt, Stefan. (2006).
\newblock Statistical mechanics of community detection.
\newblock {\em Physical review e}, {\bf 74}(1).

\bibitem[\protect\citename{Riolo {\em et~al.}\relax, }2017]{riolo:2017a}
Riolo, Maria~A., Cantwell, George~T., Reinert, Gesine, \& Newman, M. E.~J.
  (2017).
\newblock Efficient method for estimating the number of communities in a
  network.
\newblock {\em Physical review e}, {\bf 96}(3).

\bibitem[\protect\citename{Rossi \& Ahmed, }2015]{rossi:2015a}
Rossi, Ryan~A., \& Ahmed, Nesreen~K. (2015).
\newblock The network data repository with interactive graph analytics and
  visualization.
\newblock  {\em Proceedings of the twenty-ninth aaai conference on artificial
  intelligence}.

\bibitem[\protect\citename{Rosvall \& Bergstrom, }2007]{rosvall:2007a}
Rosvall, M., \& Bergstrom, C.~T. (2007).
\newblock An information-theoretic framework for resolving community structure
  in complex networks.
\newblock {\em Proceedings of the national academy of sciences}, {\bf 104}(18),
  7327--7331.

\bibitem[\protect\citename{{Rosvall} {\em et~al.}\relax, }2009]{rosvall:2009a}
{Rosvall}, M., {Axelsson}, D., \& {Bergstrom}, C.~T. (2009).
\newblock The map equation.
\newblock {\em European physical journal special topics}, {\bf 178}(Nov.),
  13--23.

\bibitem[\protect\citename{Schaub {\em et~al.}\relax, }2017]{schaub:2017a}
Schaub, Michael~T., Delvenne, Jean-Charles, Rosvall, Martin, \& Lambiotte,
  Renaud. (2017).
\newblock The many facets of community detection in complex networks.
\newblock {\em Applied network science}, {\bf 2}(1).

\bibitem[\protect\citename{Silverman, }1986]{silverman:1986a}
Silverman, B.~W. (1986).
\newblock {\em Density estimation for statistics and data analysis}.
\newblock London New York: Chapman and Hall.

\bibitem[\protect\citename{Traag {\em et~al.}\relax, }2011]{traag:2011a}
Traag, V.~A., Dooren, P.~Van, \& Nesterov, Y. (2011).
\newblock Narrow scope for resolution-limit-free community detection.
\newblock {\em Physical review e}, {\bf 84}(1).

\bibitem[\protect\citename{Traag {\em et~al.}\relax, }2015]{traag:2015a}
Traag, V.~A., Aldecoa, R., \& Delvenne, J.-C. (2015).
\newblock Detecting communities using asymptotical surprise.
\newblock {\em Physical review e}, {\bf 92}(2).

\bibitem[\protect\citename{Vinh {\em et~al.}\relax, }2010]{vinh:2010a}
Vinh, Nguyen~Xuan, Epps, Julien, \& Bailey, James. (2010).
\newblock Information theoretic measures for clusterings comparison: Variants,
  properties, normalization and correction for chance.
\newblock {\em J. mach. learn. res.}, {\bf 11}(Dec.), 2837--2854.

\bibitem[\protect\citename{Wasserman, }1994]{wasserman:1994a}
Wasserman, Stanley. (1994).
\newblock {\em Social network analysis : methods and applications}.
\newblock Cambridge New York: Cambridge University Press.

\bibitem[\protect\citename{Xie \& Szymanski, }2012]{xie:2012a}
Xie, Jierui, \& Szymanski, Boleslaw~K. (2012).
\newblock Towards linear time overlapping community detection in social
  networks.
\newblock {\em Pages  25--36 of:} {\em Advances in knowledge discovery and data
  mining}.
\newblock Springer Berlin Heidelberg.

\bibitem[\protect\citename{Yang \& Leskovec, }2013]{yang:2013a}
Yang, Jaewon, \& Leskovec, Jure. (2013).
\newblock Defining and evaluating network communities based on ground-truth.
\newblock {\em Knowledge and information systems}, {\bf 42}(1), 181--213.

\end{thebibliography}

\label{lastpage}

\end{document}